\documentclass[a4paper, 12pt]{article}
\usepackage{setspace}
\usepackage{lipsum}
\usepackage[hmargin=0.9in,vmargin=1.1in]{geometry}
\usepackage{amssymb,amsmath,amsthm}
\usepackage{hyperref}
\usepackage{graphicx}
\usepackage[utf8]{inputenc}
\usepackage{algorithmic}
\usepackage{algorithm}
\usepackage{feynmf}
\usepackage[dvipsnames]{xcolor}
\usepackage[english]{babel}
\usepackage[autostyle]{csquotes}
\usepackage[italicdiff]{physics}
\usepackage{cancel}
\usepackage{graphicx}
\usepackage{mathtools} 
\usepackage{eqparbox}
\usepackage{enumerate}
\usepackage{svg}
\usepackage{wrapfig}
\usepackage{tikz}
\usepackage{rotating}
\usepackage[maxbibnames=6,style=numeric-comp,sorting=none]{biblatex}
\usepackage{mathrsfs} 
\usepackage{cleveref}
\usepackage[T1]{fontenc}
\usepackage{pdfpages}
\usepackage{adjustbox}
\usepackage{soul}
\newcommand{\Disc}[2]{\underset{#2}{\text{Disc}}\left({#1}\right)}
\newcommand{\disct}[2]{\underset{#2}{\widetilde{\text{disc}}}\left({#1}\right)}
\newcommand{\Disct}[2]{\underset{#2}{\widetilde{\text{Disc}}}\left({#1}\right)}
\newcommand{\disc}[2]{\underset{#2}{\text{disc}}\left({#1}\right)}

\def\a{\hat{a}}
\def\ad{\a^{\dagger}}
\def\ophi{\hat{\varphi}}

\def\mk{\mathbf{k}}
\def\mmp{\mathbf{p}}

\def\p{\mathbf{p}}

\newcommand{\expvalbar}[1]{\left\langle \overline{ #1 } \right\rangle}

\numberwithin{equation}{section}

\definecolor{ForestGreen}{rgb}{0.13, 0.55, 0.13}
\definecolor{airforceblue}{rgb}{0.36, 0.54, 0.66}
\definecolor{orange}{rgb}{1.0, 0.5, 0.0}
\definecolor{amethyst}{rgb}{0.6, 0.4, 0.8}
\definecolor{awesome}{rgb}{1.0, 0.13, 0.32}
\definecolor{chromeyellow}{rgb}{1.0, 0.65, 0.0}

\begin{document}

\title{\vspace{-18mm}\Large\textbf{Schwinger-Keldysh Cosmological Cutting Rules}}
\author{\normalsize Francisco Colip\'i-Marchant$^{a,b}$, Gabriel Marin$^{b,c,d}$, Gonzalo A. Palma$^{b}$, and Francisco Rojas$^{d}${}\\[2mm]
\normalsize{$^a$\emph{\small{Facultad de Ingenier\'ia, Universidad San Sebasti\'an, Santiago 8420524, Chile.}}}\\
\normalsize{$^b$\emph{\small{Departamento de F\'isica, FCFM, Universidad de Chile,
Blanco Encalada 2008, Santiago, Chile.}}}\\
\normalsize{\emph{\small{
\begin{tabular}{c}
$^c$Dipartimento di Fisica, Universit\`a di Torino, and INFN Sezione di Torino, \\ Via P.~Giuria 1, I-10125 Torino, Italy.
\end{tabular}
}}}\\
\normalsize{$^d$\emph{\small{Facultad de Ingenier\'ia y Ciencias, Universidad Adolfo Ib\'a\~nez, Santiago, Chile.}}}\\}
\date{}
\maketitle
\begin{abstract}
    In this work, we study the realisation of unitarity-based cutting rules for primordial cosmological correlators computed within the Schwinger-Keldysh path integral formalism. While cutting rules have been previously derived for wavefunction coefficients, here we examine them directly at the level of cosmological observables expressed diagrammatically. The resulting rules closely resemble those familiar from flat-space scattering amplitudes, but with an additional subtlety: in order to express the discontinuity of a correlator as the product of lower-order correlators, one must introduce a specific combinations of diagrams which do not appear in the computation of observables themselves. We explicitly verify these rules for several classes of correlators, both at tree level and with loop corrections, arising from theories involving different types of interactions.
\end{abstract}


\newpage
\tableofcontents

\section{Introduction}

Cosmological correlators play a central role in the theoretical description of the early universe, as they encode the statistical properties of primordial fluctuations generated during inflation. Quantum fluctuations of primordial fields (most notably the curvature perturbation $\zeta$ and tensor fluctuations $\gamma_{ij}$) are produced during this epoch and ultimately seed the large-scale structures observed today \cite{Guth:1980zm,Linde:1981mu,Starobinsky:1982ee}. Observations of the cosmic microwave background (CMB) indicate that these primordial fluctuations are well described, to leading order, by a Gaussian probability distribution, fully characterized by the two-point function of  $\zeta$, or equivalently, by its power spectrum in Fourier space. Nevertheless, higher-order correlations arising from departures from Gaussianity offer a powerful probe of new physics beyond the simplest inflationary scenarios \cite{Achucarro:2022qrl,Allen:1987vq,Baumann:2014nda,Maldacena:2002vr,Dalal:2007cu,Chen:2009zp,Green:2020whw,Meerburg:2019qqi,Palma:2019lpt}. A prominent example is a non-vanishing three-point correlation function (bispectrum) of CMB temperature fluctuations, which provides a direct observational signature of primordial non-Gaussianity \cite{Bartolo:2004if}.

A powerful framework to compute cosmological correlators is provided by the in-in, or Schwinger-Keldysh (SK), formalism \cite{Weinberg:2005vy}. While conceptually well suited to cosmology, this approach often leads to technically challenging analytical expressions, primarily due to the presence of nested time integrals. An alternative approach is offered by the cosmological bootstrap program, which aims to determine correlation functions without reference to a specific model of inflation. Instead, this approach relies on basic physical principles such as locality, unitarity, and consistency with spacetime symmetries \cite{Maldacena:2011nz,Mata:2012bx,Bzowski:2011ab,Bzowski:2012ih,Kundu:2014gxa,Kundu:2015xta,Arkani-Hamed:2015bza,Shukla:2016bnu,Arkani-Hamed:2018kmz,Baumann:2019oyu,Sleight:2019hfp,Pajer:2020wnj,Pajer:2020wxk,Pimentel:2022fsc,Baumann:2020dch,Baumann:2021fxj,Meltzer:2021zin,Hogervorst:2021uvp,DiPietro:2021sjt,Green:2020ebl,Cabass:2021fnw,Green:2023ids,Baumann:2022jpr}. Unitarity, in particular, imposes powerful constraints on perturbative observables in any spacetime setting. In flat Minkowski space, where well-defined asymptotic in- and out-states exist, the consequences of unitarity are encoded in the optical theorem. Within perturbation theory, this theorem relates the discontinuity of loop amplitudes to sums over products of lower-order on-shell amplitudes, allowing higher-order corrections to be systematically reconstructed from lower-order data. This procedure is formalised by Cutkosky’s cutting rules \cite{Cutkosky:1960sp,Veltman:1994wz,tHooft:1973wag}. A particularly powerful extension is known as generalised unitarity, in which multiple internal propagators are simultaneously placed on shell. This method applies to both non-supersymmetric and supersymmetric theories, including non-planar contributions, and has proven instrumental in simplifying loop calculations in quantum field theory \cite{Bern:1994zx,Bern:1996je,Bern:1997sc,Britto:2004nc,Bern:2011qt}.

In primordial cosmology, the implications of unitarity \cite{Cespedes:2020xqq,Goodhew:2020hob,Jazayeri:2021fvk,Melville:2021lst,Goodhew:2021oqg,Baumann:2021fxj,Meltzer:2021zin,DiPietro:2021sjt,Cabass:2021fnw,Tong:2021wai,Qin:2023bjk,AguiSalcedo:2023nds,Albayrak:2023hie,Stefanyszyn:2023qov,Ghosh:2024aqd,DuasoPueyo:2024rsa,Ema:2024hkj,Donath:2024utn,Stefanyszyn:2024msm,Ghosh:2025pxn,Jain:2025maa,Lee:2025kgs,Thavanesan:2025ibm, Palma:2026qgn} are less developed than in flat space. This is largely due to the fact that cosmological spacetimes of interest are approximately de Sitter, where time-translation invariance is absent and there are no non-dynamical asymptotic boundaries with which to define an $S$-matrix in the traditional sense. Nevertheless, as emphasised in \cite{Melville:2021lst,Goodhew:2021oqg}, unitarity still implies a systematic set of cosmological cutting rules that constrain wavefunction coefficients for an arbitrary number of fields and at any loop order. These rules relate the discontinuity of an $n$-loop diagram to lower-loop contributions, and determine the discontinuity of tree-level diagrams in terms of those with fewer external legs.

Subsequent work translated these relations into statements for in-in correlation functions, either by working within an equivalent in-out formalism \cite{Donath:2024utn} or by adopting a retarded/advanced ($r/a$) basis for the Schwinger-Keldysh path integral \cite{Ema:2024hkj}. Approximate cutting relations for correlation functions, including reducible effective field theory truncation errors as well as irreducible errors arising from specific integrals, were also explored in \cite{Tong:2021wai}. Beyond their conceptual significance, making these cuts explicit within the in-in formalism offers clear practical advantages: once the cuts are extracted, time integrals factorize, dramatically simplifying perturbative calculations \cite{Qin:2022lva}.

In this work, we aim to deepen the understanding of the role of cutting rules in cosmological correlators by deriving relations that apply directly to in-in correlators computed using the Schwinger-Keldysh formalism. The paper is organised as follows. In Section~\ref{sec:sk}, we briefly review the Schwinger-Keldysh formalism for computing cosmological correlators. Section~\ref{sec:cot} provides a concise overview of the implications of unitarity in both flat and de Sitter spacetimes. In Section~\ref{sec:toy-model}, we explicitly demonstrate the cutting behaviour of an $s$-channel diagram, whose analytic treatment requires an algebraic tool referred to as barred correlators. Section~\ref{sec:deriv} extends this analysis to interactions involving spatial and temporal derivatives. In Section~\ref{sec:general-tree-lvl-rule}, we generalise the results to arbitrary tree-level diagrams with a linear $V$-vertex configuration, providing a systematic recipe for performing cuts in this class of diagrams. In this context, we also extend the definition of barred correlators required for general tree-level topologies and present a representative diagrammatic example. Section~\ref{sec:cutting-loops} analyses general one-loop correlators with $V$ vertices, revealing a cutting structure analogous to the tree-level case. Finally, we illustrate these results by explicitly computing a one-loop example with $V=2$.

\paragraph{Note added:} While completing this article, we became aware of the very recent work~\cite{Das:2025qsh}, which exhibits significant conceptual and technical overlap with the approach developed here. Our analysis is based on the Master’s research theses available in~\cite{MarinMacedo:2025jco} and \cite{ColipiMarchant}. We thank the authors of~\cite{Das:2025qsh} for acknowledging the overlap between their work and~\cite{MarinMacedo:2025jco}.


\section{Schwinger-Keldysh Formalism} \label{sec:sk}
We begin with a review of the essential elements that we will need from the SK formalism for a real scalar field $\varphi$, a more complete and detailed explanation can be found in reference \cite{Chen:2017ryl}.

Cosmological correlation functions are calculated in a similar way to amplitudes in quantum field theory. An immediate subtlety is that Poincaré symmetry is explicitly broken by the evolving background metric, this is quantitatively seen by the presence of the time dependent scale factor $a(t)$, leading to an explicit temporal dependence to cosmological observables. Furthermore, in cosmology there is no defined \enquote{out} states, nonetheless it is possible to obtain correlation functions in a diagrammatic way, although with some subtleties. The quantity of interest, namely equal time correlation function, can be expressed as,
\begin{equation}
    \expval{Q(\tau)} \equiv \matrixel{\Omega}{\mathcal{O}_{1}(\tau,\vb{x}_{1}) \cdots \mathcal{O}_{n}(\tau,\vb{x}_{n})}{\Omega},
\end{equation}
where $\{\mathcal{O}_{i}\}$ are a set of operators constructed from the field operators in the Lagrangian $\mathcal{L}$, and $\ket{\Omega}$ is usually taken to be the vacuum state. Consequently, it is possible to construct a path integral representation for $\expval{Q}$ as follows,
\begin{multline}
    \expval{Q}=\int\mathcal{D}\varphi_{+} \mathcal{D}\varphi_{-} \, \prod_{j=1}^{N}\varphi^{A_{j}}_{+}(\tau,\vb{x}_{j})
    \exp\bigg[i\int_{-\infty}^{\tau_f}\dd\tau\dd[3]\vb{x}\,\Big(\mathcal{L}_\text{cl}[\varphi_+]-\mathcal{L}_{\text{cl}}[\varphi_-]\Big)\bigg]\times \\
    \times\prod_{A,\vb{x}}\delta\Big(\varphi_{+}^{A}(\tau_f,\vb{x})-\varphi_{-}^{A}(\tau_f,\vb{x})\Big).
    \label{correlation-sk}
\end{multline}
Where $\varphi_{+}^{A}(\tau_{f},\vb{x})$ and $\varphi_{-}^{A}(\tau_{f},\vb{x})$ correspond to time-ordered and anti-time-ordered field configurations, respectively, and are stitched at $\tau_{f}$ by the delta product in the last line.

At this point, quantum field theory methods are applied to write the path integral and to perform perturbative calculations by expanding the Lagrangian density into its free $\mathcal{L}_{0}$ and interacting part $\mathcal{L}_{\text{int}}$, to obtain the correlation functions, the main difference being that we now require two fields $\varphi_{\pm}$, and thus two sources $J_{\pm}$, allowing to write the generating functional as follows,
\begin{align}
    & Z[J_{+},J_{-}] = \exp[i\int\dd\tau\dd[3]\vb{x}\,\left( \mathcal{L}_{\text{int}}\left[ \dfrac{\delta}{i\delta J_{+}} \right] - \mathcal{L}_{\text{int}}\left[ \dfrac{\delta}{i\delta J_{-}} \right]Z_{0}[J_{+},J_{-}] \right)] \\
    & Z_{0}[J_{+},J_{-}] \equiv \int\mathcal{D}\varphi_{+} \mathcal{D}\varphi_{-}\,\exp[i\int\dd\tau\dd[3]\vb{x}\,\Big(\mathcal{L}_\text{0}[\varphi_+]-\mathcal{L}_{\text{0}}[\varphi_-] + J_{+}\varphi_{+} - J_{-}\varphi_{-} \Big)] &
\end{align}
The corresponding tree-level propagators, as a function of time and space coordinates $\tau_{1}, \tau_{2}, \vec{x}_{1}, \vec{x_{2}}$ are defined as,
\begin{equation}
    -i\Delta_{ab}(\tau_{1},\vec{x}_{1};\tau_{2},\vec{x}_{2}) = \dfrac{1}{ia}\fdv{J_{a}(\tau_{1},\vec{x}_{1})}\dfrac{1}{ib}\fdv{J_{b}(\tau_{2},\vec{x}_{2})}\eval{Z_{0}[J_{+},J_{-}]}_{J_{\pm}=0}.
\end{equation}
Given that $a,b=\pm$, there are 4 types of propagators:
\begin{subequations}
    \begin{align}
        -i\Delta_{++} & = \expval{T\{\varphi(\tau_{1},\vec{x}_{1})\varphi(\tau_{2},\vec{x}_{2})\}}{\Omega}, \\
        -i\Delta_{--} & = \expval{\bar{T}\{\varphi(\tau_{1},\vec{x}_{1})\varphi(\tau_{2},\vec{x}_{2})\}}{\Omega}, \\
        -i\Delta_{+-} & = \expval{\varphi(\tau_{2},\vec{x}_{2})\varphi(\tau_{1},\vec{x}_{1})}{\Omega}, \\
        -i\Delta_{-+} & = \expval{\varphi(\tau_{1},\vec{x}_{1})\varphi(\tau_{2},\vec{x}_{2})}{\Omega},
    \end{align}
    \label{sk-prop-1}
\end{subequations}
where $T$ and $\bar{T}$ provide temporal and anti-temporal ordering, respectively.

It is practical to go to the 3-momentum space due to translational and rotational symmetries on each time slice, the propagators will then be,
\begin{equation}
    G_{ab}(k;\tau_{1},\tau_{2}) = -i\int\dd[3]x\,e^{-i\vb{k}\vb{x}}\Delta_{ab}(\tau_{1},\vec{x}_{1};\tau_{2},\vb{0}).
\end{equation}
Note that momentum dependence $k = |\vb{k}|$ exists, yet there is no dependence on the directions, resulting from rotational symmetry. We can then expand the propagators,
\begin{subequations}
    \begin{align}
        G_{++} & = G_{>}(k;\tau_{1},\tau_{2})\Theta(\tau_{1}-\tau_{2}) + G_{<}(k;\tau_{1},\tau_{2})\Theta(\tau_{2}-\tau_{1}), \\
        G_{+-} & = G_{<}(k;\tau_{1},\tau_{2}), \\
        G_{-+} & = G_{>}(k;\tau_{1},\tau_{2}), \\
        G_{--} & = G_{<}(k;\tau_{1},\tau_{2})\Theta(\tau_{1}-\tau_{2}) + G_{>}(k;\tau_{1},\tau_{2})\Theta(\tau_{2}-\tau_{1}),
    \end{align}\label{bulk-to-bulk}
\end{subequations}
where the functions $G_{\lessgtr}(k;\tau_{1},\tau_{2})$ are a combination of the mode functions $u(\tau,k)$ and its complex conjugate $u(\tau,k)^{\ast}$,
\begin{subequations}
    \begin{align}
        G_{>}(k;\tau_{1},\tau_{2}) & \equiv u(\tau_{1},k)u^{\ast}(\tau_{2},k), \\
        G_{<}(k;\tau_{1},\tau_{2}) & \equiv u^{\ast}(\tau_{1},k)u(\tau_{2},k).
    \end{align}
    \label{prop-prop}
\end{subequations}
An advantage of using this formalism is that we do have a diagrammatic representation that is quite similar to those derived in usual flat space quantum field theories, with the added twist of not performing a Fourier transform on the temporal coordinate $\tau$. This is because we are in a de Sitter background. Setting up the convention of denoting temporally ordered vertices by a black dot and anti-temporally ordered vertices by a white dot, the four bulk-to-bulk propagators then look as follows
\begin{figure}[H]
\centering
\def\svgwidth{0.45\linewidth}
    \adjustbox{trim=0.35cm 0cm 0cm 0cm,clip}{
    \includegraphics[width=0.45\textwidth]{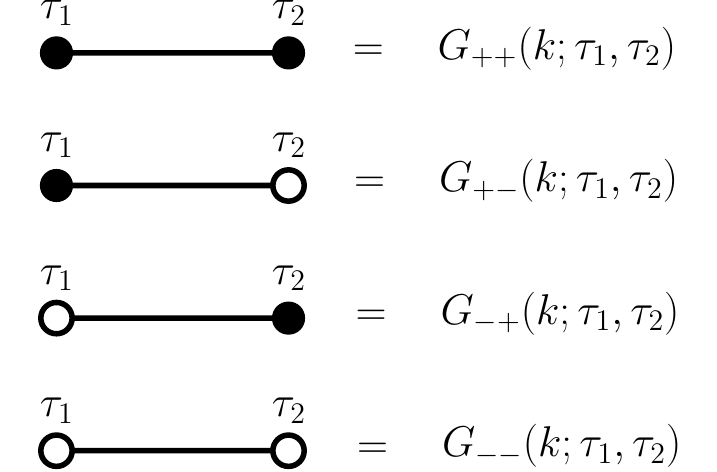}
        }
\end{figure}
\noindent External legs, that are on the temporal boundary, are denoted with a white square. Note that no distinction is made between $+$ and $-$ labels at $\tau_{f}$, due to the delta functions in (\ref{correlation-sk}).
\begin{figure}[H]
\centering
    \adjustbox{trim=0cm 0.3cm 0cm 0.3cm,clip}{
\includegraphics[width=0.6\textwidth]{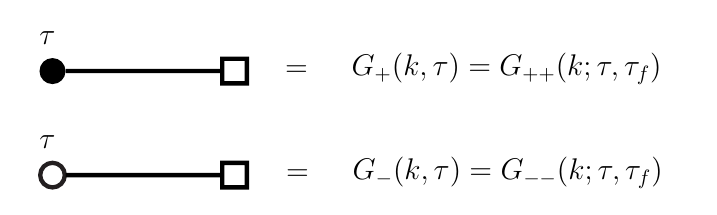}
    }
\end{figure}
\noindent Analysing the last diagrams while recalling the fact that $\Theta(x)=0$ if $x<0$, and $\Theta(x)=1$ if $x>0$, it is then possible to rewrite $G_{\lessgtr}(k;\tau_{1},\tau_{2})$ in terms of $G_{\pm}(k,\tau)$ and time-constant quantities
\begin{subequations}
    \begin{align}
        G_{+}(k,\tau) & = G_{>}(k;\tau,\tau_{f})\Theta(\tau-\tau_{f}) + G_{<}(k;\tau,\tau_{f})\Theta(\tau_{f}-\tau) = u^{\ast}(\tau,k)u(\tau_{f},k), \label{bulk-to-bound-plus} \\ 
        G_{-}(k,\tau) & = G_{<}(k;\tau,\tau_{f})\Theta(\tau-\tau_{f}) + G_{>}(k;\tau,\tau_{f})\Theta(\tau_{f}-\tau) = u(\tau,k)u^{\ast}(\tau_{f},k), \label{bulk-to-bound-minus}
    \end{align}
\end{subequations}
and similarly for the bulk-to-bulk propagators in momentum space,
\begin{subequations}
    \begin{align}
        G_{++}(k,\tau_{1};\tau_{2}) & = \dfrac{1}{|u(\tau_{f};k)|^{2}}\Big[ G_{-}(k;\tau_{1})G_{+}(k;\tau_{2})\Theta(\tau_{1}-\tau_{2}) \nonumber \\
        & \hspace{6cm}+ G_{+}(k;\tau_{1})G_{-}(k;\tau_{2})\Theta(\tau_{2}-\tau_{1})\Big], \label{btbprop++}\\
        G_{+-}(k,\tau_{1};\tau_{2}) & = \dfrac{1}{|u(\tau_{f};k)|^{2}}G_{+}(k;\tau_{1})G_{-}(k;\tau_{2}), \label{btbprop+-}\\
        G_{-+}(k,\tau_{1};\tau_{2}) & = \dfrac{1}{|u(\tau_{f};k)|^{2}}G_{-}(k;\tau_{1})G_{+}(k;\tau_{2}), \label{btbprop-+}\\
        G_{--}(k,\tau_{1};\tau_{2}) & =  \dfrac{1}{|u(\tau_{f};k)|^{2}}\Big[G_{+}(k;\tau_{1})G_{-}(k;\tau_{2})\Theta(\tau_{1}-\tau_{2}) \nonumber \\ 
        &\hspace{6cm}+ G_{-}(k;\tau_{1})G_{+}(k;\tau_{2})\Theta(\tau_{2}-\tau_{1})\Big], \label{btbprop--}
    \end{align}
    \label{btbprop}
\end{subequations}
To study interaction terms diagrammatically, start by considering the following toy model as a simple cubic interaction given by the action\footnote{Which can be easily generalised to, for instance, $\varphi^{n}$ theories.},
\begin{equation}
    S_{\text{int}} = -\int \dd \tau \dd[3] x \, a^{4}(\tau)\dfrac{1}{3!}\lambda\varphi^{3},
\end{equation}
where the term $a^{4}(\tau)$ is present because of the factor $\sqrt{-g}$. Then, the vertex interaction in 3-momentum space is given by:
\begin{figure}[H]
    \centering
    \begin{minipage}{0.45\textwidth}
        \centering
        \includegraphics[width=1\textwidth]{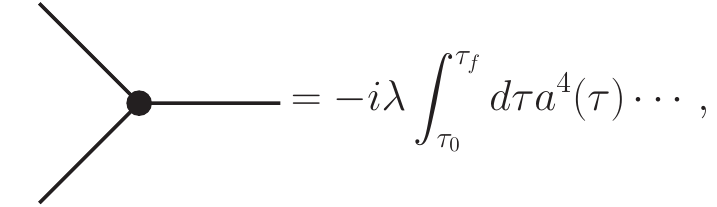}
    \end{minipage}
    \begin{minipage}{0.45\textwidth}
        \centering
        \includegraphics[width=1\textwidth]{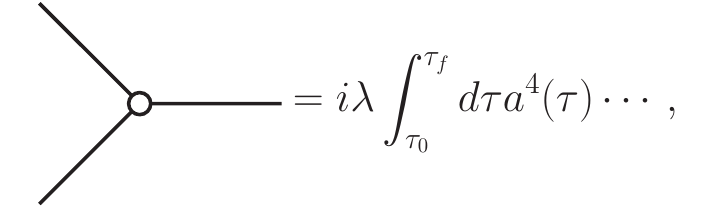}
    \end{minipage}
\end{figure}
\noindent where the temporal integral is from the initial slice $\tau_{0}$, which is taken to be $\tau_{0} \rightarrow -\infty$, up to $\tau_{f}$, the evaluation time of the correlation function. The ellipsis represents temporal-dependent variables associated to propagators. Notice that one is the complex conjugate of the other. Consequently, we have the following properties for the propagators,
\begin{align}
    [G_{ab}(k; \tau_{1}, \tau_{2})]^{\ast} & = G_{(-a)(-b)}(k; \tau_{1}, \tau_{2}), \label{bulk-to-bulk-complex-conjugate-prop}\\
    [G_{a}(k; \tau)]^{\ast} & = G_{(-a)}(k; \tau). \label{bulk-to-bound-complex-conjugate-prop}
\end{align}
As a summary for the diagrammatic rules, we calculate the correlation function by generating all topologically inequivalent diagrams with interacting vertices determined by $\mathcal{L}_{int}$ to the desired perturbative order. For each vertex, we can apply either a black or white colour in all possible combinations. Each internal line connecting two vertices is then associated with a bulk-to-bulk propagator matching its respective colour. Likewise, each external line is associated to a bulk-to-boundary propagator, where a square is introduced to explicitly denote the boundary. Additionally, total momentum conservation must  be conserved at each vertex. For instance, a tree level 3-point vertex diagrammatically correspond to:
\newlength{\figurehshift}
\setlength{\figurehshift}{-16em}
\begin{figure}[ht]
\centering
\makebox[\textwidth][c]{%
  \hspace{\figurehshift}%
  \begin{minipage}[c]{0.9\textwidth}
    \centering
    \begin{minipage}[c]{0.48\textwidth}
      \centering
      \raisebox{-0.5ex}{%
        $\displaystyle
        \langle
        \varphi(\vb{k}_{1})
        \varphi(\vb{k}_{2})
        \varphi(\vb{k}_{3})
        \rangle^{\prime}_{a} =
        $
      }
    \end{minipage}%
    \hspace{-5em}%
    \begin{minipage}[c]{0.48\textwidth}
      \centering
      \includegraphics[width=1.8\textwidth]{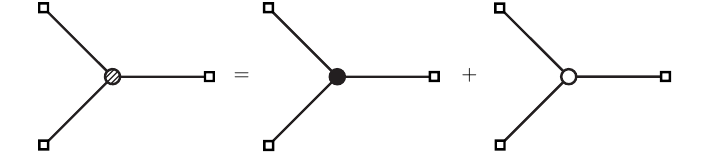}
    \end{minipage}
  \end{minipage}
}
\end{figure}
\noindent Where $\langle \varphi(\vb{k}_{1})\varphi(\vb{k}_{2})\varphi(\vb{k}_{3})\rangle'$ is the term containing the diagrammatic summation. Notice the computation of the above correlation function can be simplified by using (\ref{bulk-to-bound-complex-conjugate-prop}), allowing to rewrite it as the imaginary part of a black vertex diagram. Let us refine this last statement. The above correlator is,
\begin{equation}
    \expval{ \varphi^{1}_{k_{1}k_{2}k_{3}} }^{\prime} \equiv \expval{ \varphi(\vb{k}_{1}) \varphi(\vb{k}_{2}) \varphi(\vb{k}_{3}) }^{\prime} = \sum_{a=\pm} (-i\lambda) a \int_{-\infty}^{0} d\tau\, G_{a}(k_{1};\tau)G_{a}(k_{2};\tau)G_{a}(k_{3};\tau),
    \label{eq:3-point-correlator}
\end{equation}
where we have introduced the notation $\expval{ \varphi^{1}_{k_{1}k_{2}k_{3}} }^{\prime}$, where the top index refers to the number of vertices. We define the following quantity
\begin{equation}
    \mathcal{A}(k_{1},k_{2},k_{3},a) \equiv (-i\lambda) a \int_{-\infty}^{0} d\tau\,G_{a}(k_{1};\tau)G_{a}(k_{2};\tau)G_{a}(k_{3};\tau).
    \label{def:A-3-point}
\end{equation}
Therefore, using (\ref{bulk-to-bound-complex-conjugate-prop}) and $\mathcal{A}$ we arrive at the following form for the three-point function:
\begin{align}
    \expval{ \varphi(\vb{k}_{1}) \varphi(\vb{k}_{2}) \varphi(\vb{k}_{3}) }^{\prime} & = \mathcal{A}(k_{1},k_{2},k_{3},a=+) + \mathcal{A}(k_{1},k_{2},k_{3},a=-) \nonumber \\
    & = \mathcal{A}(k_{1},k_{2},k_{3},a=+) - \mathcal{A}^{\ast}(k_{1},k_{2},k_{3},a=+) \nonumber \\
    & = 2i\Im(\mathcal{A}(k_{1},k_{2},k_{3},a=+)).
\end{align}
This inspires to explore the remaining complex structure of $\mathcal{A}$, that is
\begin{align}
    2\Re(\mathcal{A}(k_{1},k_{2},k_{3},a=+)) & = \mathcal{A}(k_{1},k_{2},k_{3},a=+) + \mathcal{A}^{\ast}(k_{1},k_{2},k_{3},a=+) \nonumber \\
    & = \mathcal{A}(k_{1},k_{2},k_{3},a=+) - \mathcal{A}(k_{1},k_{2},k_{3},a=-) \nonumber \\
    & = \sum_{a=\pm} a \,\mathcal{A}(k_{1}, k_{2}, k_{3}, a) \equiv \expvalbar{ \varphi^{1}_{k_{1}k_{2}k_{3}} }^{\prime}_{a}.
    \label{eq:3-point-barred-correlator}
\end{align}
Diagrammatically we can display this as 
\setlength{\figurehshift}{-4em}
\begin{figure}[ht]
\centering
\makebox[\textwidth][c]{%
  \hspace{\figurehshift}%
  \begin{minipage}[c]{0.9\textwidth}
    \centering
    \begin{minipage}[c]{0.48\textwidth}
      \centering
      \raisebox{-0.5ex}{%
        $\displaystyle
        \langle\overline{
        \varphi(\vb{k}_{1})
        \varphi(\vb{k}_{2})
        \varphi(\vb{k}_{3})}
        \rangle^{\prime}_{a} =
        $
      }
    \end{minipage}%
    \hspace{-5em}%
    \begin{minipage}[c]{0.48\textwidth}
      \centering
      \includegraphics[width=1.1\textwidth]{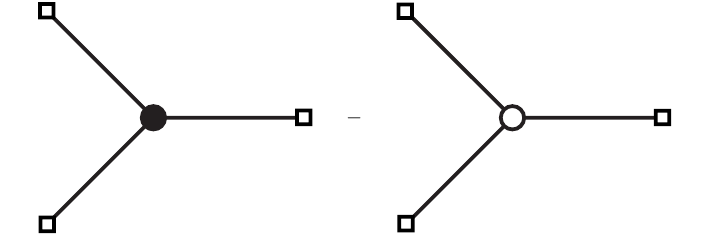}
    \end{minipage}
  \end{minipage}
}
\end{figure}
\noindent We will name this object \textit{barred} correlator and utilise it when exploring the cutting of a two-vertex four-point $s-$channel correlation function and further generalise it when exploring the $V$ vertex case.

\section{Cosmological Optical Theorem} \label{sec:cot}

\subsection{Unitarity in Flat Space}

If the time evolution of a quantum field theory is generated by an operator $U(t_f, t_i)$ satisfying: 
\begin{equation}
    UU^{\dagger}=1,
    \label{eq:unit-cond}
\end{equation}
then such a quantum field theory is said to be \textit{unitary}. This idea, although very simple, allows to construct constraints about the theory and its relevant interactions. The \textit{optical theorem} in flat space, depends on the unitarity of the $S-$matrix, defined as: 
\begin{equation}
    \langle f|i \rangle = S_{i\rightarrow f} =\lim_{\substack{t_i\rightarrow-\infty \\ t_f\rightarrow\infty}} U(t_f, t_i).
\end{equation}
Moreover, we expand the $S-$matrix as follows,
\begin{equation}
    S=1+i\mathcal{T},
    \label{S_matrix_expansion}
\end{equation}
where the identity corresponds to the \textit{forward scattering}, representing states propagating without any interaction, and $\mathcal{T}$ is the interacting part, typically calculated perturbatively through a diagrammatic expansion. Imposing the unitarity condition \eqref{eq:unit-cond}, then \eqref{S_matrix_expansion} yields,
\begin{equation}
    SS^{\dagger}=1 \Rightarrow \mathcal{T}-\mathcal{T^{\dagger}} = i\mathcal{T}\mathcal{T^{\dagger}}.
\end{equation}
By operating with the final and initial states $\bra{f}$ and $\ket{i}$ on the left and right, respectively, whilst inserting a complete set of intermediate states on the right we obtain,
\begin{equation}
    \mathcal{M}(i \rightarrow f) - \mathcal{M}^{\ast}(f\rightarrow i) = i\sum_{X}\int d\Pi_{X}(2\pi)^{4}\delta^{(4)}(k_{\text{in}} - k_{X}) \mathcal{M}^{\ast}(f \rightarrow X)\mathcal{M}(i \rightarrow X),
    \label{eq:gen-optical-theorem-qft}
\end{equation}
usually referred to as the generalised optical theorem. This relationship imposes a constraint on the value of the amplitudes. Moreover, by performing a perturbative analysis, one finds a set of cutting rules \cite{Cutkosky:1960sp}, where we diagrammatically slice loops to obtain trees or a lower number of loops.

\subsection{Unitarity in de Sitter}

An $S-$matrix definition in de Sitter spacetimes remains an open problem. Amongst the reasons is that the boundary geometry of de Sitter spacetimes obstructs an unambiguous free particle preparation, impeding the usual formulation of scattering amplitudes as transition amplitudes between in and out states \cite{Marolf:2012kh,Bousso:2004tv}. Nonetheless, recent progress proposes a de Sitter $S-$matrix for massive scalar fields via on‑shell limits of time‑ordered or amputated cosmological correlators, relating directly to in‑in equal‑time correlators and boundary wavefunction coefficients \cite{Melville:2023kgd,Melville:2024ove}. From a particle physics perspective, in \cite{Taylor:2024vdc,Taylor:2025spp} the authors develop a scattering amplitude formalism in global de Sitter space using a Dyson series approach, where geodesic observers are related by de Sitter symmetry transformations with asymptotically flat space amplitudes in the high-mass or short-wavelength limit.

Despite these limitations, it is possible to elude them in order to obtain a cosmological version for the generalised optical theorem. In \cite{Goodhew:2020hob}, the authors begin by using the Dyson series in the interaction picture,
\begin{align}
    \mathcal{U}(\tau, \tau_0) & = \mathcal{T}\left \{\exp\left(-i\int_{\tau_0}^{\tau} \tilde{H}_I(\tau')d\tau' \right)\right \} \nonumber \\
    & = 1-i \int_{\tau_0}^{\tau}d\tau_1 \tilde{H}_I(\tau_1) - \int_{\tau_0}^{\tau} d\tau_1 d\tau_2 \tilde{H}_I(\tau_1)\tilde{H}_I(\tau_2)\Theta(\tau_1-\tau_2) \nonumber \\
    & \phantom{=} + \frac{i}{2}\int_{\tau_0}^{\tau} d\tau_1 d\tau_2 d\tau_3 \tilde{H}_I(\tau_1)\tilde{H}_I(\tau_2)\tilde{H}_I(\tau_3)\Theta(\tau_1-\tau_2)\Theta(\tau_2-\tau_3) + \cdots .
    \label{DysonSeries}
\end{align}
Expanding $\mathcal{U} = 1 + \delta \mathcal{U}$, similarly to the $S-$matrix expansion whilst imposing the unitarity condition we obtain,
\begin{equation}
    \mathcal{U}\mathcal{U^{\dagger}} = 1 \Rightarrow \delta \mathcal{U}+\delta \mathcal{U^{\dagger}}=-\delta \mathcal{U}\delta \mathcal{U^{\dagger}}.
    \label{cot_raw}
\end{equation}
Analogously to flat space this corresponds to a non-perturbative relation. The structure of $\delta \mathcal{U}$ as products of the interaction Hamiltonian $H_{\text{int}}$ suggests a diagrammatic expansion in orders of the coupling constants $g$. Consequently, equation \eqref{cot_raw}, expanded perturbatively in orders in $g$, takes the schematic form:
\begin{equation}
    \delta \mathcal{U}_g+\delta \mathcal{U}_g^{\dagger} = 0, \qquad \delta \mathcal{U}_{g^2}+\delta \mathcal{U}_{g^2}^{\dagger} =-\delta \mathcal{U}_g\delta \mathcal{U}_g^{\dagger}, \qquad \cdots.
    \label{orderbyorder}
\end{equation}
This result is quite valuable as it provides a similar relationship to the $S-$matrix one in flat space. Furthermore, the analogue version for transition amplitude to $\delta \mathcal{U}$ can be obtained by operating between two particle states of the free theory. Specifically, these corresponds to the $n$-particle states of the free theory $\ket{\left \{\mathbf{k}, \alpha  \right \}_n}$, where the $\alpha$'s characterise the particle type and $\ket{0}$, the vacuum of the free theory. The result reads,
\begin{multline}
    \langle \left \{\mathbf{k}, \alpha \right \}_{n}|\delta\mathcal{U}(0, -\infty)|\Omega\rangle + \langle \left \{\mathbf{k}, \alpha \right \}_{n}|\delta\mathcal{U}^{\dagger}(0, -\infty)|\Omega\rangle = -\sum_{m=0}^{\infty}\sum_{\beta_1,\dots,\beta_m} \int \frac{d^3\mathbf{l}_1}{(2\pi)^3} \cdots \frac{d^3\mathbf{l}_m}{(2\pi)^3} \times \\ \times\langle \left \{\mathbf{k}, \alpha \right \}_{n}|\delta\mathcal{U}(0, -\infty)| \left \{\mathbf{l}, \beta \right \}_{m} \rangle \langle \left \{\mathbf{l}, \beta \right \}_{m}|\delta\mathcal{U}^{\dagger}(0, -\infty)|\Omega\rangle,
    \label{cot}
\end{multline}
where the $m$-particle states $\ket{\left \{\mathbf{l}, \beta \right \}_{m}}$ are introduced as a complete basis of states for an identity matrix. This is the cosmological version for the optical theorem.

Implications of the cosmological optical theorem (COT) for the coefficients of the wavefunction of the universe formalism are analysed in \cite{Goodhew:2020hob}, leading to unitarity constraints.

\section{Tree-level Cutting Rule}\label{sec:toy-model}

From a practical standpoint, one of the main obstacles to compute correlation functions within the Schwinger-Keldysh path integral formalism are evaluating temporal nested integrals. This is due to the presence of Heaviside step functions at the propagators \cref{btbprop++,btbprop+-,btbprop-+,btbprop--}, \textit{e.g.} see section 3 of \cite{Chen:2017ryl} in which the authors considered a quasi-single-field example, where a numerical analysis for the bispectrum was done.

Clearly, the number of nested integrals increases when considering higher-order interactions. Thus, an operator inspired from the unitarity condition \ref{cot_raw} and \cite{Melville:2021lst} is proposed, reducing the number of nested temporal integrals, allowing for a diagrammatic cut interpretation consisting on multiplying lower-point diagrams.

We begin by studying the simplest possible cut utilising a $\varphi^{3}$ theory, and then generalize it to $\varphi^{n}$ interactive theories. Consider the following $s-$channel contribution for a 4-point correlation function\footnote{To be more precise, we need to include $t$ and $u-$ channels, however we can easily account for these by performing momentum permutations. We will refrain from an explicit computation, as the procedure is analogous.}, which can be represented diagrammatically as:
\begin{figure}[H]
\centering
\adjustbox{trim=0.3cm 0cm 0cm 0cm,clip}{
    \includegraphics[width=0.85\textwidth]{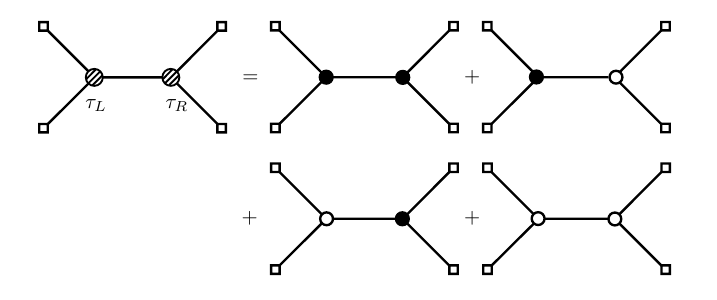}
        }
\end{figure}
\noindent Defining the following discontinuity operator\footnote{Note that this operation it is not written in a standard manner, due to the plus sign. For a more complete discussion and examples on this operator see \ref{sec:disc}.}:
\begin{equation}
    \disc{f(k_{1},\dots,k_{n};k_{I};\{\vb{k}\})}{k_{I}} = f(k_{1},\dots,k_{n};k_{I};\{\vb{k}\}) + f^{\ast}(-k_{1},\dots,-k_{n};k_{I};-\{\vb{k}\}),
    \label{disc-op}
\end{equation}
where $f$ is an arbitrary function depending on external momenta $k_{i}$, internal momenta $k_{I}$, and spatial momenta $\{\vb{k}\}$. We will denote, similar to (\ref{def:A-3-point})
\begin{equation}
    \expval{ \varphi^{2}_{k_{1}k_{2}k_{3}k_{4};s} }^{\prime} \equiv \expval{ \varphi(\tau,\vb{k}_{1})\varphi(\tau,\vb{k}_{2})\varphi(\tau,\vb{k}_{3})\varphi(\tau,\vb{k}_{4}) }^{\prime},
\end{equation}
where the top index indicates the number of vertices, whilst the sub indices $k_{i}$ refers to the external energy dependence and $s$ displays that we are considering the $s-$channel contribution. Mathematically, we can write down the value of the diagram following the Feynman rules described in section \ref{sec:sk},
\begin{multline}
    \expval{ \varphi^{2}_{k_{1}k_{2}k_{3}k_{4}} }^{\prime}_{s} = \\
    (-i\lambda)^{2}\sum_{a,b}ab\int_{-\infty}^{0}\dd \tau_{L}\,\dd \tau_{R}\,G_{a}(k_{1};\tau_{L})G_{a}(k_{2};\tau_{L})G_{ab}(s;\tau_{L},\tau_{R})G_{b}(k_{3};\tau_{R})G_{b}(k_{4};\tau_{R}).
    \label{4-point-corr-funct}
\end{multline}
Note that the above expression neglects the factors $a^{4}(\tau_{L})a^{4}(\tau_{R})$ from $\sqrt{-g}$ from both integrals; we will abstain from including them in the text as their relevance to the current analysis is minimal. Applying the discontinuity operation (\ref{disc-op}) to the correlation function (\ref{4-point-corr-funct}),
\begin{align}
    & \disc{\expval{ \varphi^{2}_{k_{1}k_{2}k_{3}k_{4}} }^{\prime}_{s}}{s} = \nonumber\\
    & \quad(-i\lambda)^{2}\sum_{a,b}ab\int_{-\infty}^{0}\dd \tau_{L}\,\dd \tau_{R}\,G_{a}(k_{1};\tau_{L})G_{a}(k_{2};\tau_{L})G_{ab}(s;\tau_{L},\tau_{R})G_{b}(k_{3};\tau_{R})G_{b}(k_{4};\tau_{R}) \nonumber \\
    & \quad + (i\lambda)^{2} \sum_{a,b}ab \int_{-\infty}^{0}\dd \tau_{L}\,\dd \tau_{R}\,G_{a}^{\ast}(\bar k_{1};\tau_{L})G_{a}^{\ast}(\bar k_{2};\tau_{L})G_{ab}^{\ast}(s;\tau_{L},\tau_{R})G_{b}^{\ast}(\bar k_{3};\tau_{R})G_{b}^{\ast}(\bar k_{4};\tau_{R}),
\label{disc-op-1}
\end{align}
where $\bar{k}_{i} = - k_{i}$ is just notation for the analytic continuation of the energies given by the discontinuity operator.
Furthermore, it is possible to reduce (\ref{disc-op-1}) by factorising bulk-to-boundary propagators using (\ref{bulk-to-bulk-complex-conjugate-prop}) (\ref{bulk-to-bound-complex-conjugate-prop}) and,
\begin{align}
    G_{a}(-k_{i},\tau_{j}) & = -G_{a}^{\ast}(k_{i},\tau_{j}) = -G_{(-a)}(k_{i},\tau_{j}), \label{property-bulk-to-boundary-prop} \\
    G_{ab}(-k_{i},\tau_{j}, \tau_{l}) & = -G_{ab}^{\ast}(k_{i},\tau_{j}, \tau_{l}) = -G_{(-a)(-b)}(k_{i},\tau_{j}, \tau_{l}). \label{property-bulk-to-bulk-prop}
\end{align}
The latter property, which we will refer as \textit{parity relations} of bulk-to-boundary propagators can be easily verified for massless and conformally coupled fields in de Sitter theories, with their respective mode functions being:
\begin{align}
    G_{>}(k;\tau_{1},\tau_{2}) & = \dfrac{H^{2}}{2k^{3}}(1+ik\tau_{1})(1-ik\tau_{2})e^{-ik(\tau_{1}-\tau_{2})}, \\
    G_{>}(k;\tau_{1},\tau_{2}) & = \dfrac{H^{2}\tau_{1}\tau_{2}}{2k}e^{-ik(\tau_{1}-\tau_{2})}.
    \label{modal_func}
\end{align}
Subsequently, we will denote bulk-to-bulk and bulk-to-boundary propagators as: $G_{a}(k_{i};\tau) = G_{a}^{k_{i}}(\tau)$ and $G_{ab}(s,\tau_{L},\tau_{R}) = G_{ab}^{s}(\tau_{L},\tau_{R})$. Then, the discontinuity reads,
\begin{multline}
    \disc{\expval{ \varphi^{2}_{k_{1}k_{2}k_{3}k_{4}} }^{\prime}_{s}}{s} = \\
    (-i\lambda)^{2}\sum_{a,b}ab\int_{-\infty}^{0}\dd \tau_{L}\,\dd \tau_{R}\,G_{a}^{k_{1}}(\tau_{L})G_{a}^{k_{2}}(\tau_{L})\big[G_{ab}^{s}(\tau_{L},\tau_{R}) + G_{ab}^{s\ast}(\tau_{L},\tau_{R}) \big]G_{b}^{k_{3}}(\tau_{R})G_{b}^{k_{4}}(\tau_{R}).
    \label{disc-op-fact}
\end{multline}
By analysing the sum $G_{ab}^{s}(\tau_{L},\tau_{R}) + G_{ab}^{s\ast}(\tau_{L},\tau_{R})$ in its four combinations, and exploiting the property for the step function $\Theta(\tau_{L}-\tau_{R}) + \Theta(\tau_{R}-\tau_{L}) = 1$, it is straightforward to prove that for all $a, b$:
\begin{equation}
    G_{ab}^{s}(\tau_{L},\tau_{R}) + G_{ab}^{s\ast}(\tau_{L},\tau_{R}) = \big[G_{+-}^{s}(\tau_{L},\tau_{R}) + G_{-+}^{s}(\tau_{L},\tau_{R})\big], \label{sum-of-bulk-to-bulk}
\end{equation}
which can be further re-expressed in terms of bulk-to-boundary propagators by using (\ref{btbprop+-}) and (\ref{btbprop-+}),
\begin{equation}
    G_{ab}^{s}(\tau_{L},\tau_{R}) + G_{ab}^{s\ast}(\tau_{L},\tau_{R}) = \dfrac{1}{|u(\tau_{f};k)|^{2}} [G_{+}(k;\tau_{1})G_{-}(k;\tau_{2}) + G_{-}(k;\tau_{1})G_{+}(k;\tau_{2})]. \label{sum-of-bulk-to-bulk-v2}
\end{equation}
It is important to highlight the fact that there are no nested integrals in terms of the conformal time $\tau_{L}$ and $\tau_{R}$ due to the absence of step functions. Thus, we may split the temporal integrals from (\ref{disc-op-fact}) and interpret the remaining as a product of two vertices followed by an integration over the internal momentum variable. However, during the expansion we will encounter terms of the following nature (taking $a=b=+$):
\begin{equation}
    \int_{\vb{s}\vb{s'}}f(s,s')\left( \int_{-\infty}^{0}d\tau_{L}\,G_{+}^{k_{1}}(\tau_{L})G_{+}^{k_{2}}(\tau_{L})G_{+}^{s}(\tau_{L}) \right)\left( \int_{-\infty}^{0}d\tau_{R}\,G_{-}^{s'}(\tau_{R})G_{+}^{k_{3}}(\tau_{R})G_{+}^{k_{4}}(\tau_{R}) \right),
    \label{first-term-disc-expansion}
\end{equation}
where $f(\tau_{f};s,s')$ corresponds to a combination of mode functions $u(\tau_{f};k)$ given by (\ref{bulk-to-bound-plus}) and (\ref{bulk-to-bound-minus}). More precisely,
\begin{equation}
    \int_{\vb{s}\vb{s'}}f(s,s') = \int d^{3}s\, d^{3}s'\, \dfrac{\delta^{(3)}(\vb{s}-\vb{s'})}{u(\tau_{f};s)u^{\ast}(\tau_{f};s')}.
    \label{def-f-factor}
\end{equation}
By examining the right-most parenthesis of (\ref{first-term-disc-expansion}) we immediately find out that such term should not have a diagrammatic interpretation, since it is composed of two $+$ and one $-$ bulk-to-boundary propagators. Fortunately, by using the property (\ref{property-bulk-to-boundary-prop}) we are able to revert the $-$ sign to a $+$, whilst adding an overall minus sign and a negative value for the momenta at one of the bulk-to-boundary propagators,
\begin{equation}
    \int_{\vb{s}\vb{s'}}f(s,s')\left( \int_{-\infty}^{0}d\tau_{L}\,G_{+}^{k_{1}}(\tau_{L})G_{+}^{k_{2}}(\tau_{L})G_{+}^{s}(\tau_{L}) \right)\left( \int_{-\infty}^{0}d\tau_{R}\,G_{+}^{-s'}(\tau_{R})G_{+}^{k_{3}}(\tau_{R})G_{+}^{k_{4}}(\tau_{R}) \right). \label{time-integral-separation}
\end{equation}

Diagrammatically, \eqref{time-integral-separation} corresponds to the multiplication of two three-point $a=+$ vertices. Therefore, we say that we performed a cut on the internal propagator and can be draw as:
\begin{figure}[H]
\centering
\adjustbox{trim=0cm 0cm 0cm 0cm,clip}{
    \includegraphics[width=0.55\textwidth]{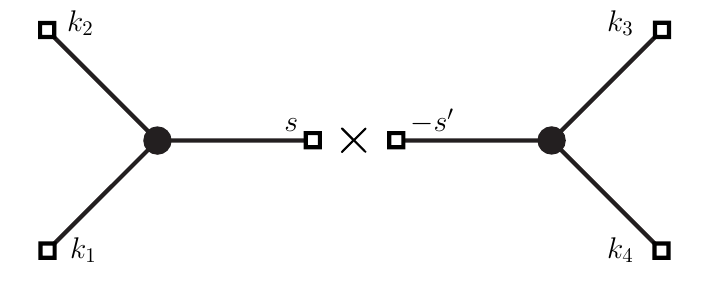}
        }
\end{figure}

Repeating this analysis for the remaining terms, and recovering the neglected factors such as the coupling and those coming from $\sqrt{-g}$, we find the following diagrammatic expression for the $s-$channel discontinuity of the four-point correlation function:
\begin{figure}[H]
\centering
\adjustbox{trim=0cm 0cm 0cm 0cm,clip}{
    \includegraphics[width=1\textwidth]{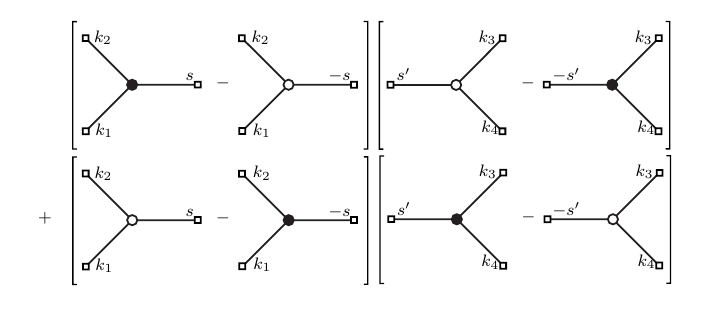}
        }
\end{figure}
\noindent Which can be conveniently written as multiplication of discontinuity operations of 3-point correlators \cref{eq:3-point-correlator} and 3-point \textit{barred} correlators \cref{eq:3-point-barred-correlator} as follows,
\begin{align}
    \disc{\expval{ \varphi^{2}_{k_{1}k_{2}k_{3}k_{4}} }^{\prime}_{s})}{s} = & \dfrac{1}{2} \int_{\vb{s}\vb{s'}}f(\tau_{f};s,s')\, i\disc{i\expval{ \varphi^{1}_{k_{1}k_{2}s} }^{\prime}}{s}i\disc{i\expval{ \varphi^{1}_{s^{\prime}k_{3}k_{4}} }^{\prime}}{s^{\prime}} \nonumber \\
    + & \dfrac{1}{2}\int_{\vb{s}\vb{s'}}f(\tau_{f};s,s')\, \disc{ \expvalbar{\varphi^{1}_{k_{1}k_{2}s}}^{\prime} }{s}\disc{ \expvalbar{\varphi^{1}_{s^{\prime}k_{3}k_{4}}}^{\prime} }{s^{\prime}}.
    \label{eq:v-2-tree-level-cutting-rule}
\end{align}
This expression is remarkably similar to the cosmological cutting rules to wavefunction coefficients \cite{Jazayeri:2021fvk,Melville:2021lst,Goodhew:2021oqg}, the main differences being that we are directly computing a \textit{discontinuity} operation for \textit{in-in} cosmological correlators, \textit{i.e.}, the observables. Nonetheless, we are presented with a twist. We require a new algebraic object, related to the correlation function through $2\Re \leftrightarrow 2i\Im$, namely $\expvalbar{\varphi^{1}}^{\prime}$. This is the price to pay, when computing \textit{in-in} cuts to cosmological correlators using the SK formalism. We will generalise the definition for \textit{barred} correlators and compute a more challenging correlator in section \ref{sec:general-tree-lvl-rule}.

Additionally, it is possible to rewrite the above expression on a compact way employing the following real and imaginary bulk-to-bulk propagator decomposition, which has been useful to understand correlators infrared behaviour \cite{Palma:2023idj, Palma:2025oux}:
\begin{subequations}
    \begin{align}
        G_{++}(k,\tau_{1};\tau_{2}) & = G_{R}(k, \tau_{1}, \tau_{2}) + iG_{I}(k, \tau_{1}, \tau_{2})I(\tau_{1}, \tau_{2}), \\
        G_{+-}(k,\tau_{1};\tau_{2}) & = G_{R}(k, \tau_{1}, \tau_{2}) - iG_{I}(k, \tau_{1}, \tau_{2}), \\
        G_{-+}(k,\tau_{1};\tau_{2}) & = G_{R}(k, \tau_{1}, \tau_{2}) + iG_{I}(k, \tau_{1}, \tau_{2}), \\
        G_{--}(k,\tau_{1};\tau_{2}) & = G_{R}(k, \tau_{1}, \tau_{2}) - iG_{I}(k, \tau_{1}, \tau_{2})I(\tau_{1}, \tau_{2}),
    \end{align}
    \label{ps-propagators}
\end{subequations}
where $I(\tau_{1}, \tau_{2}) \equiv \Theta(\tau_{1}-\tau_{2}) - \Theta(\tau_{2}-\tau_{1})$. Or in a single line,
\begin{equation}
    G_{ab}(k,\tau_{1};\tau_{2}) = G_{R}(k, \tau_{1}, \tau_{2}) + ibG_{I}(k, \tau_{1}, \tau_{2})(1+\delta_{ab}\left[I(\tau_{1}, \tau_{2})-1\right]),
    \label{bulk-to-bulk-prop-gs}
\end{equation}
where $\delta_{ab}$ is the Kronecker delta. By diagrammatically denoting the real part $G_{R}(k, \tau_{1}, \tau_{2})$ with a double line, and the imaginary part $G_{I}(k, \tau_{1}, \tau_{2})$ with a dashed line, we easily find using (\ref{sum-of-bulk-to-bulk}) that (\ref{disc-op-fact}) has a straightforward diagrammatic representation
\begin{figure}[H]
\centering
\adjustbox{trim=0cm 0cm 0cm 0cm,clip}{
    \includegraphics[width=0.4\textwidth]{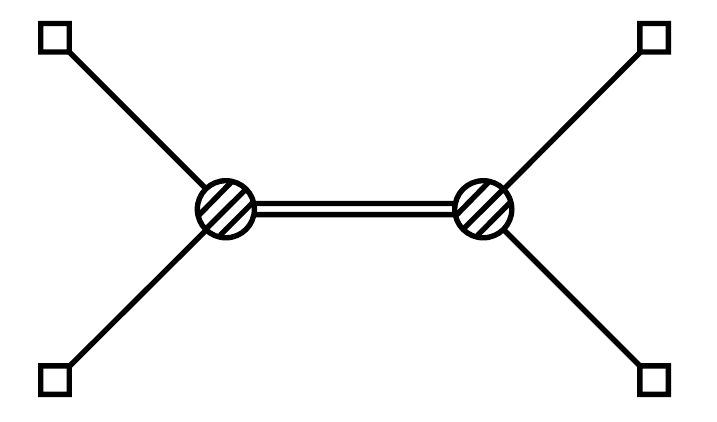}
        }
\end{figure}
\noindent Note that what looks schematically as a connected diagram is rather factorisable as multiplications of lower-point diagrams because there are no nested temporal integrals, as per (\ref{time-integral-separation}).

Additionally, an interesting task is to demonstrate if this operation is related with unitarity. By expanding the matrix elements of \eqref{cot}, we match elements in the canonical formalism to SK diagrams. More details can be found in Appendix \ref{chap:cot}.

\section{Derivative Couplings}\label{sec:deriv}
In cosmological contexts, it is quite frequent to encounter derivative couplings, for instance, when taking an EFT approach to study inflation \cite{cheung_effective_2008}. Consequently, adding different prescriptions to the Feynman rules, depending if they are spatial or temporal derivatives.

\subsection{Spatial Derivatives at Tree Level}

As an example, consider the following interactive Lagrangian,
\begin{equation}
    \mathcal{L}_{\text{int}} = - \dfrac{\lambda}{6} (\partial_{i}\varphi)^2 \varphi,
\end{equation}
where $\partial_{i}$ correspond to spatial derivatives with $i= 1,2,3$. The action of these derivatives on propagators generates three-momentum products factors. The $s-$channel contribution reads,
\begin{multline}
    \expval{ \varphi^{2}_{k_{1}k_{2}k_{3}k_{4}} }^{\prime}_{s} = 
    (-i\lambda)^{2}(\mathbf{k}_1 \cdot \mathbf{k}_2 + \mathbf{k}_1 \cdot \mathbf{s} + \mathbf{s} \cdot \mathbf{k}_2)(\mathbf{k}_3 \cdot \mathbf{k}_4 + \mathbf{k}_3 \cdot \mathbf{s} + \mathbf{s} \cdot \mathbf{k}_4)\sum_{a,b}ab\int_{-\infty}^{0}\dd \tau_{L}\,\dd \tau_{R}\, \times  \\
    \times G_{a}(k_{1};\tau_{L})G_{a}(k_{2};\tau_{L})G_{ab}(s;\tau_{L},\tau_{R})G_{b}(k_{3};\tau_{R})G_{b}(k_{4};\tau_{R}).\label{amp-spatial}
\end{multline}
We then obtain the discontinuity by inserting \eqref{amp-spatial} into \eqref{disc-op}, obtaining,
\begin{multline}
    \disc{\expval{ \varphi^{2}_{k_{1}k_{2}k_{3}k_{4}} }^{\prime}_{s}}{s}= (-i\lambda)^{2} 
    (\mathbf{k}_1 \cdot \mathbf{k}_2 + \mathbf{k}_1 \cdot \mathbf{s} + \mathbf{s} \cdot \mathbf{k}_2)
    (\mathbf{k}_3 \cdot \mathbf{k}_4 + \mathbf{k}_3 \cdot \mathbf{s} + \mathbf{s} \cdot \mathbf{k}_4)\times \\ 
    \times \sum_{a,b} ab 
    \int_{-\infty}^{0} \dd \tau_{L}\, \dd \tau_{R}\,
    G_{a}(k_{1}; \tau_{L}) G_{a}(k_{2}; \tau_{L}) 
    G_{ab}(s; \tau_{L}, \tau_{R}) 
    G_{b}(k_{3}; \tau_{R}) G_{b}(k_{4}; \tau_{R}) \\
    \quad + (i\lambda)^{2} 
    (\bar{\mathbf{k}}_1 \cdot \bar{\mathbf{k}}_2 + \bar{\mathbf{k}}_1 \cdot \bar{\mathbf{s}} + \bar{\mathbf{s}} \cdot \bar{\mathbf{k}}_2)
    (\bar{\mathbf{k}}_3 \cdot \bar{\mathbf{k}}_4 + \bar{\mathbf{k}}_3 \cdot \bar{\mathbf{s}} + \bar{\mathbf{s}} \cdot \bar{\mathbf{k}}_4)\times \\
    \times \sum_{a,b} ab 
    \int_{-\infty}^{0} \dd \tau_{L}\, \dd \tau_{R}\,
    G_{a}^{\ast}(\bar{k}_{1}; \tau_{L}) G_{a}^{\ast}(\bar{k}_{2}; \tau_{L}) 
    G_{ab}^{\ast}(s; \tau_{L}, \tau_{R}) 
    G_{b}^{\ast}(\bar{k}_{3}; \tau_{R}) G_{b}^{\ast}(\bar{k}_{4}; \tau_{R}).
\label{disc-op-spatial-1}
\end{multline}
Note that the spatial part of the internal momenta $\vb{s}$ is also affected by the discontinuity operation ($\bar{\mathbf{s}}=-\vb{s}$). Therefore, the product $\bar{\vb{k}}_i \cdot \bar{\vb{s}}$ remains invariant.  From this point, the computation is straightforward and the discontinuity yields a similar result as in section \ref{sec:toy-model}, with the addition of 3-momentum product combinations,
\begin{align}
    & \dfrac{1}{2} \int_{\vb{s}\vb{s'}}f(\tau_{f};s,s')\, \mathcal{K}(\vb{k}_{1},\vb{k}_{2},\vb{s})\,i\disc{i\expval{ \varphi^{1}_{k_{1}k_{2}s} }^{\prime}}{s}\mathcal{K}(\vb{s}',\vb{k}_{3},\vb{k}_{4})\,i\disc{i\expval{ \varphi^{1}_{s^{\prime}k_{3}k_{4}} }^{\prime}}{s^{\prime}} \nonumber \\
    & + \dfrac{1}{2}\int_{\vb{s}\vb{s'}}f(\tau_{f};s,s')\, \mathcal{K}(\vb{k}_{1},\vb{k}_{2},\vb{s})\,\disc{ \expvalbar{\varphi^{1}_{k_{1}k_{2}s}}^{\prime} }{s}\mathcal{K}(\vb{s}',\vb{k}_{3},\vb{k}_{4})\,\disc{ \expvalbar{\varphi^{1}_{s^{\prime}k_{3}k_{4}}}^{\prime} }{s^{\prime}},
\end{align}
where $\mathcal{K}$ are the momentum products:
\begin{align}
    \mathcal{K}(\vb{k}_{1},\vb{k}_{2},\vb{s}) & = \mathbf{k}_1 \cdot \mathbf{k}_2 + \mathbf{k}_1 \cdot \mathbf{s} + \mathbf{s} \cdot \mathbf{k}_2, & \mathcal{K}(\vb{s}',\vb{k}_{3},\vb{k}_{4}) & = \mathbf{k}_3 \cdot \mathbf{k}_4 + \mathbf{k}_3 \cdot \mathbf{s}' + \mathbf{s}' \cdot \mathbf{k}_4.
\end{align}
The connection to the cosmological optical theorem considering any number of spatial derivatives is reviewed in Appendix \ref{chap:cot}.

\subsection{Temporal Derivatives at Tree Level}

Temporal derivatives in an interactive Lagrangian should be treated carefully due to the Heaviside step functions in the propagators. To illustrate this, consider the following interactive Lagrangian,
\begin{equation}
\mathcal{L}_{\text{int}}=-\frac{\lambda_C}{6}\varphi \varphi'^{2},
\label{toy_lag_2}
\end{equation}
where the prime denotes derivatives respect to conformal time $\tau$. The vertex value then becomes,
\begin{align}
    \begin{minipage}{0.2\textwidth}
            \centering
            \adjustbox{trim=0cm 0cm 0cm 0cm,clip}{
    \includegraphics[width=1\textwidth]{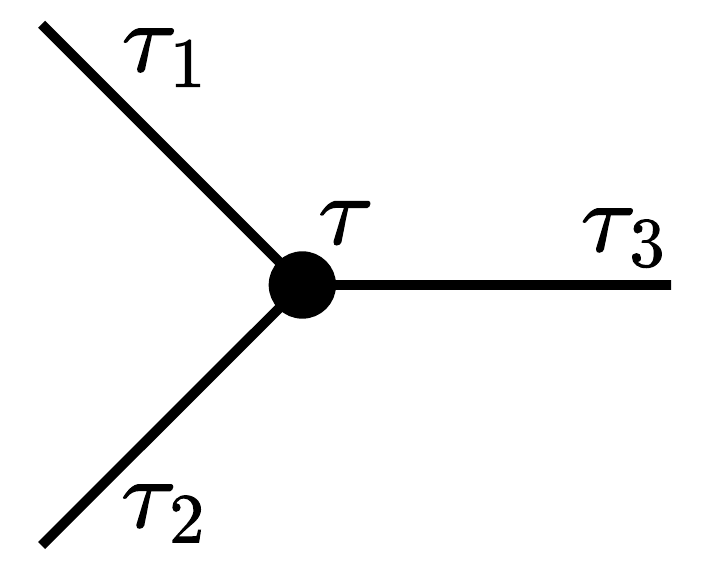}
        }
    \end{minipage} &
    =-\frac{i\lambda_C}{3}\int_{-\infty}^{0}d\tau\,a^2(\tau)\big[\partial_{\tau}G_{+a_1}(k_1;\tau,\tau_1)\big]\big[\partial_{\tau}G_{+a_2}(k_2;\tau,\tau_2)\big]G_{+a_3}(k_3;\tau,\tau_3) \nonumber\\
    & \phantom{=} + \text{2 perm}, \\
    \begin{minipage}{0.2\textwidth}
            \centering
            \adjustbox{trim=0cm 0cm 0cm 0cm,clip}{
    \includegraphics[width=1\textwidth]{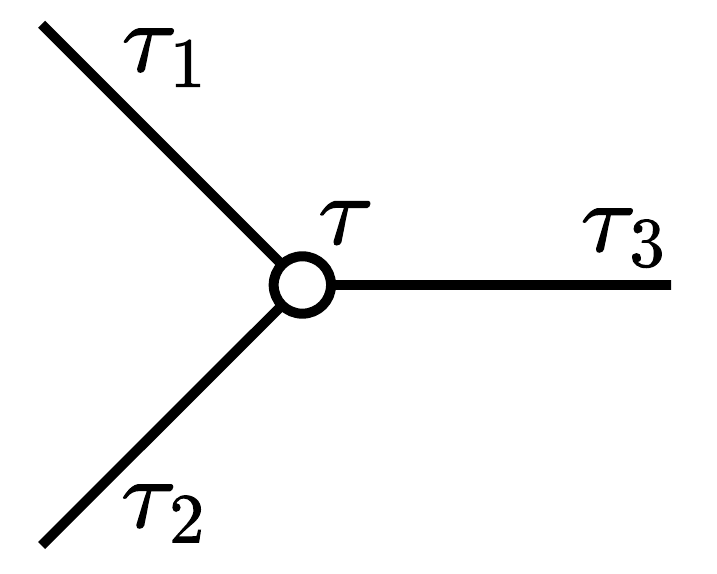}
        }
    \end{minipage} &
    =\frac{i\lambda_C}{3} \int_{-\infty}^{0}d\tau\,a^2(\tau)\big[\partial_{\tau}G_{-a_1}(k_1;\tau,\tau_1)\big]\big[\partial_{\tau}G_{-a_2}(k_2;\tau,\tau_2)\big]G_{-a_3}(k_3;\tau,\tau_3) \nonumber\\
    & \phantom{=} + \text{2 perm},
\end{align}
where the indices \(a_i\) are the indicators of the vertex where the propagator ends. Thus, the following property for the propagators, demonstrated in \cite{Chen:2017ryl}, becomes relevant,
\begin{align}
    \partial_{\tau_L} \partial_{\tau_R} G_{++}\left(k; \tau_L, \tau_R\right) & =\theta\left(\tau_L-\tau_R\right) \partial_{\tau_L} \partial_{\tau_R}G_{>}(k; \tau_L, \tau_R)+ \theta\left(\tau_R-\tau_L\right) \partial_{\tau_L} \partial_{\tau_R} G_{<}(k; \tau_L, \tau_R) \nonumber \\
    & \phantom{=} +i\delta\left(\tau_L-\tau_R\right) \nonumber \\
    &=\theta\left(\tau_L-\tau_R\right) \partial_{\tau_L} \partial_{\tau_R}(u^{*}(\tau_L, \mathbf{k})u(\tau_R, \mathbf{k})) \nonumber \\
    & \phantom{=} +  \theta\left(\tau_R-\tau_L\right) \partial_{\tau_L} \partial_{\tau_R} (u(\tau_L, \mathbf{k})u^{*}(\tau_R, \mathbf{k})) +i\delta\left(\tau_L-\tau_R\right).
\end{align}
To visualise the cuts in a theory with this type of interactions, consider the following interactive Lagrangian,
\begin{equation}
\mathcal{L}_{\text{int}}=-\frac{\lambda_D}{3!}\varphi'^{3}.
\label{toy_lag_3}
\end{equation}
Consequently, the $s-$channel contribution for a $4-$point correlation function now reads,
\begin{multline}
    \expval{ \varphi^{2}_{k_{1}k_{2}k_{3}k_{4}} }^{\prime}_{s}= (-i\lambda_D)^{2} \sum_{a,b} ab 
    \int_{-\infty}^{0} \dd \tau_{L} \dd \tau_{R} \Bigg[
    \partial_{\tau_L} G_{a}(k_{1}; \tau_{L}) 
    \partial_{\tau_L} G_{a}(k_{2}; \tau_{L}) \\
    \times \partial_{\tau_L}\partial_{\tau_R}G_{ab}(s; \tau_{L}, \tau_{R}) 
    \partial_{\tau_R}G_{b}(k_{3}; \tau_{R}) 
    \partial_{\tau_R}G_{b}(k_{4}; \tau_{R}) 
    \Bigg].
    \label{4-point-corr-funct_time}
\end{multline}
Inserting the above into \eqref{disc-op}, we obtain:
\begin{align}
    \disc{\expval{ \varphi^{2}_{k_{1}k_{2}k_{3}k_{4}} }^{\prime}_{s}}{s} = 
    (-i\lambda_D)^{2} \sum_{a,b} ab 
    \int_{-\infty}^{0} \dd \tau_{L} \dd \tau_{R} \Bigg[\partial_{\tau_L} G_{a}(k_{1}; \tau_{L}) \partial_{\tau_L} G_{a}(k_{2}; \tau_{L}) \times \nonumber \\
    \times \partial_{\tau_L} \partial_{\tau_R} G_{ab}(s; \tau_{L}, \tau_{R})\partial_{\tau_R} G_{b}(k_{3}; \tau_{R}) \partial_{\tau_R} G_{b}(k_{4}; \tau_{R}) 
    \Bigg] \nonumber \\
    + (i\lambda_D)^{2} \sum_{a,b} ab 
    \int_{-\infty}^{0} \dd \tau_{L} \dd \tau_{R} \Bigg[\partial_{\tau_L} G^{*}_{a}(\bar{k}_{1}; \tau_{L}) \partial_{\tau_L} G^{*}_{a}(\bar{k}_{2}; \tau_{L}) \times \nonumber \\
    \times \partial_{\tau_L} \partial_{\tau_R} G^{*}_{ab}(s; \tau_{L}, \tau_{R}) \partial_{\tau_R} G^{*}_{b}(\bar{k}_{3}; \tau_{R}) \partial_{\tau_R} G^{*}_{b}(\bar{k}_{4}; \tau_{R}) 
    \Bigg],
\label{disc-op-temporal-1}
\end{align}
where we expect that \eqref{property-bulk-to-bulk-prop} holds, even if propagators have time derivatives. Taking temporal derivatives to bulk-to-boundary propagators \eqref{modal_func} yields in the massless case,
\begin{subequations}
\begin{align}
    \partial_\tau G_{+}(k;\tau)&=\frac{1}{2k}H^2\tau (1+ik\tau_f)e^{ik(\tau-\tau_f)}=-\partial_{\tau} G_{+}^*(-k;\tau), \\
    \partial_\tau G_{-}(k;\tau)&=\frac{1}{2k}H^2\tau (1-ik\tau_f)e^{-ik(\tau-\tau_f)}=-\partial_{\tau} G_{-}^*(-k;\tau),
\end{align}
\end{subequations}
whilst for the conformally coupled case,
\begin{subequations}
\begin{align}
    \partial_\tau G_{+}(k;\tau)&=\frac{1}{2k} H^2 \tau_0 e^{ik(\tau-\tau_0)}(1+ik\tau)=-\partial_t G_{+}^*(-k;\tau),\\
    \partial_{\tau} G_{-}(k;\tau)&= \frac{1}{2k} H^2 \tau_0 e^{-ik(\tau-\tau_0)}(1-ik\tau) =-\partial_{\tau} G_{-}^*(-k;\tau).
\end{align}
\end{subequations}
Then, we extend \eqref{property-bulk-to-bulk-prop} to consider propagators with time derivatives:
\begin{align}
    \partial_\tau G_{a}(-k_{i},\tau_{j}) & = -\partial_\tau G_{a}^{\ast}(k_{i},\tau_{j}) = -\partial_\tau G_{(-a)}(k_{i},\tau_{j}), \label{property-bulk-to-boundary-prop_time} \\
    \partial_\tau G_{ab}(-k_{i},\tau_{j}, \tau_{l}) & = - \partial_\tau G_{ab}^{\ast}(k_{i},\tau_{j}, \tau_{l}) = -\partial_\tau G_{(-a)(-b)}(k_{i},\tau_{j}, \tau_{l}). \label{property-bulk-to-bulk-prop_time}
\end{align}
Applying these properties to \eqref{disc-op-temporal-1}, we obtain:
\begin{multline}
    \disc{\expval{ \varphi^{2}_{k_{1}k_{2}k_{3}k_{4}} }^{\prime}_{s}}{s} =(-i\lambda_D)^{2} \sum_{a,b} ab 
    \int_{-\infty}^{0} \dd \tau_{L} \dd \tau_{R} \Bigg[
        \partial_{\tau_L} G_{a}(k_{1}; \tau_{L}) 
        \partial_{\tau_L} G_{a}(k_{2}; \tau_{L}) \\
        \times [\partial_{\tau_L} \partial_{\tau_R} G_{ab}(s; \tau_{L}, \tau_{R}) + \partial_{\tau_L} \partial_{\tau_R} G^*_{ab}(s; \tau_{L}, \tau_{R}) ] 
        \partial_{\tau_R} G_{b}(k_{3}; \tau_{R}) 
        \partial_{\tau_R} G_{b}(k_{4}; \tau_{R}) 
    \Bigg].
\label{disc-op-temporal-2}
\end{multline}
Furthermore, utilising \ref{sec:two_dt}, it is straightforward to see that
\begin{multline}
    \partial_{\tau_L}\partial_{\tau_R}G_{ab}^{s}(\tau_{L},\tau_{R}) + \partial_{\tau_L}\partial_{\tau_R}G_{ab}^{s\ast}(\tau_{L},\tau_{R}) = \big[\partial_{\tau_L}\partial_{\tau_R}G_{+-}^{s}(\tau_{L},\tau_{R}) + \partial_{\tau_L}\partial_{\tau_R}G_{-+}^{s}(\tau_{L},\tau_{R})\big]\\
    =\dfrac{1}{|u(\tau_{f};k)|^{2}} [\partial_{\tau_L}G_{+}(k;\tau_{L})\partial_{\tau_R}G_{-}(k;\tau_{R}) + \partial_{\tau_L}G_{-}(k;\tau_{L})\partial_{\tau_R}G_{+}(k;\tau_{R})].
    \label{sum-of-bulk-to-bulk_time}
\end{multline}
In the above, time derivatives are applied to every $G_{\pm}$ in $G_{+-}$ and $G_{-+}$. Then, to visualise cuts, consider the case $\mathcal{A}(k_{1},k_{2},k_{3},k_{4};s;a=+,b=+) = \mathcal{A}_{++}$, in other words, the diagrammatic contribution consisting on every vertex being black. After some algebra we obtain,
\begin{multline}
    (-i\lambda_D)^{2}  \int_{\vb{s}\vb{s'}}f(s,s')\bigg(\int_{-\infty}^{0} \dd \tau_{L} \ \partial_{\tau_L} G_{+}(k_{1}; \tau_{L}) \partial_{\tau_L} G_{+}(k_{2}; \tau_{L}) \partial_{\tau_L}G_{+}(-s;\tau_{L})\Bigg) \times \\
    \times \Bigg(\int_{-\infty}^{0}\dd \tau_{R} \ \partial_{\tau_R}G_{+}(s;\tau_{R})\partial_{\tau_R} G_{+}(k_{3}; \tau_{R})\partial_{\tau_R} G_{+}(k_{4}; \tau_{R})\Bigg),
\end{multline}
exhibiting the same structure as \eqref{time-integral-separation}. Therefore, we interpret the above equation diagrammatically as cutting through the internal propagator, now in the presence of temporal derivatives, in a similar manner.
 
\section{General Tree-Level Rule}\label{sec:general-tree-lvl-rule}

Let us begin the analysis considering an $N-$point correlation function associated to a $V-$vertex graph. Since we desire generality, the interactive Lagrangian to be studied is,
\begin{equation}
    \mathcal{L}_{\text{int}} = \sum_{k=3}^{N} \dfrac{\lambda_{k}}{k!}\varphi^{k},
\end{equation}
for arbitrary $N$. Diagrammatically, a general tree-level linear chain graph configuration looks:
\begin{figure}[H]
\centering
\includegraphics[width=0.85\textwidth]{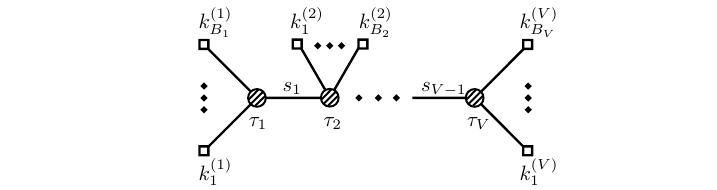}
\end{figure}
\noindent Every vertex $i$, with $i=1,\dots, V$ on the graph contains $B_{i}$ external lines with individual three-momentum $\vb{k}^{(i)}_{j}$, where $j=1,\dots, B_{i}$. The total momentum and energy flowing into the vertex, and the total number of external legs is given by,
\begin{align}
    \vb{K}_{i} & = \sum_{j=1}^{B_{i}}k^{(i)}_{j}, & E_{i} & = \sum_{j=1}^{B_{i}}|k^{(i)}_{j}|, & B = \sum_{j=1}^{V} B_{j}.
\end{align}

\subsection{Barred Correlation Function}

In order to proceed, let us properly define the \textit{barred} correlation function used in section \ref{sec:toy-model} for a general tree-level linear chain graph. As denoted in section \ref{sec:sk}, we are going to denote $\mathcal{A}$ to be a function depending on a group of external momenta $\{k_{\text{ext}}\}$, internal momenta $\{s_{\text{int}}\}$, and vertex sign indices $\{a\}$:
\begin{multline}
    \mathcal{A}^{V} \equiv \mathcal{A}(\{k_{\text{ext}}\};\{s_{\text{int}}\};\{ a \}) = \\
    \prod_{j=1}^{V}\left[\int_{-\infty}^{0}d\tau_{j}(-i\lambda^{(j)})\,a_{j}\prod_{l=1}^{B_{j}}G_{a_{j}}\left(k^{(j)}_{l};\tau_{j}\right)\right]\prod_{m=1}^{V-1}G_{a_{m}a_{m+1}}(s_{m};\tau_{m},\tau_{m+1}),
    \label{eq:n-point}
\end{multline}
corresponding to the usual correlation function prior to the overall summation on the vertex sign indices $a$. Thus, by construction, it is related to the $V-$vertex correlator by:
\begin{equation}
    \expval{ \prod_{j=1}^{V}\prod_{l=1}^{B_{j}}\varphi(k_{l}^{(j)}) }^{\prime} = \sum_{a_{1},\dots,a_{V}} \mathcal{A}(\{k_{\text{ext}}\};\{s_{\text{int}}\};\{ a \}).
    \label{def:correlator}
\end{equation}
In this context, the prime notation on the correlators is simply a matter of not writing the Dirac delta function of momentum conservation, and the factor $\lambda^{(j)}$ is the coupling constant which depends on the numbers of bulk-to-bulk and bulk-to-boundary propagators attached to the $j$th vertex. Then, \textit{e.g.}, if there are four bulk-to-boundary propagators and two bulk-to-bulk propagators attached to the $j$th vertex, then $\lambda^{(j)} = \lambda_{6}$. Diagrammatically, this example is seen as:
\begin{figure}[H]
\centering
\includegraphics[width=0.85\textwidth]{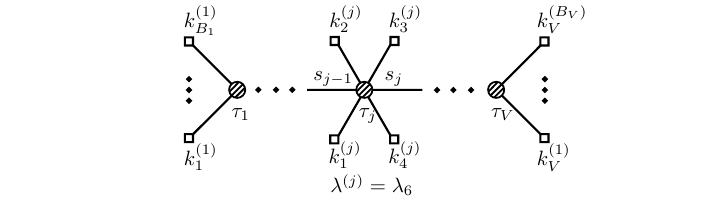}
\end{figure}
\noindent Let us perform one of the $V$ summations over the vertex indices:
\begin{align}
    & \expval{ \prod_{j=1}^{V}\prod_{l=1}^{B_{j}}\varphi(k_{l}^{(j)}) }^{\prime} = \sum_{\substack{a_{j} \\
    j\neq n}} \mathcal{A}(\{k_{\text{ext}}\};\{s_{\text{int}}\};\{ a \}, a_{n}=+) + \mathcal{A}(\{k_{\text{ext}}\};\{s_{\text{int}}\};\{ a \}, a_{n}=-) \nonumber \\
    & \phantom{ \prod_{j=1}^{V}\prod_{l=1}^{B_{j}} } = \sum_{\substack{a_{j} \\
    j\neq n}} \mathcal{A}(\{k_{\text{ext}}\};\{s_{\text{int}}\};\{ a \}, a_{n}=+) + (-1)^{V}\mathcal{A}^{\ast}(\{k_{\text{ext}}\};\{s_{\text{int}}\};\{ - a \}, a_{n}=+),
    \label{eq:corr-one-sum-performed}
\end{align}
where $a_{j}$ is just short notation for $a_{1},\dots,a_{V}$. In the second line we have used the complex conjugate of $\mathcal{A}$, providing an overall $(-1)^{V}$ factor due to the $i^{V}$ imaginary unit, whilst interchanging every $a\rightarrow -a$. Therefore, depending on $V$, we will have two possible results:
\begin{equation}
    \expval{ \prod_{j=1}^{V}\prod_{l=1}^{B_{j}}\varphi(k_{l}^{(j)}) }^{\prime} =
    \begin{cases}
        \displaystyle 2 \Re\left(\sum_{\substack{a_{j} \\ j\neq n}}\mathcal{A}(\{k_{\text{ext}}\};\{s_{\text{int}}\};\{ a \}, a_{n}=+)\right), & \text{ even } $V$, \\
        \displaystyle 2 i \Im\left(\sum_{\substack{a_{j} \\ j\neq n}}\mathcal{A}(\{k_{\text{ext}}\};\{s_{\text{int}}\};\{ a \}, a_{n}=+)\right), & \text{ odd } $V$.
    \end{cases}
    \label{eq:correlation-fn-fixing-a}
\end{equation}
Note that there are multiple ways of writing \cref{eq:correlation-fn-fixing-a} since $n$ can range from $1$ to $V$. We are also going to define a general \textit{barred} correlator as:
\begin{equation}
    \expvalbar{ \prod_{j=1}^{V}\prod_{l=1}^{B_{j}}\varphi(k_{l}^{(j)}) }^{\prime}_{a_{I_{n}}} = \sum_{a_{j}}\prod_{i\in I}a_{i}\mathcal{A}(\{k_{\text{ext}}\};\{s_{\text{int}}\};\{ a \}),
    \label{def:barred-corr}
\end{equation}
where $I_{n} = \{i_{1},\dots, i_{n}\}$ is an index set where $i_{j-1}\leq i_{j} \leq i_{j+1}$ ranges from $1$ to $V$. Bear in mind that if two indices are equal, $i_{m}=i_{m+1}$, the corresponding correlator will be unaffected by these indices, mathematically:
\begin{equation}
    \expvalbar{ \prod_{j=1}^{V}\prod_{l=1}^{B_{j}}\varphi(k_{l}^{(j)}) }^{\prime}_{a_{I_{n}}} = \sum_{a_{j}}\prod_{\substack{i \in I \\
        i \neq m, m+1}}a_{i}\mathcal{A}(\{k_{\text{ext}}\};\{s_{\text{int}}\};\{ a \}) = \expvalbar{ \prod_{j=1}^{V}\prod_{l=1}^{B_{j}}\varphi(k_{l}^{(j)}) }^{\prime}_{a_{I^{\prime}_{n}}},
\end{equation}
where $I^{\prime}_{n} = I_{n}\backslash \{i_{m}, i_{m+1}\}$, the above is true because $a^{2}_{m}=1$ $\forall m$. Note that \cref{def:barred-corr} is a generalisation of a typical correlation function, which is recovered by taking $I=\varnothing$. This definition will be of assistance when computing discontinuity operations on correlation functions. In order to grasp intuition about this new correlator, let us start analysing a particular case when the index set consists of a single element, $I_{1}=\{n\}$, then by performing the $a_{n}$ sum:
\begin{align}
    & \expvalbar{ \prod_{j=1}^{V}\prod_{l=1}^{B_{j}}\varphi(k_{l}^{(j)}) }^{\prime}_{a_{n}} = \sum_{a_{j}}a_{n}\mathcal{A}(\{k_{\text{ext}}\};\{s_{\text{int}}\};\{ a \}) \nonumber \\
    & \phantom{ \prod_{j=1}^{V}\prod_{l=1}^{B_{j}} } = \sum_{\substack{a_{j} \\
    j\neq n}} \mathcal{A}(\{k_{\text{ext}}\};\{s_{\text{int}}\};\{ a \}, a_{n}=+) - \mathcal{A}(\{k_{\text{ext}}\};\{s_{\text{int}}\};\{ a \}, a_{n}=-) \nonumber \\
    & \phantom{ \prod_{j=1}^{V}\prod_{l=1}^{B_{j}} } = \sum_{\substack{a_{j} \\
    j\neq n}} \mathcal{A}(\{k_{\text{ext}}\};\{s_{\text{int}}\};\{ a \}, a_{n}=+) - (-1)^{V}\mathcal{A}^{\ast}(\{k_{\text{ext}}\};\{s_{\text{int}}\};\{ - a \}, a_{n}=+),
\end{align}
where the steps are similar to the ones employed in \cref{eq:corr-one-sum-performed}, with an extra minus sign. Similarly, depending on $V$,
\begin{equation}
    \expvalbar{ \prod_{j=1}^{V}\prod_{l=1}^{B_{j}}\varphi(k_{l}^{(j)}) }^{\prime}_{a_{n}} =
    \begin{cases}
        \displaystyle 2 i \Im\left(\sum_{\substack{a_{j} \\ j\neq n}}\mathcal{A}(\{k_{\text{ext}}\};\{s_{\text{int}}\};\{ a \}, a_{n}=+)\right), & \text{ even } $V$, \\
        \displaystyle 2 \Re\left( \sum_{\substack{a_{j} \\ j\neq n}} \mathcal{A}(\{k_{\text{ext}}\};\{s_{\text{int}}\};\{ a \}, a_{n}=+)\right), & \text{ odd } $V$.
    \end{cases}
    \label{eq:bar-correlation-fn-fixing-a}
\end{equation}
Hence, the \textit{barred} correlator (\ref{eq:bar-correlation-fn-fixing-a}) is defined by swapping $2\Re$ and $2i\Im$ on the correlation function (\ref{eq:correlation-fn-fixing-a}). It is quite straightforward to see that we will have similar relations when considering the cardinality of the index set $I$ to be greater than one. In general,
\begin{align}
    & \expvalbar{ \prod_{j=1}^{V}\prod_{l=1}^{B_{j}}\varphi(k_{l}^{(j)}) }^{\prime}_{a_{I_{n}}} = \sum_{a_{j}}\prod_{i\in I_{n}} \mathcal{A}(\{k_{\text{ext}}\};\{s_{\text{int}}\};\{ a \}) \nonumber \\
    & = \sum_{\substack{a_{j} \\
    j\neq i_{n}}} \prod_{i\in I_{n-1}}a_{i} \mathcal{A}(\{k_{\text{ext}}\};\{s_{\text{int}}\};\{ a \}, a_{i_{n}}=+) - \prod_{i\in I_{n-1}}a_{i}\mathcal{A}(\{k_{\text{ext}}\};\{s_{\text{int}}\};\{ a \}, a_{i_{n}}=-) \nonumber \\
    & = \sum_{\substack{a_{j} \\
    j\neq i_{n}}} \prod_{i\in I_{n-1}}a_{i} \mathcal{A}(\{k_{\text{ext}}\};\{s_{\text{int}}\};\{ a \}, a_{i_{n}}=+) \nonumber \\ 
    & \phantom{ = \sum_{\substack{a_{j} \\
    j\neq i_{n}}} \prod_{i\in I_{n-1}}a_{i} \mathcal{A}(\{k_{\text{ext}}\};\{s_{\text{int}}\} } + (-1)^{V+|I|}\prod_{i\in I_{n-1}}a_{i} \mathcal{A}^{\ast}(\{k_{\text{ext}}\};\{s_{\text{int}}\};\{ -a \}, a_{i_{n}}=+),
    \label{eq:corr-fn-general-a_In}
\end{align}
where $I_{n-1} = I_{n}\backslash \{i_{n}\}$ is the $I_{n}$ set with the last element $i_{n}$ removed. Similarly,
\begin{align}
    & \expvalbar{ \prod_{j=1}^{V}\prod_{l=1}^{B_{j}}\varphi(k_{l}^{(j)}) }^{\prime}_{a_{I_{n-1}}} = \sum_{a_{j}}\prod_{i\in I_{n-1}}a_{i} \mathcal{A}(\{k_{\text{ext}}\};\{s_{\text{int}}\};\{ a \}) \nonumber \\
    & \phantom{\expvalbar{ \prod_{j=1}^{V}\prod_{l=1}^{B_{j}}\varphi(k_{l}^{(j)}) }^{\prime}_{a_{I_{n-1}}}} = \sum_{\substack{a_{j} \\
    j\neq i_{n}}} \prod_{i\in I_{n-1}}a_{i} \mathcal{A}(\{k_{\text{ext}}\};\{s_{\text{int}}\};\{ a \}, a_{i_{n}}=+) \nonumber \\ 
    & \phantom{ = \sum_{\substack{a_{j} \\
    j\neq i_{n}}} \prod_{i\in I_{n-1}}a_{i} \mathcal{A}(\{k_{\text{ext}}\};\{s_{\text{int}}\} } - (-1)^{V+|I|}\prod_{i\in I_{n-1}}a_{i} \mathcal{A}^{\ast}(\{k_{\text{ext}}\};\{s_{\text{int}}\};\{ -a \}, a_{i_{n}}=+).
    \label{eq:corr-fn-general-a_In-1}
\end{align}
Hence, (\ref{eq:corr-fn-general-a_In}) and (\ref{eq:corr-fn-general-a_In-1}) are related by a $2\Re$ and $2i\Im$ swapping:
\begin{equation}
    \expvalbar{ \prod_{j=1}^{V}\prod_{l=1}^{B_{j}}\varphi(k_{l}^{(j)}) }^{\prime}_{a_{I_{n}}} \xleftrightarrow{2\Re\leftrightarrow 2i\Im\,} \expvalbar{ \prod_{j=1}^{V}\prod_{l=1}^{B_{j}}\varphi(k_{l}^{(j)}) }^{\prime}_{a_{I_{n-1}}}.
\end{equation}
We will make use of these \textit{barred} correlators when computing general correlators discontinuities. For the linear chain graph case, it will suffice to use a barred correlator with $0\leq |I|\leq 2$, this is, an index set $I$ with at most two elements.

Additionally, it is worth noting that the structure of barred correlators resembles the form of the advanced fields from the mixed $r/a$ (Keldysh basis). By definition, these fields are related to the original SK fields on the forwards ($+$) and backward ($-$) contours by the linear combinations  \cite{Ema:2024hkj}:
\begin{equation}
    \varphi_{r} \equiv \dfrac{1}{2}\left( \varphi_{+} + \varphi_{-} \right), \qquad \varphi_{a} \equiv \varphi_{+} - \varphi_{-}.
\end{equation}
A connection between the cutting rules derived in \cite{Ema:2024hkj} and this work may be tempting. The former work shows a cutting rule for the in-in correlator itself in the Keldysh $r/a$ basis, rewriting diagrams into fully retarded functions and cut propagators (Wightman functions) absent of time-ordering Heaviside functions, hence conformal-time factorisable. However, our approach focuses around applying an explicit algebraic operation tailored by cosmological unitarity into the in-in correlators and how these are written in terms of lower-point correlators under the same operation, where barred correlators were introduced, which will serve to obtain a closed expression. Since the former objects should not contribute to the physical correlation function, the formulation in a different Schwinger–Keldysh basis is therefore not straightforward.

To investigate the cutting behaviour of \cref{def:correlator}, we take the discontinuity operation defined on (\ref{disc-op}) with every $V-1$ internal momenta $s_{1},\dots, s_{V-1}$ as subscript of the operation.
\begin{multline}
    \disc{\sum_{a_{1}, \dots, a_{V}}\mathcal{A}^{V}}{\vb{s}_{1}\cdots \vb{s}_{V-1}} = \sum_{a_{1}, \dots, a_{V}}\prod_{j=1}^{V}\left[\int_{-\infty}^{0}d\tau_{j}(-i\lambda^{(j)})\,a_{j}\prod_{l=1}^{B_{j}}G_{a_{j}}\left(k^{(j)}_{l};\tau_{j}\right)\right] \times \\
    \times \left[\prod_{m=1}^{V-1}G_{a_{m}a_{m+1}}(s_{m};\tau_{m},\tau_{m+1}) + (-1)^{B+2V}\prod_{m=1}^{V-1}G_{(-a_{m})(-a_{m+1})}(s_{m};\tau_{m},\tau_{m+1})\right], \label{eq:V-vertex-disc}
\end{multline}
where the extra $(-1)^B$ comes from (\ref{property-bulk-to-boundary-prop}) while $(-1)^{2V}=1$ is from the imaginary unit and the sign factor $a$. Thus, we need to study the structure of the last line in the above equation. 
\subsection{Fully Positive Diagram}
Let us start gaining intuition by considering the \textit{fully positive diagram}, that is, where $a_{i}=+$ for every $i$:
\begin{equation}
    \prod_{m=1}^{V-1}G_{++}(s_{m};\tau_{m},\tau_{m+1}) + (-1)^{B}\prod_{m=1}^{V-1}G_{--}(s_{m};\tau_{m},\tau_{m+1}),
\end{equation}
exchanging every Heaviside function from $\Theta(\tau_{j+1}-\tau_{j})$ to $1 -\Theta(\tau_{j}-\tau_{j+1})$ on the last product, we can write $G_{--}$ in terms of $G_{++}$ while adding extra factors depending on $G_{+}$ and $G_{-}$,
\begin{multline}
    \prod_{m=1}^{V-1}G_{++}(s_{m};\tau_{m},\tau_{m+1}) \\ + (-1)^{B}\prod_{m=1}^{V-1}\left[ 2\Re\left(G_{+}(s_{m};\tau_{m})G_{-}(s_{m};\tau_{m+1})\right) - G_{++}(s_{m};\tau_{m},\tau_{m+1})\right].
    \label{eq:g--intog++prop}
\end{multline}
Expanding the last product,
\begin{multline}
(1 + (-1)^{B+V-1})\prod_{m=1}^{V-1}G_{++}(s_{m};\tau_{m},\tau_{m+1}) + (-1)^{B}\prod_{m=1}^{V-1}2\Re\left(G_{+}(s_{m};\tau_{m})G_{-}(s_{m};\tau_{m+1})\right) \\
+ (-1)^{B} \sum_{\substack{J\in P([V-1]) \\
                    |J|\neq 0, V-1}}\left( \prod_{j\notin J}2\Re\left(G_{+}(s_{j};\tau_{j})G_{-}(s_{j};\tau_{j+1})\right)\right)\left(\prod_{j\in J} -G_{++}(s_{j};\tau_{j},\tau_{j+1}) \right),
                    \label{eq:n-point-cut-1}
\end{multline}
where $P([V-1])$ corresponds to the power set of a set $[V-1]$, a shorthand notation for $\{1,\dots, V-1\}$, and $|J|$ is the cardinality of the subset $J$, an index set. Here we have purposely extracted two terms corresponding to $|J|=0, V-1$ displayed on the first line in the above equation. We then can use the following fact for the real part:
\begin{multline}
    \Re\left(G_{+}(s_{m};\tau_{m})G_{-}(s_{m};\tau_{m+1})\right) = \Re\left(G_{+}(s_{m};\tau_{m})\right)\Re\left(G_{+}(s_{m};\tau_{m+1})\right) \\
    + \Im\left(G_{+}(s_{m};\tau_{m})\right)\Im\left(G_{+}(s_{m};\tau_{m+1})\right),
    \label{eq:real-part-bulk-to-bulk-prop-not-gen}
\end{multline}
where the identities $\Re\left(G_{+}\right) = \Re\left(G_{-}\right)$ and $\Im\left(G_{+}\right) = - \Im\left(G_{-}\right)$ have been used. In order to have a \textit{cut} diagram, the first term of (\ref{eq:n-point-cut-1}) should vanish, that is because it corresponds to a totally connected diagram\footnote{This is just referring to the fact that it possesses $V-1$ step functions, therefore, every temporal integral is nested.}. To ensure it is zero, when performing the discontinuity operation, where  in general, two options are available: taking $\disc{\mathcal{A}}{s}$ or $i\disc{i\mathcal{A}}{s}$, we select the appropriate one such that we have the following factors associated to the fully connected term:
\begin{align}
    (1 + (-1)^{B+V}), & = 0 \qquad \text{for odd } B+V, \label{eq:cancel-cond-odd} \\
    (1 + (-1)^{B+V-1}), & = 0 \qquad \text{for even } B+V. \label{eq:cancel-cond-even}
\end{align}
Then, (\ref{eq:n-point-cut-1}) reads:
\begin{align}
    & (-1)^{B} \sum_{J\in P([V-1])} \Bigg( \prod_{j\notin J} 2\Re\left(G_{+}(s_{j};\tau_{j})\right) \Re\left(G_{+}(s_{j};\tau_{j+1})\right)\Bigg) \times \nonumber \\ 
    & \phantom{(-1)^{B} \sum_{J\in P([V-1])}} \times \left(\prod_{j\in J} 2\Im\left(G_{+}(s_{j};\tau_{j})\right)\Im\left(G_{+}(s_{j};\tau_{j+1})\right) \right) \nonumber \\
    + & (-1)^{B} \sum_{\substack{J\in P([V-1]) \\
    |J|\neq 0, V-1}}(-1)^{|J|}\Bigg[ \sum_{L\subseteq J^{C}} \Bigg( \prod_{l\notin L} 2\Re\left(G_{+}(s_{l};\tau_{l})\right)\Re\left(G_{+}(s_{l};\tau_{l+1})\right)\Bigg) \times \nonumber \\ 
    & \phantom{(-1)^{B}} \times \left(\prod_{l\in L} 2\Im\left(G_{+}(s_{l};\tau_{l})\right)\Im\left(G_{+}(s_{l};\tau_{l+1})\right) \right)\Bigg]\left(\prod_{j\in J} G_{++}(s_{j};\tau_{j},\tau_{j+1}) \right),
    \label{eq:disc-g++expansion}
\end{align}
where $J^{C}$ corresponds to the complement set of $J$. At this point we are able to roughly interpret the above expression diagrammatically by inspecting the absence (or not) of bulk-to-bulk propagators. The first sum will correspond to $2^{V-1}$ terms, which can then be expressed as products of one-vertex discontinuity (\textit{e.g.}, a three-point one-vertex in a $\varphi^{3}$ theory)\footnote{More details on the one-vertex discontinuity computation can be found in Appendix \ref{sec:disc}.}. The second term corresponds to $3^{V-1}-2^{V-1}-1$ elements\footnote{Specifically, this number arises from the cardinality of the power set of $V-1$ elements, which is: 
\begin{equation}
    |P([V-1])| = \sum_{i=0}^{V-1} \binom{V-1}{i} = 2^{V-1}.
\end{equation}
Therefore, the total number of elements in the sum is given by, 
\begin{equation}
    \sum_{k=1}^{V-2} \binom{V-1}{k} 2^{V-1-k} = 3^{V-1} - 2^{V-1} - 1,
\end{equation}
where $2^{V-1-k}$ corresponds to the cardinality of the power set $P(J^{C})$, with $k$ being summed from 1 to $V-2$, since $|J| \neq 0, V-1$. The final result is obtained applying the binomial theorem.} that will individually contribute to the sum of $n-$vertex correlation function discontinuity, with $n = 2, \dots, V-1$. For instance, the term on the sum associated to $J=\{ 1, \dots, V-2 \}$:
\begin{multline}
    (-1)^{B+2V-2} \prod_{j=1}^{V-2}G_{++}(s_{j}; \tau_{j}, \tau_{j+1}) 2\Re\left(G_{+}(s_{V-1};\tau_{V-1})\right)\Re\left(G_{+}(s'_{V-1};\tau_{V})\right) \\
    + (-1)^{B+2V-2} \prod_{j=1}^{V-2}G_{++}(s_{j}; \tau_{j}, \tau_{j+1}) 2\Im\left(G_{+}(s_{V-1};\tau_{V-1})\right)\Im\left(G_{+}(s'_{V-1};\tau_{V})\right),
\end{multline}
will be written, when considering bulk-to-boundary propagators and remaining factors, as a product of the discontinuity of a one-vertex correlation function (given the real and imaginary parts of the bulk-to-boundary propagator at time $\tau_{V}$) times a discontinuity of an \textit{all-plus} $(V-1)-$diagram contribution, which will become the $(V-1)-$vertex correlation function when adding the remaining diagrams\footnote{This statement is just to gain intuition, we will provide a more formal statement in the next computation.}. In a picture the above is represented as:
\begin{figure}[H]
\centering
\includegraphics[width=0.72\textwidth]{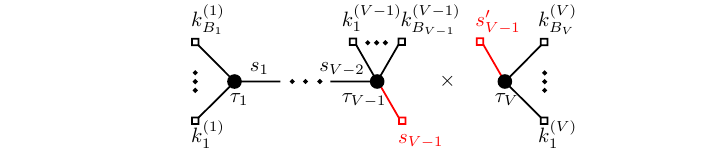}
\end{figure}
\subsection{The Complete Correlator}
Let us then consider the full discontinuity (\ref{eq:V-vertex-disc}). We start by noting the following property of bulk-to-bulk propagators,
\begin{equation}
    G^{\ast}_{ab}(s;\tau_{i},\tau_{j}) = 2\Re\left( G_{a}(s;\tau_{i})G_{(-a)}(s;\tau_{j}) \right) - G_{ab}(s;\tau_{i},\tau_{j}),
\end{equation}
corresponding to a generalisation of the one used in \cref{eq:g--intog++prop}. Then, by expanding the $V-1$ multi-binomial term provided by the above property we obtain: 
\begin{align}
& (1 + (-1)^{B+V-1})\prod_{m=1}^{V-1}G_{a_{m}a_{m+1}}(s_{m};\tau_{m},\tau_{m+1}) \nonumber\\
+ & (-1)^{B}\prod_{m=1}^{V-1}2\Re\left( G_{a_{m}}(s_{m};\tau_{m})G_{(-a_{m})}(s_{m};\tau_{m+1}) \right) \nonumber\\
+ & (-1)^{B} \sum_{\substack{J\in P([V-1]) \\
                    |J|\neq 0, V-1}}\left( \prod_{j\notin J}2\Re\left( G_{a_{m}}(s_{m};\tau_{m})G_{(-a_{m})}(s_{m};\tau_{m+1}) \right)\right)\times \nonumber \\ 
&\phantom{ (-1)^{B} \sum_{\substack{J\in P([V-1]) \\
                    |J|\neq 0, V-1}} \Bigg( \prod_{j\notin J}2\Re ( G_{a_{m}}(s_{m};\tau_{m}) }\times\left(-\prod_{j\in J} G_{a_{m}a_{m+1}}(s_{m};\tau_{m},\tau_{m+1}) \right).
    \label{eq:sum-term-disc-op}
\end{align}
Furthermore, we can eliminate the first term by the same argument used in (\ref{eq:cancel-cond-odd}) and (\ref{eq:cancel-cond-even}) for the \textit{fully positive diagram} case. To further interpret the remaining terms let us expand the real part of the two bulk-to-boundary product, in other words, a generalisation of \cref{eq:real-part-bulk-to-bulk-prop-not-gen},
\begin{multline}
    \Re\left( G_{a_{m}}(s_{m};\tau_{m})G_{(-a_{m})}(s_{m};\tau_{m+1}) \right) = \Re\left( G_{a_{m}}(s_{m};\tau_{m}) \right)\Re\left( G_{a_{m+1}}(s_{m};\tau_{m+1}) \right) \\
    + a_{m}a_{m+1}\Im\left( G_{a_{m}}(s_{m};\tau_{m}) \right)\Im\left( G_{a_{m+1}}(s_{m};\tau_{m+1}) \right).
    \label{real-property-with-two-a}
\end{multline}
It is important to stress that on the right-hand side of the above property we have an expression dependent of $a_{m+1}$, whilst the left-hand side is not. This can be obtained by recalling the bulk-to-boundary identities $\Re\left(G_{+}\right) = \Re\left(G_{-}\right)$ and $\Im\left(G_{+}\right) = - \Im\left(G_{-}\right)$. Thus, \cref{eq:sum-term-disc-op} is now expanded,
\begin{align}
    & (-1)^{B} \sum_{J\in P([V-1])} \Bigg( \prod_{j\notin J} 2 \Re\left( G_{a_{j}}(s_{j};\tau_{j}) \right)\Re\left( G_{a_{j+1}}(s_{j};\tau_{j+1}) \right) \Bigg) \times \nonumber \\
    & \phantom{ (-1)^{B} \sum_{J\in P([V-1])} } \times \Bigg( \prod_{j\in J} 2 a_{j} \Im\left( G_{a_{j}}(s_{j};\tau_{j}) \right) a_{j+1} \Im\left( G_{a_{j+1}}(s_{j};\tau_{j+1}) \right) \Bigg) \nonumber \\
+ & (-1)^{B} \sum_{\substack{J\in P([V-1]) \\
    |J|\neq 0, V-1}}(-1)^{|J|}\Bigg[ \sum_{L\subseteq J^{C}} \Bigg( \prod_{l\notin L} 2\Re\left(G_{a_{l}}(s_{l};\tau_{l})\right) \Re\left(G_{a_{l+1}}(s_{l};\tau_{l+1})\right) \Bigg) \times \nonumber \\ 
    & \times \left(\prod_{l\in L} 2 a_{l} \Im\left(G_{a_{l}}(s_{l};\tau_{l})\right) a_{l+1} \Im\left(G_{a_{l+1}}(s_{l};\tau_{l+1})\right) \right)\Bigg]\left(\prod_{j\in J} G_{a_{j}a_{j+1}}(s_{j};\tau_{j},\tau_{j+1}) \right),
    \label{eq:disc-two-sum-expansion-general-a}
\end{align}
which is essentially the same as \cref{eq:disc-g++expansion} however, extended to account every vertex ($+$ or $-$) combination. Now comes the task of writing this expression in terms of correlation functions possessing $V-1, \dots, 1$ number of vertices, or equivalently, lower-point correlation functions. Let us start by inspecting the simpler terms, that is, extracting from the above equation, the ones \textit{exclusively} involving the real part. We have to consider contributions from both sums. Then we analyse each case individually before combining the results\footnote{We already have some intuition about the first sum, since, as pointed out in the \textit{fully positive diagram} case, it will correspond to products of one-vertex correlation functions.}. Taking one of the elements on the second sum $J=\{ 1, \dots, V-2 \}$,
\begin{equation}
    (-1)^{B+V}\prod_{j=1}^{V-2}G_{a_{j}a_{j+1}}(s_{j};\tau_{j},\tau_{j+1}) 2\Re\left( G_{a_{V-1}}(s_{V-1};\tau_{V-1}) \right)\Re\left( G_{a_{V}}(s_{V-1};\tau_{V}) \right),
\end{equation}
where we have used $(-1)^{B+V-2}=(-1)^{B+V}$. Reinstating back the temporal integrals and overall constants,
\begin{multline}
    (-1)^{B-B_{V}+V-1}\sum_{a_{1}, \dots, a_{V-1}}\prod_{j=1}^{V-1}\left[\int_{-\infty}^{0}d\tau_{j}(-i\lambda^{(j)})\,a_{j}\prod_{l=1}^{B_{j}}G_{a_{j}}\left(k^{(j)}_{l};\tau_{j}\right)\right] \times \\
    \times \prod_{m=1}^{V-2}G_{a_{m}a_{m+1}}(s_{m};\tau_{m},\tau_{m+1})
    2\Re\left( G_{a_{V-1}}(s_{V-1};\tau_{V-1}) \right) \times \\
    \times (-1)^{B_{V}+1} \sum_{a_{V}} \int_{-\infty}^{0}d\tau_{V}(-i\lambda^{(V)})\,a_{V}\prod_{l=1}^{B_{V}}G_{a_{V}}\left(k^{(V)}_{l};\tau_{V}\right)\Re\left( G_{a_{V}}(s_{V-1};\tau_{V}) \right).
\end{multline}
It can be noted that, if $B-B_{V-1}-1$ and $B_{V}+1$ are even numbers, then we can utilise \cref{disc-(v-1)-vertex,disc-1-vertex} to rewrite the above in terms of discontinuities of $V-1$ and one-point correlators as follows\footnote{If $B-B_{V-1}-1$ and $B_{V}-1$ are odd numbers or combinations between even and odd, is analogue. In that case we use instead $i\disc{i\,\cdot}{}$.},
\begin{multline}
    \dfrac{(-1)^{V-|J|}}{2^{|J|}} \int_{\vb{s_{V-1}}\vb{s_{V-1}^{\prime}}}\disc{\sum_{a_{1}, \dots, a_{V-1}}\mathcal{A}^{V-1}(\{ k^{\text{ext}}_{i}, s_{V-1}\}; \{s_{1},\dots,s_{V-2}\}; \{a_{i}\})}{s_{V-1}}\times \\
    \times \disc{\sum_{a_{V}}\mathcal{A}^{1}(\{s_{V-1}^{\prime}, k^{\text{ext}}_{V} \};\{\}; \{ a_{V}\}) }{s_{V-1}^{\prime}},
    \label{eq:first-amplitude-disc-factorization}
\end{multline}
where $|J|=1$, is the instruction to cut through the propagator with internal momentum $s_{V-1}$ and $k^{\text{ext}}_{i}$ corresponds to the external momenta $k^{(1)}_{i},\dots,k^{(B_{i})}_{i}$, where $i$ is the associated vertex number, thus, ranging from $1$ to $V-1$. Additionally, the first integral was already introduced,
\begin{equation}
    \int_{\vb{s}\vb{s}^{\prime}}f(s,s^{\prime}) = \int d^{3}s\, d^{3}s'\, \dfrac{\delta^{(3)}(\vb{s}-\vb{s'})}{u(\tau_{f};s)u^{\ast}(\tau_{f};s')}.
\end{equation}
The empty set in \cref{eq:first-amplitude-disc-factorization} is just informing that the diagram does not posses any internal momenta. Note that $s_{V-1}$ is no longer an internal energy, yet an external one. Diagrammatically, this statement translates to:
\begin{figure}[H]
\centering
\includegraphics[width=0.85\textwidth]{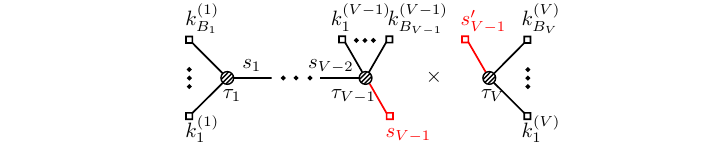}
\end{figure}
\noindent Where red lines correspond to taking a discontinuity operation associated to that momenta. Furthermore, let us study one of the terms appearing at the first sum in \cref{eq:disc-two-sum-expansion-general-a} consisting entirely of real products,
\begin{equation}
    (-1)^{B+V}\prod_{j=1}^{V-1}2\Re\left( G_{a_{j}}(s_{j};\tau_{j}) \right)\Re\left( G_{a_{j+1}}(s_{j};\tau_{j+1}) \right).
    \label{eq:disc-v-vertex-real-contribution}
\end{equation}
Reinstating the $V$ temporal integrals and overall constants:
\begin{multline}
    (-1)^{B_{1}+1}\sum_{a_{1}}\int_{-\infty}^{0}d\tau_{1}(-i\lambda^{(1)})\,a_{1}\prod_{l=1}^{B_{1}}G_{a_{1}}(k_{l}^{(1)};\tau_{1})2\Re\left( G_{a_{1}}(s_{1};\tau_{1}) \right) \times \\
    \times \prod_{j=2}^{V-1}(-1)^{B_{j}+1}\sum_{a_{j}}\int_{-\infty}^{0}d\tau_{j}(-i\lambda^{(j)})\,2a_{j}\Re\left( G_{a_{j}}(s_{j};\tau_{j}) \right)\prod_{l=1}^{B_{j}}G_{a_{j}}(k_{l}^{(j)};\tau_{j})\Re\left( G_{a_{j+1}}(s_{j};\tau_{j}) \right) \times \\
    \times (-1)^{B_{V}+1}\sum_{a_{V}}\int_{-\infty}^{0}d\tau_{V}(-i\lambda^{(V)})\,a_{V}\prod_{l=1}^{B_{V}}G_{a_{V}}(k_{l}^{(V)};\tau_{V})\Re\left( G_{a_{V}}(s_{V-1};\tau_{V}) \right).
    \label{eq:disc-v-vertex-real-contribution-t-int}
\end{multline}
Subsequently, we can rewrite the later in terms of one-vertex point discontinuity operations using \cref{disc-1-vertex}. Taking $B_{i}+1$ to be even numbers\footnote{The odd number discussion is similar to the already examined case, we just have to switch $\disc{\cdot}{}\leftrightarrow i\disc{i\,\cdot}{}$. } for all $i=1,\dots,V$,
\begin{multline}
    \dfrac{(-1)^{V-1-|J|}}{2^{|J|}}\int_{\vb{s_{1}}\cdots\,\vb{s_{V-1}^{\prime}}}\disc{\sum_{a_{1}}\mathcal{A}^{1}(\{ k_{1}, s_{1}\};\{\};\{a_{1}\})}{s_{1}} \times \\ \times \prod_{j=2}^{V-1}\disc{ \disc{\sum_{a_{j}}\mathcal{A}^{1}( \{ s_{j-1}^{\prime}, k_{j}, s_{j} \}; \{\}; \{ a_{j} \})}{s_{j}}}{s_{j-1}^{\prime}} \times \\
    \times \disc{\sum_{a_{V}}\mathcal{A}^{1}(\{ s_{V-1}^{\prime}, k_{V} \}; \{\}; \{ a_{V} \})}{s_{V-1}^{\prime}},
\end{multline}
where $|J|=V-1$, corresponding to cutting through every propagator. The integral is written as,
\begin{equation}
    \int_{\vb{s_{1}}\cdots\,\vb{s_{n}^{\prime}}} = \prod_{i=1}^{n}\int d^{3}s_{i}\, d^{3}s_{i}^{\prime}\, \dfrac{\delta^{(3)}(\vb{s}_{i}-\vb{s}_{i}^{\prime})}{u(\tau_{f};s_{i})u^{\ast}(\tau_{f};s_{i}^{\prime})}.
\end{equation}
Diagrammatically we can interpret these cuts as follows:
\begin{figure}[H]
\centering
\includegraphics[width=0.45\textwidth]{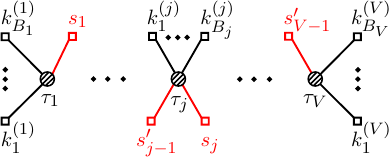}
\end{figure}
At this point, we already know how to deal with the real part contributions from \cref{eq:disc-two-sum-expansion-general-a}. Furthermore, let us extend the analysis when imaginary parts are present. We will observe the necessity of defining a different type of correlator in order to write it in a compact manner. For this, consider the imaginary contribution of the second sum in \cref{eq:disc-two-sum-expansion-general-a} when taking $J=\{ 1,\dots,V-2 \}$,
\begin{equation}
    (-1)^{B+V}\prod_{j=1}^{V-2}G_{a_{j}a_{j+1}}(s_{j};\tau_{j},\tau_{j+1}) 2a_{V-1}\Im\left( G_{a_{V-1}}(s_{V-1};\tau_{V-1}) \right)a_{V}\Im\left( G_{a_{V}}(s_{V-1};\tau_{V}) \right).
\end{equation}
Note the presence of $a_{V-1}$ and $a_{V}$. Restoring the temporal integrals:
\begin{multline}
    (-1)^{B-B_{V}+V-1}\sum_{a_{1}, \dots, a_{V-1} }\prod_{j=1}^{V-1}\left[\int_{-\infty}^{0}d\tau_{j}(-i\lambda^{(j)})\,a_{j}\prod_{l=1}^{B_{j}}G_{a_{j}}\left(k^{(j)}_{l};\tau_{j}\right)\right] \times \\
    \times \prod_{m=1}^{V-2}G_{a_{m}a_{m+1}}(s_{m};\tau_{m},\tau_{m+1})
    \,2a_{V-1}\Im\left( G_{a_{V-1}}(s_{V-1};\tau_{V-1}) \right) \times \\
    \times (-1)^{B_{V}+1}\sum_{a_{V}} \int_{-\infty}^{0} d\tau_{V}(-i\lambda^{(V)}) \,a_{V} \prod_{l=1}^{B_{V}}G_{a_{V}}\left(k^{(V)}_{l};\tau_{V}\right)a_{V}\Im\left( G_{a_{V}}(s_{V-1};\tau_{V}) \right).
\end{multline}
Since we are considering the parity of $B-B_{V}-1$ and $B_{V}+1$ to be even, we use \cref{idisci-1-vertex-a,idisci-(v-1)-vertex-a} to show that,
\begin{multline}
    \dfrac{(-1)^{V-1-|J|}}{2^{|J-1|}}\int_{\vb{s_{V-1}}\vb{s_{V-1}^{\prime}}}i\disc{\sum_{a_{1}, \dots, a_{V-1}} i a_{V-1} \mathcal{A}^{V-1}(\{ k^{\text{ext}}_{i}, s_{V-1}\} ; \{ s_{1},\dots,s_{V-2}\} ; \{a_{i}\}) }{s_{V-1}}\times \\ 
    \times i\disc{\sum_{a_{V} } i a_{V} \mathcal{A}^{1}(\{s_{V-1}^{\prime}, k^{\text{ext}}_{V} \}; \{ \}; \{ a_{V} \})}{s_{V-1}^{\prime}},
\end{multline}
where $|J|=1$. In parallel to \cref{eq:disc-v-vertex-real-contribution} we also have its imaginary contribution,
\begin{equation}
    (-1)^{B+V}\prod_{j=1}^{V-1}2a_{j}\Im\left( G_{a_{j}}(s_{j};\tau_{j}) \right)a_{j+1}\Im\left( G_{a_{j+1}}(s_{j};\tau_{j+1}) \right).
    \label{eq:disc-v-vertex-imaginary-contribution}
\end{equation}
Furthermore, reinstating the temporal integrals the above becomes:
\begin{multline}
    (-1)^{B_{1}+1}\int_{-\infty}^{0}d\tau_{1}(-i\lambda^{(1)})\,a_{1}\prod_{l=1}^{B_{1}}G_{a_{1}}(k_{l}^{(1)};\tau_{1})\,2a_{1}\Im\left( G_{a_{1}}(s_{1};\tau_{1}) \right) \times \\
    \times \prod_{j=2}^{V-1}(-1)^{B_{j}+1}\int_{-\infty}^{0}d\tau_{j}(-i\lambda^{(j)})\,a_{j}2a_{j}\Im\left( G_{a_{j}}(s_{j};\tau_{j}) \right)\prod_{l=1}^{B_{j}}G_{a_{j}}(k_{l}^{(j)};\tau_{j})\,a_{j}\Im\left( G_{a_{j+1}}(s_{j};\tau_{j}) \right) \times \\
    \times (-1)^{B_{V}+1}\int_{-\infty}^{0}d\tau_{V}(-i\lambda^{(V)})\,a_{V}\prod_{l=1}^{B_{V}}G_{a_{V}}(k_{l}^{(V)};\tau_{V})\,a_{V}\Im\left( G_{a_{V}}(s_{V-1};\tau_{V}) \right),
    \label{eq:disc-v-vertex-imaginary-contribution-t-int}
\end{multline}
which, in essence, looks exactly like \cref{eq:disc-v-vertex-real-contribution-t-int} interchanging $\Re \leftrightarrow \Im$, whilst adding extra sign factors $a_{1}$, $a_{j}^{2}=1$, and $a_{V}$. Therefore, \cref{eq:disc-v-vertex-imaginary-contribution-t-int} can be expressed as a product of $V$ one-vertex correlators by using \cref{idisci-1-vertex-a}:
\begin{multline}
    \dfrac{(-1)^{V-1-|J|}}{2^{|J|}}\int_{\vb{s_{1}}\cdots\,\vb{s_{V-1}^{\prime}}}i\disc{ \sum_{a_{1}} i a_{1} \mathcal{A}^{1}(\{ k_{1}, s_{1}\}; \{\}; \{a_{1}\})}{s_{1}} \times \\ \times \prod_{j=2}^{V-1}i\disc{i^{2} \disc{ \sum_{a_{j}} i\mathcal{A}^{1}(\{ s_{j-1}^{\prime}, k_{j}, s_{j} \}; \{\}; \{ a_{j} \})}{s_{j}}}{s_{j-1}^{\prime}} \times \\
    \times i\disc{\sum_{a_{V}}i a_{V} \mathcal{A}^{1}( \{s_{V-1}^{\prime}, k_{V} \}; \{\}; \{a_{V}\})}{s_{V-1}^{\prime}}.
\end{multline}
Note that there is a sign cancellation in the above equation, given by the minus sign at \cref{idisci-1-vertex-a}. Specifically, there are $2V$ negative signs inserted, one for each discontinuity: $2V-2$ internal double discontinuities and two for the subdiagrams located on the edges.

\subsection{Cutting Recipe}

Analysing the above cases allows us to define the complete discontinuity operation for a $V-$vertex correlation function in a linear chain graph configuration. The general rule for these types of diagrams is summarised as follows:
\begin{enumerate}
    \item \label{step:1} Identify the number of vertices of a given diagram. Compute the number $B = \sum_{j=1}^{V}B_{j}$, in other words, the total number of external legs. This determines the appropriate discontinuity operation to take:
    \begin{align}
        & \disc{\expval{\varphi^{V}}}{\substack{\text{internal} \\ \text{momenta}}}, \qquad \text{for odd } B+V, \label{step:eq:odd} \\
        i & \disc{i\expval{\varphi^{V}}}{\substack{\text{internal} \\ \text{momenta}}}, \qquad \text{for even } B+V. \label{step:eq:even}
    \end{align}
    \item Draw and sum over every possible diagrammatic cut to the internal channels. Mathematically this corresponds to taking the sum over the power set $P([V-1])$ while eliminating $\varnothing$, since the fully-connected term is guaranteed to be absent because of step \ref{step:1}. \label{step:2}
    \item \label{step:3} For each cut diagram, draw its connected components, or equivalently, its connected subdiagrams. Identify the number of vertices and external legs. This will indicate if we are going to compute $\disc{\expval{\varphi^{\text{sub}}}}{}$ or $i\disc{i\expval{\varphi^{\text{sub}}}}{}$, similarly to step \ref{step:1}.
    \item \label{step:4} Each cut modifies a bulk-to-bulk propagator into products of real and imaginary bulk-to-boundary components, with additional sign indices. This is just a manifestation of \cref{real-property-with-two-a}. Diagrammatically we can interpret this as:
    \begin{figure}[H]
    \centering
    \includegraphics[width=0.9\textwidth]{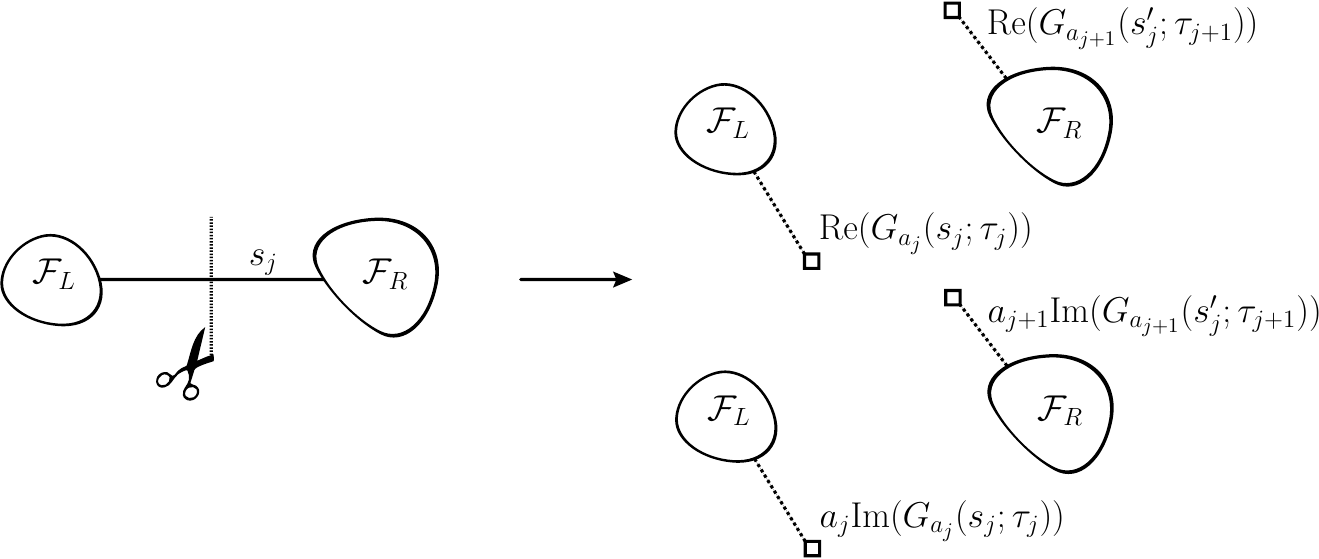}
    \end{figure}
    Where $\mathcal{F}_{L}$ and $\mathcal{F}_{R}$ are representing blobs, namely, indicating that these regions contain subdiagrams that are not being cut. We also are replacing each internal cut momenta $s$ by a pair $\{s,s^{\prime}\}$ whilst adding the integral: 
    \begin{equation}
        \int_{\vb{s}\vb{s}^{\prime}} = \int d^{3}s\,d^{3}s^{\prime} \dfrac{\delta^{(3)}(\vb{s}-\vb{s}^{\prime})}{u(\tau_{f};s)u^{\ast}(\tau_{f};s^{\prime})}.
        \label{eq:internal-mom-cut-integration} 
    \end{equation}
    \item \label{step:5} Identify and replace each subdiagram as a discontinuity operation of a correlation function or barred correlator (depending on the presence of extra sign indices), taking as an input the momenta that is being cut.
\end{enumerate}
Mathematically, we can write the above rule, previous summing over the vertex indices, for a $V-$vertex correlator\footnote{This is considering the parity of the sub-diagram to be consistent with the use of $\text{disc}(\cdot)$. If this is not true we simply need to interchange it with $i\text{disc}(i\,\cdot)$.}:
\begin{align}
    & \disc{\mathcal{A}^{V}}{s_{1}\cdots s_{V-1}} = \sum_{\substack{J\in P([V-1]) \\
    |J|\neq 0}} \sum_{b_{j}=0,1} \dfrac{(-1)^{|J|-1}}{2^{|J|}}\int_{\vb{s_{J}}\vb{s_{J}^{\prime}}} \left[\overline{\mathcal{I}}_{0}^{j_{1}}\,\disc{\mathcal{I}_{0}^{j_{1}}\left(\delta_{0b_{j_{1}}} + ia_{j_{1}}\delta_{1b_{j_{1}}}\right) \mathcal{A}^{j_{1}}}{s_{j_{1}}}\right] \times \nonumber \\
    & \times \prod_{\substack{L\subseteq J_{n-1} \\
    |J| > 1}}\Bigg[ \delta_{0b_{l_{k}}}\overline{\mathcal{I}}^{l_{k+1}}_{l_{k}+1}\,  \disc{\mathcal{I}^{l_{k+1}}_{l_{k}+1}\overline{\mathcal{I}}^{l_{k+1}}_{l_{k+1}}\,\disc{\mathcal{I}^{l_{k+1}}_{l_{k+1}}\left(\delta_{0b_{l_{k+1}}}+ia_{l_{k+1}}\delta_{1b_{l_{k+1}}}\right)\mathcal{A}^{l_{k+1}-l_{k}}}{s_{l_{k+1}}}}{s_{l_{k}}^{\prime}} \nonumber \\
    &  + i\delta_{1b_{l_{k}}}\overline{\mathcal{I}}^{l_{k+1}}_{l_{k}+1} \,  \disc{i\,\mathcal{I}^{l_{k+1}}_{l_{k}+1}\overline{\mathcal{I}}^{l_{k+1}}_{l_{k+1}}\,\disc{\mathcal{I}^{l_{k+1}}_{l_{k+1}}a_{l_{k}+1}\left(\delta_{0b_{l_{k}}}+ia_{l_{k+1}}\delta_{1b_{l_{k+1}}}\right)\mathcal{A}^{l_{k+1}-l_{k}}}{s_{l_{k+1}}}}{s_{l_{k}}^{\prime}} \Bigg] \times \nonumber \\
    & \phantom{ \disc{\sum_{a}\mathcal{A}^{V}}{s_{1}\cdots s_{V-1} = }} \times \left[\overline{\mathcal{I}}^{V}_{j_{n}+1}\,\disc{\mathcal{I}^{V}_{j_{n}+1}\left(\delta_{0b_{j_{n}}}+ia_{j_{n}+1}\delta_{1b_{j_{n}}}\right)\mathcal{A}^{V-j_{n}}}{s^{\prime}_{j_{n}}}\right],
    \label{eq:cutting-rule}
\end{align}
where we interpret $l_{k} \coloneqq j_{k}$ and $l_{k+1} \coloneqq j_{k+1}$, and the imaginary factors $\mathcal{I}^{n}_{m}, \overline{\mathcal{I}}^{n}_{m}$ are defined as:
\begin{align}
    \mathcal{I}^{n}_{m} & = (+i)^{\sum_{j=m}^{n}B_{j}+(j_{n}-j_{m})+ (j_{n}-j_{m}-1)} & \overline{\mathcal{I}}^{n}_{m} & = (-i)^{\sum_{j=m}^{n}B_{j}+(j_{n}-j_{m})+ (j_{n}-j_{m}-1)},
    \label{eq:imaginary-factors}
\end{align}
where we have defined the notation $B_{0}=0$, and $J_{n-1} = \{j_{1},\dots,j_{n-1}\}$. In the above, $j_{i}$ is the $i$th element of the index set $J$. Including the summation over the sign indices $a_{1},\dots, a_{V}$, we can rewrite the above in terms of correlation functions and \textit{barred} correlators as per (\ref{def:correlator}) and (\ref{def:barred-corr}),
\begin{align}
    & \disc{\expval{ \varphi^{V} }^{\prime}}{s_{1}\cdots s_{V-1}} = \sum_{\substack{J\in P([V-1]) \\
    |J|\neq 0}} \sum_{b_{j}=0,1} \dfrac{(-1)^{V-1-|J|}}{2^{|J|}} \int_{\vb{s_{j_{1}}}\cdots\,\vb{s_{j_{n}}^{\prime}}} \times \nonumber \\
    & \phantom{ \disc{\expval{ \varphi^{V} }^{\prime}}{s_{1}\cdots s_{V-1}} = \sum_{\substack{J\in P([V-1]) \\
    |J|\neq 0}} } \times \left[ \delta_{0b_{j_{1}}} \overline{\mathcal{I}}_{0}^{j_{1}} \, \disc{ \mathcal{I}_{0}^{j_{1}} \expval{ \varphi^{j_{1}} }^{\prime} }{s_{j_{1}}} + i \delta_{1b_{j_{1}}} \overline{\mathcal{I}}_{0}^{j_{1}}\, \disc{ i \mathcal{I}_{0}^{j_{1}} \expvalbar{ \varphi^{j_{1}} }^{\prime}_{a_{j_{1}}} }{s_{j_{1}}} \right] \times \nonumber \\
    & \times \sum_{\substack{L\subseteq J_{n-1} \\
    |J| > 1}} \prod_{l_{k}\notin L} \Bigg[ \delta_{0b_{l_{k}}} \overline{\mathcal{I}}^{l_{k+1}}_{l_{k+1}} \, \disc{ \mathcal{I}^{l_{k+1}}_{l_{k+1}} \overline{\mathcal{I}}^{l_{k+1}}_{l_{k+1}}\,\disc{ \delta_{0b_{l_{k+1}}} \mathcal{I}^{l_{k+1}}_{l_{k+1}} \expval{\varphi^{l_{k+1}-l_{k}}}^{\prime} }{s_{l_{k+1}}}}{s_{l_{k}}^{\prime}} + \nonumber \\
    & \phantom{ \sum_{\substack{L\subseteq J_{n-1} \\
    |J| > 1}} \prod_{l_{k}\notin L} \Bigg[ \delta_{0b_{l_{k}}}\overline{\mathcal{I}}^{l_{k+1}}_{l_{k+1}} \, }
    + \delta_{0b_{l_{k}}} \overline{\mathcal{I}}^{l_{k+1}}_{l_{k+1}} \, \disc{i\disc{ \mathcal{I}^{l_{k+1}}_{l_{k+1}} \overline{\mathcal{I}}^{l_{k+1}}_{l_{k+1}} \,i\delta_{1b_{l_{k+1}}} \mathcal{I}^{l_{k+1}}_{l_{k+1}} \expvalbar{\varphi^{l_{k+1}-l_{k}}}^{\prime}_{a_{l_{k+1}}} }{s_{l_{k+1}}}}{s_{l_{k}}^{\prime}}  \Bigg] \times \nonumber \\
    & \times \prod_{l_{k}\in L} \Bigg[ i\delta_{1b_{l_{k}}}\overline{\mathcal{I}}^{l_{k+1}}_{l_{k+1}} \,\disc{ \mathcal{I}^{l_{k+1}}_{l_{k+1}}  \overline{\mathcal{I}}^{l_{k+1}}_{l_{k+1}} \,i\disc{ \delta_{0b_{l_{k+1}}} \mathcal{I}^{l_{k+1}}_{l_{k+1}} \expvalbar{\varphi^{l_{k+1}-l_{k}}}^{\prime}_{a_{l_{k}+1}} }{s_{l_{k+1}}}}{s_{l_{k}}^{\prime}} + \nonumber \\
    & \phantom{ \prod_{l_{k}\in L} \Bigg[ i\delta_{1b_{l_{k}}}\overline{\mathcal{I}}^{l_{k+1}}_{l_{k+1}} \, } + i\delta_{1b_{l_{k}}} \overline{\mathcal{I}}^{l_{k+1}}_{l_{k+1}} \, \disc{  \mathcal{I}^{l_{k+1}}_{l_{k+1}} \overline{\mathcal{I}}^{l_{k+1}}_{l_{k+1}} \,  i^{2}\disc{ i \delta_{1b_{l_{k+1}}} \mathcal{I}^{l_{k+1}}_{l_{k+1}}\expvalbar{\varphi^{l_{k+1}-l_{k}}}^{\prime}_{a_{l_{k}+1}a_{l_{k+1}}} }{s_{l_{k+1}}}}{s_{l_{k}}^{\prime}}  \Bigg] \times \nonumber \\
    & \phantom{\prod} \times \left[ \delta_{0b_{j_{n}}}\overline{\mathcal{I}}^{V}_{j_{n}+1}\,\disc{ \mathcal{I}^{V}_{j_{n}+1}\, \expval{\varphi^{V-j_{n}}}^{\prime} }{s_{j_{n}}^{\prime}} + i\delta_{1b_{j_{n}}}\overline{\mathcal{I}}^{V}_{j_{n}+1}\,\disc{i \mathcal{I}^{V}_{j_{n}+1} \, \expvalbar{\varphi^{V-j_{n}}}^{\prime}_{a_{j_{n}+1}} }{s_{j_{n}}^{\prime}} \right],
    \label{eq:disc-final-not-reduced}
\end{align}
where we have used the notation,
\begin{equation}
    \varphi^{n} \equiv \prod_{j=1}^{n}\prod_{l=1}^{B_{n}}\varphi(k^{(j)}_{l}).
\end{equation}
For aesthetic purposes, we can further reduce the lines of \cref{eq:disc-final-not-reduced} by defining the following discontinuity operators, depending on momenta $s_{j}$ and a number $n$:
\begin{equation}
    \disct{  \mathcal{A} }{s_{j};\,n} = \delta_{\delta_{0b_{j}}}\disc{\mathcal{A}}{s_{j}} + i \delta_{1b_{j}}\disc{ia_{n}\mathcal{A}}{s_{j}}.
\end{equation}
With this definition we can rewrite \cref{eq:disc-final-not-reduced}, performing the index summation,
\begin{multline}
    \sum_{\substack{J\in P([V-1]) \\
    |J|\neq 0}}  \sum_{b_{j}=0,1} \dfrac{(-1)^{|J|-1}}{2^{|J|}} \int_{\vb{s_{j_{1}}}\cdots\,\vb{s_{j_{n}}^{\prime}}}  \disct{ \expval{\varphi^{j_{1}}}^{\prime} }{s_{j_{1}};\,j_{1}} \times \\ 
    \times \prod_{\substack{j_{k}\in J \\
    |J|>1, j_{k}\neq j_{n}}} 
    \disct{\disct{ \expval{\varphi^{j_{k+1}-j_{k}}}^{\prime} }{{s_{j_{k+1}};\,j_{k+1}}}}{s_{j_{k}}^{\prime};\,j_{k}}\disct{ \expval{\varphi^{V-j_{n}}}^{\prime} }{s_{j_{n}}^{\prime};\,j_{n}}.
\end{multline}
Diagrammatically, using the convention set on \ref{step:4}, the cuts can be seen as
\begin{figure}[H]
\centering
\includegraphics[width=0.9\textwidth]{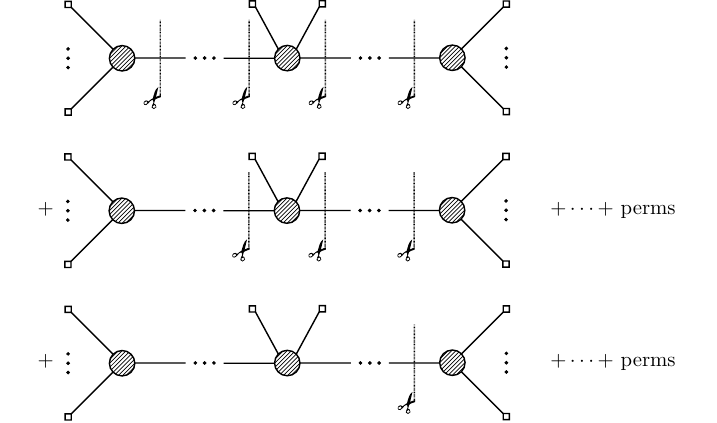}
\caption{Schematic visualisation for the discontinuity operation with respect to every internal energy. The result corresponds to every possible cut to the original diagram. Dashed lines with a scissor is in accordance to step \ref{step:4} of the cutting recipe.}
\end{figure}

\subsection{Three-vertex example}

Before analysing non-chain diagrams, let us compute explicitly a $V=3$ example, following the master formulas of \cref{eq:cutting-rule,eq:disc-final-not-reduced}, as well as the cutting recipe. Diagrammatically, this corresponds to 
\begin{figure}[H]
    \centering
    \includegraphics[width=0.85\textwidth]{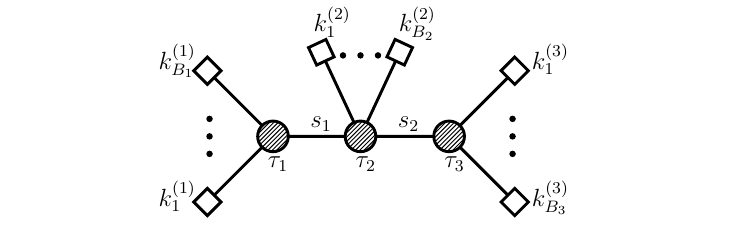}
\end{figure}
\noindent Setting up the notation, we define the bulk-to-boundary pieces as
\begin{align}
    & \Phi_{1} = \prod_{l=1}^{B_{1}}\varphi(k_{l}^{(1)}), & & \Phi_{2} = \prod_{l=1}^{B_{2}}\varphi(k_{l}^{(2)}), & & \Phi_{3} = \prod_{l=1}^{B_{3}}\varphi(k_{l}^{(3)}).
\end{align}
Furthermore, we identify the number of vertices and the number of external legs. The resulting parity of the addition of these quantities determines if the discontinuity operation possesses an extra imaginary number or not. Consider: $B_{1}$ and $B_{3}$ to be even numbers, and $B_{2}$ to be an odd one. Hence, according to step \ref{step:1}, we require to use \cref{step:eq:even}. Proceeding with step \ref{step:2}, we then draw every possible diagrammatic cut of the $V=3$ diagram, given by the power set $P([V-1]) \backslash  \{ \varnothing \} = \{ \{1\},\{2\},\{1,2\} \}$, where each number represents which internal line is being cut. This will correspond to the following three schematic graphs labelled by $c_{i}$, where $i$ indicates the cut position:
\begin{figure}[H]
    \centering
    \includegraphics[width=1\textwidth]{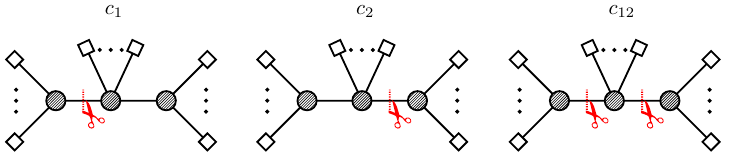}
\end{figure}
\noindent Moreover, the connected components (or connected subdiagrams) identification of step \ref{step:3} are realised as follows: $c_{1}$ and $c_{2}$ are the product of a one-vertex and a two-vertex subdiagrams, whilst $c_{12}$ is the product of three one-vertex subdiagrams. The imaginary unit content in the discontinuity operation is being bookkept by the $\mathcal{I}_{m}^{n}$ and $\overline{\mathcal{I}}_{m}^{n}$ factors of \cref{eq:imaginary-factors} in the master formula. Step \ref{step:4} indicates that we must replace each bulk-to-bulk propagator associated with the internal line cut with a product of two real and imaginary bulk-to-boundary functions, the latter also including an additional sign factor $a_{i}$, where $i$ is in correspondence with which vertex is the new bulk-to-boundary propagator attached to. In addition, we must replace each internal cut momenta $s_{i}$ by a pair $\{s_{i},s_{i}^{\prime}\}$ and add the integration of \cref{eq:internal-mom-cut-integration}. Furthermore, using \cref{disc-1-vertex,idisci-1-vertex,disc-1-vertex-a,idisci-1-vertex-a} we can recast the resulting expression in terms of discontinuity operations of regular or barred correlation functions, depending on their sign factor $a$ content. For $c_{1}$, corresponding to $J=\{1\}$:
\begin{multline}
    c_{1} = \int_{\vb{s}_{1}\vb{s}_{1}^{\prime}} \sum_{b_{1}=0,1}\dfrac{-1}{2}\left[ -\delta_{0b_{1}}i\disc{i\expval{\Phi_{1}\varphi(s_{1})}}{s_{1}} - \delta_{1b_{1}} \disc{\expvalbar{\Phi_{1}\varphi(s_{1})}_{a_{1}}}{s_{1}} \right] \times \\
    \times \left[ -\delta_{0b_{1}}i\disc{i\expval{\varphi(s_{1}^{\prime})\Phi_{2}\Phi_{3}}}{s_{1}^{\prime}} - \delta_{1b_{1}}\disc{\expvalbar{\varphi(s_{1}^{\prime})\Phi_{2}\Phi_{3}}}{s_{1}^{\prime}}  \right],
\end{multline}
by collapsing the $b_{1}$ delta, it yields,
\begin{multline}
    c_{1} = -\dfrac{1}{2}\int_{\vb{s}_{1}\vb{s}_{1}^{\prime}}  i\disc{i\expval{\Phi_{1}\varphi(s_{1})}}{s_{1}}i\disc{i\expval{\varphi(s_{1}^{\prime})\Phi_{2}\Phi_{3}}}{s_{1}^{\prime}} \\
    - \dfrac{1}{2} \int_{\vb{s}_{1}\vb{s}_{1}^{\prime}} \disc{\expvalbar{\Phi_{1}\varphi(s_{1})}_{a_{1}}}{s_{1}}\disc{\expvalbar{\varphi(s_{1}^{\prime})\Phi_{2}\Phi_{3}}_{a_{2}}}{s_{1}^{\prime}}.
\end{multline}
In addition, $c_{2}$ will share a similar result:
\begin{multline}
    c_{2} = -\dfrac{1}{2} \int_{\vb{s}_{2}\vb{s}_{2}^{\prime}} i\disc{i\expval{\Phi_{1}\Phi_{2}\varphi(s_{2})}}{s_{2}}i\disc{i\expval{\varphi(s_{2}^{\prime})\Phi_{3}}}{s_{2}^{\prime}} \\
    - \dfrac{1}{2} \int_{\vb{s}_{2}\vb{s}_{2}^{\prime}}\disc{\expvalbar{\Phi_{1}\Phi_{2}\varphi(s_{2})}_{a_{2}}}{s_{2}}\disc{\expvalbar{\varphi(s_{2}^{\prime})\Phi_{3}}_{a_{3}}}{s_{2}^{\prime}}.
\end{multline}
As a consequence of the double cut, $c_{12}$ will have double discontinuities correlator terms. In the master formula of \cref{eq:disc-final-not-reduced}, this is imprinted in the $\sum_{L\subseteq J_{n-1}}$ sum given by the cardinality two set $J=\{1,2\}$, yielding:
\begin{multline}
    c_{12} = \int_{\vb{s}_{1}\vb{s}_{1}^{\prime}\vb{s}_{2}\vb{s}_{2}^{\prime}}\sum_{b_{1},b_{2}=0,1}\dfrac{1}{4}\left[ -\delta_{0b_{1}}i\disc{i\expval{\Phi_{1}\varphi(s_{1})}}{s_{1}} - \delta_{1b_{1}} \disc{\expvalbar{\Phi_{1}\varphi(s_{1})}_{a_{1}}}{s_{1}} \right] \times \\
    \times \left[ \delta_{0b_{1}}\disc{\disc{\delta_{0b_{2}}\expval{\varphi(s_{1}^{\prime}) \Phi_{2}\varphi(s_{2})}}{s_{2}}}{s_{1}^{\prime}} + \delta_{0b_{1}}\disc{i\disc{i\delta_{1b_{2}}\expvalbar{\varphi(s_{1}^{\prime}) \Phi_{2}\varphi(s_{2})}_{a_{2}}}{s_{2}}}{s_{1}^{\prime}} \right] \times \\
    \times \left[ \delta_{1b_{1}}i\disc{i\disc{\delta_{0b_{2}}\expvalbar{\varphi(s_{1}^{\prime}) \Phi_{2}\varphi(s_{2})}_{a_{2}}}{s_{2}}}{s_{1}^{\prime}} + \delta_{1b_{1}}i\disc{i^{2}\disc{i\delta_{1b_{2}}\expval{\varphi(s_{1}^{\prime}) \Phi_{2}\varphi(s_{2})}}{s_{2}}}{s_{1}^{\prime}} \right] \times \\
    \times \left[ -\delta_{0b_{2}}i\disc{i\expval{\varphi(s_{2}^{\prime})\Phi_{3}}}{s_{2}^{\prime}} - \delta_{1b_{2}} \disc{\expvalbar{\varphi(s_{2}^{\prime})\Phi_{3}}_{a_{3}}}{s_{2}^{\prime}} \right],
\end{multline}
where, for the last term in the third line we have used $\expvalbar{\varphi}_{a_{2}a_{2}} = \expval{\varphi}$, since $a^{2}=1$. Then, performing the $b_{1},b_{2}$ sum:
\begin{multline}
    c_{12} = \dfrac{1}{4}\int_{\vb{s}_{1}\vb{s}_{1}^{\prime}\vb{s}_{2}\vb{s}_{2}^{\prime}} i\disc{i\expval{\Phi_{1}\varphi(s_{1})}}{s_{1}}\disc{\disc{\expval{\varphi(s_{1}^{\prime}) \Phi_{2}\varphi(s_{2})}}{s_{2}}}{s_{1}^{\prime}}i\disc{i\expval{\varphi(s_{2}^{\prime})\Phi_{3}}}{s_{2}^{\prime}} \\
    + \dfrac{1}{4} \int_{\vb{s}_{1}\vb{s}_{1}^{\prime}\vb{s}_{2}\vb{s}_{2}^{\prime}} i\disc{i\expval{\Phi_{1}\varphi(s_{1})}}{s_{1}}\disc{i\disc{i\expvalbar{\varphi(s_{1}^{\prime}) \Phi_{2}\varphi(s_{2})}_{a_{2}}}{s_{2}}}{s_{1}^{\prime}}\disc{\expvalbar{\varphi(s_{2}^{\prime})\Phi_{3}}_{a_{3}}}{s_{2}^{\prime}} \\
    + \dfrac{1}{4} \int_{\vb{s}_{1}\vb{s}_{1}^{\prime}\vb{s}_{2}\vb{s}_{2}^{\prime}} \disc{\expvalbar{\Phi_{1}\varphi(s_{1})}_{a_{1}}}{s_{1}} i\disc{i\disc{\expvalbar{\varphi(s_{1}^{\prime}) \Phi_{2}\varphi(s_{2})}_{a_{2}}}{s_{2}}}{s_{1}^{\prime}} i\disc{i\expval{\varphi(s_{2}^{\prime})\Phi_{3}}}{s_{2}^{\prime}} \\
    + \dfrac{1}{4} \int_{\vb{s}_{1}\vb{s}_{1}^{\prime}\vb{s}_{2}\vb{s}_{2}^{\prime}} \disc{\expvalbar{\Phi_{1}\varphi(s_{1})}_{a_{1}}}{s_{1}} i\disc{i^{2}\disc{i\expval{\varphi(s_{1}^{\prime}) \Phi_{2}\varphi(s_{2})}}{s_{2}}}{s_{1}^{\prime}} \disc{\expvalbar{\varphi(s_{2}^{\prime})\Phi_{3}}_{a_{3}}}{s_{2}^{\prime}}.
\end{multline}
The discontinuity operation for the $V=3$ example is then simply the addition of the above results,
\begin{equation}
    \disc{\expval{\Phi_{1}\Phi_{2}\Phi_{3}}}{s_{1}s_{2}} = c_{1}+c_{2}+c_{12}.
\end{equation}

\subsection{Cutting a General Diagram}

Furthermore, the rules listed above can be utilised for a general tree-level diagram, not only exclusively to the linear chain graph example previously analysed. We then provide the general expression for the discontinuity operation taken to every internal momenta, ignoring number factors and signs:
\begin{equation}
    \disc{\expval{\varphi^{\text{tree-level}}}}{\substack{\text{internal}\\\text{momenta}}} = \sum_{c\in\mathcal{C}}\sum_{b_{f(c)}=0,1}\int_{\mathcal{F}(\mathcal{V})}\prod_{d\in \mathcal{D}_{c}}\Disct{\expval{\varphi^{\mathcal{V}}}^{\prime}}{\mathcal{F}(\mathcal{V});\,f(c)},
\end{equation}
where $\mathcal{C}$ is the set of all possible diagrammatic cuts\footnote{In analogy to the linear chain graph, mathematically this corresponds to the power set minus the empty set.}, $c$ an element of this set, $f(c)$ is a function that provides a cut index number, $b_{f(c)}=0,1$ is needed to provide barred and regular correlators, $\mathcal{D}_{c}$ is the set containing the connected components (connected diagrams) following a cut $c\in\mathcal{C}$, $d$ an individual element (a connected diagram) of $\mathcal{D}_{c}$, $\mathcal{V}=|d|$ is the number of vertices of a the diagram $d$, and $\mathcal{F}(\mathcal{V})$ is a function that provides the (originally internal) momenta that is being cut, \textit{e.g.}, $s_{i}$ in the studied example of this section. We have also used a Disc operator notation defined in \cref{eq:mayusc-disc-def}. Let us illustrate each element visually. Consider the following correlator:
\begin{eqnarray*}
\parbox{25mm}{\includegraphics[width=0.2\textwidth]{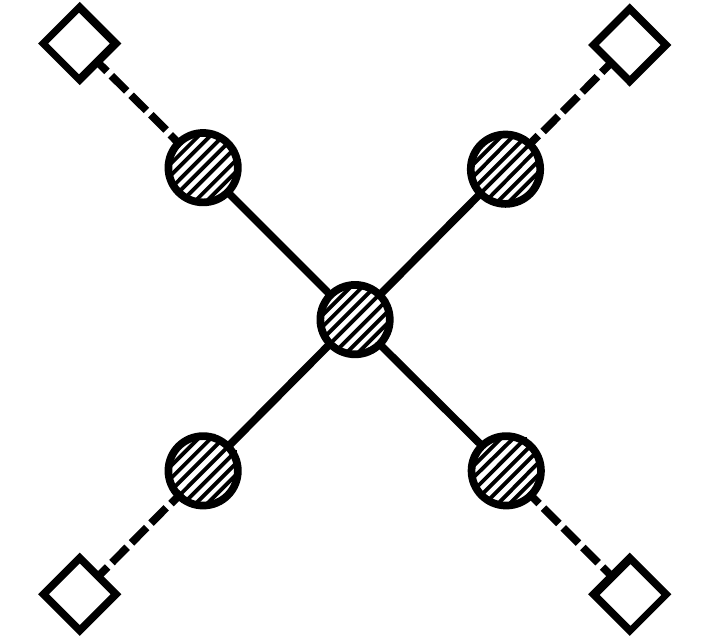}}~&=&~ \displaystyle \left\langle \prod_{j=1}^{4}\prod_{l=1}^{B_{j}}\varphi(k_{l}^{(j)}) \right\rangle^{\prime}
\end{eqnarray*}
\noindent where the dashed lines are representing an arbitrary number of bulk-to-boundary propagators, $B_{j}$ for each vertex. Then, the set $\mathcal{C}$ is:
\begin{figure}[H]
\centering
\includegraphics[width=1\textwidth]{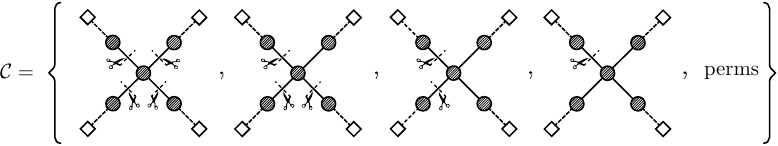}
\end{figure}
\noindent A cut diagram $c$ (an element of $\mathcal{C}$) and its connected components are:
\begin{figure}[H]
\centering
\includegraphics[width=1\textwidth]{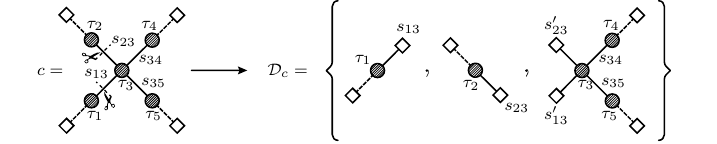}
\end{figure}
\noindent Furthermore a subdiagram $d\in\mathcal{D}_{c}$ with $\mathcal{V}_{i}$ for $i=1,2,3$ has:
\begin{align}
    \mathcal{F}(\mathcal{V}_{1}) & = \{ s_{13} \}, & \mathcal{F}(\mathcal{V}_{2}) & = \{ s_{23} \}, & \mathcal{F}(\mathcal{V}_{3}) & = \{ s^{\prime}_{13}, s^{\prime}_{23} \}.
\end{align}

The general norm regarding to the use of barred correlators is that for a connected subdiagram $d\in \mathcal{D}_{\text{cut}}$, obtained by cutting $|I_{n}|$ internal propagators of a connected diagram $c$, the discontinuity operation of the correlation function can be written in terms of the objects $\expvalbar{\varphi}^{\prime}_{I_{n}}, \dots, \expvalbar{\varphi}^{\prime}_{I_{1}}, \expval{\varphi}^{\prime}$. The reason lies in the use of \cref{real-property-with-two-a}. This argument is true regardless of the graph structure and exclusively depends on the number of cut internal propagators. To observe this behaviour schematically, consider $|I_{n}|=6$ in the following diagram:
\begin{figure}[H]
\centering
\includegraphics[width=1\textwidth]{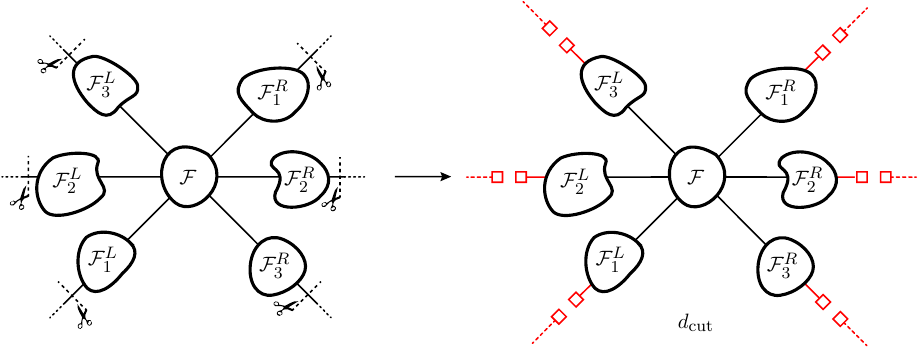}
\end{figure}
\noindent Where $\mathcal{F}$, $\mathcal{F}^{L}$, and $\mathcal{F}^{R}$ are blobs consisting on an arbitrary number of interactions. The colour red indicates that its respective bulk-to-boundary propagators are replaced by $\Re(G)$ and $\Im(G)$, accordingly to step \ref{step:4} of the cutting recipe. Then, $d_{\text{cut}}$ will be proportional to:
\begin{equation}
    d_{\text{cut}} = \sum_{a}\sum_{n,m=0}^{3}\prod_{j=1}^{n}\prod_{k=1}^{m}a_{j}^{L}a_{k}^{R}\mathcal{A}(\mathcal{F}_{L},\mathcal{F},\mathcal{F}_{R})
\end{equation}
In this notation, $\mathcal{A}(\mathcal{F},\mathcal{F},\mathcal{F}_{R})$ stands for the value of the connected subdiagram. In the above example, as per \cref{def:barred-corr},
\begin{align}
    n=m=3 \quad & \longrightarrow \quad \expvalbar{\varphi^{\mathcal{F}}}_{a^{L}_{1}a^{L}_{2}a^{L}_{3} a^{R}_{1}a^{R}_{2}a^{R}_{3}}, \label{ex:n3m3} \\
    n=3, m=2 \quad & \longrightarrow \quad \expvalbar{\varphi^{\mathcal{F}}}_{a^{L}_{1}a^{L}_{2}a^{L}_{3} a^{R}_{1}a^{R}_{2}}, \label{ex:n3m2} \\
    & \, \cdots \nonumber \\
    n=0, m=1 \quad & \longrightarrow \quad \expvalbar{\varphi^{\mathcal{F}}}_{a^{R}_{1}}, \label{ex:n0m1}\\
    n=m=0 \quad & \longrightarrow \quad \expval{\varphi^{\mathcal{F}}} \label{ex:n0m0}.
\end{align}
Hence, the discontinuity operation will consist of a sum containing \cref{ex:n3m3,ex:n3m2,ex:n0m1,ex:n0m0} multiplied by the associated cuts from blobs $\mathcal{F^{L}}$ and $\mathcal{F^{R}}$.

\section{Cutting Loops}
\label{sec:cutting-loops}

In this section, we consider the following general $V$-vertex one-loop correlator:
\begin{figure}[H]
\centering
\includegraphics[width=0.6\textwidth]{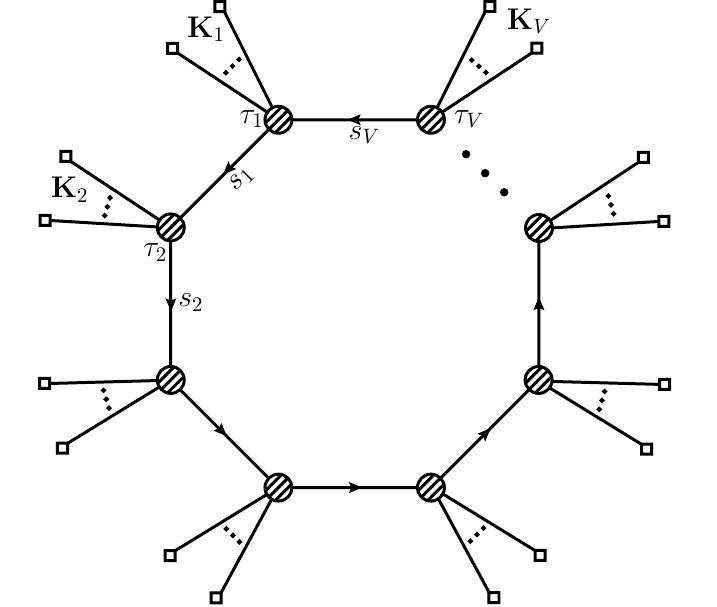}
\end{figure}
\noindent which has also been studied in \cite{Qin:2023bjk}. Similar to the tree-level case, every vertex $i$, with $i=1,\dots, V$ contains $B_{i}$ external lines with individual three-momentum $\vb{k}^{(i)}_{j}$, where $j=1,\dots, B_{i}$. The total momentum and energy flowing into the vertex, and the total number of external legs is given by:
\begin{align}
    \vb{K}_{i} & = \sum_{j=1}^{B_{i}}k^{(i)}_{j}, & E_{i} & = \sum_{j=1}^{B_{i}}|k^{(i)}_{j}| & B = \sum_{j=1}^{V} B_{j}.
\end{align}
In accordance to section \ref{sec:general-tree-lvl-rule}, we define the following object:
\begin{multline}
    \mathcal{A}^{V}_{\text{1-loop}} \equiv \mathcal{A}^{V}_{\text{1-loop}}( \{k_{\text{ext}}\}; \{k_{\text{int}}\}; \{a\} ) = \\
    \prod_{i=1}^{V}\left[\int_{-\infty}^{0}d\tau_{i}(-i\lambda)a_{i}\prod_{j=1}^{B_{i}}G_{a_{i}}\left(k^{(i)}_{j};\tau_{i}\right)\right]\int_{1- \text{loop}}\prod_{m=1}^{V}G_{a_{m}a_{m+1}}(s_{m};\tau_{m},\tau_{m+1}),
\end{multline}
with index identification $\tau_{V+1} = \tau_{1}$ and $a_{V+1} = a_{V}$. The loop internal momentum integral is defined,
\begin{equation}
    \int_{1- \text{loop}} = \int d^{3}\vb{s}_{1} \prod_{m=2}^{V}\int d^{3}\vb{s}_{m}\, \delta^{(3)}(\vb{s}_{m-1}-\vb{s}_{m}+\vb{K}_{m}),
    \label{eq:loop-integral}
\end{equation}
taking $\vb{s}_{1}$ to be the unconstrained momentum of the loop. The motive behind this integral definition arises because it is simpler to perform the discontinuity operation, as we will need to select each individual internal line with such operation.
The operation (\ref{disc-op}) will, by using (\ref{property-bulk-to-boundary-prop}), reduce to,
\begin{multline}
    \disc{\mathcal{A}^{V}_{\text{1-loop}}}{\vb{s}_{1}\cdots \vb{s}_{V}} = \prod_{j=1}^{V}\left[\int_{-\infty}^{0}d\tau_{j}(-i\lambda) a_{j} \prod_{l=1}^{B_{j}} G_{a_{j}}\left(k^{(j)}_{l};\tau_{j}\right)\right]\times \\
    \times \int_{1- \text{loop}}\left[ \prod_{m=1}^{V}G_{a_{m}a_{m+1}}(s_{m};\tau_{m},\tau_{m+1}) + (-1)^{B}\prod_{m=1}^{V}G^{\ast}_{a_{m}a_{m+1}}(s_{m};\tau_{m},\tau_{m+1}) \right].
    \label{eq-disc-op-1-loop-n-vert}
\end{multline}
The above expression is obtained since the discontinuity operation flips the sign of every three-momenta, leaving the Dirac delta in the loop integral invariant. Moreover, the following steps are essentially the same as in the previous section with the addition of an $V+1=1$ index identification and a loop integral. Then, the discontinuity will be written purely in terms of tree-level correlators, schematically this is seen as:
\begin{equation}
    \disc{\expval{\varphi^{V}_{1-\text{loop}}}^{\prime}}{s_{1}\cdots s_{V}} = \sum_{\text{cuts}}\prod_{\text{momenta}} \int_{\vb{s}\vb{s}^{\prime}} \int_{\vb{s_{1}}} \Disct{\expval{\varphi^{\mathcal{V}}_{\text{tree}}}}{\text{internal cuts}},
\end{equation}
where $\mathcal{V}$ takes values from 1 to $V$. This expression, in a nutshell, is informing that the discontinuity operation for a $V-$vertex one-loop correlator corresponds to the sum of all possible cuts to the internal energies in the form of discontinuities of lower vertex, tree-level correlators (\textit{barred} and regular ones). The first integral in the above equation has been introduced before in \cref{def-f-factor} whilst the second one corresponds to \cref{eq:loop-integral} posterior to the Dirac delta integration.

\subsection{Loop Example}

Let us look at the following one-loop diagram in a $\varphi^{4}$ theory to see its cutting behaviour
\begin{figure}[H]
\centering
\includegraphics[width=0.9\textwidth]{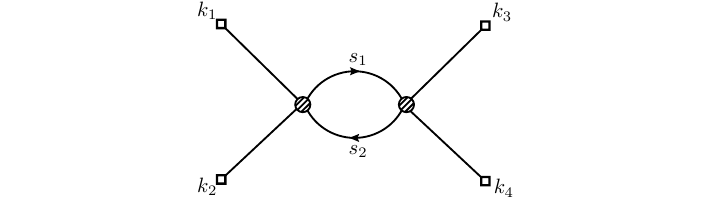}
\end{figure}
\noindent Using (\ref{eq-disc-op-1-loop-n-vert}),
\begin{multline}
    i\disc{i\mathcal{A}^{2}_{\text{loop}}}{\vb{s}_{1}\vb{s}_{2}} = \prod_{j=1}^{2}\left[\int_{-\infty}^{0}d\tau_{j}(-i\lambda_{4}) a_{j}\prod_{l=1}^{2}G_{a_{j}}\left(k^{(j)}_{l};\tau_{j}\right)\right]\times \\
    \times \int_{\vb{s}_{1}\vb{s}_{2}}\left[ G_{a_{1}a_{2}}(s_{1};\tau_{1},\tau_{2})G_{a_{2}a_{1}}(s_{2};\tau_{2},\tau_{1}) - G^{\ast}_{a_{1}a_{2}}(s_{1};\tau_{1},\tau_{2})G^{\ast}_{a_{2}a_{1}}(s_{2};\tau_{2},\tau_{1}) \right].
\end{multline}
Furthermore, the above can be written in terms of discontinuity operations of tree-level correlation and \textit{barred} correlation functions. Ignoring overall constant factors it yields,
\begin{align}
    & i\disc{i\mathcal{A}^{2}_{\text{loop}}}{\vb{s}_{1}\vb{s}_{2}} =  -\dfrac{1}{2}\int_{\vb{s}_{1}\vb{s}^{\prime}_{1}} \int_{\vb{s}_{1}\vb{s}_{2}} i\disc{i^{2}\disc{i\mathcal{A}^{2}_{\text{tree}}}{s^{\prime}_{1}}}{s_{1}} - \dfrac{1}{2}\int_{\vb{s}_{1}\vb{s}^{\prime}_{1}} \int_{\vb{s}_{1}\vb{s}_{2}} \disc{\disc{a_{1}a_{2}\mathcal{A}^{2}_{\text{tree}}}{s^{\prime}_{1}}}{s_{1}} \nonumber \\
    & \phantom{i\disc{i\mathcal{A}^{2}_{\text{loop}}}{\vb{s}_{1}\vb{s}_{2}}=} -\dfrac{1}{2} \int_{\vb{s}_{2}\vb{s}^{\prime}_{2}} \int_{\vb{s}_{1}\vb{s}_{2}} i\disc{i^{2}\disc{i\mathcal{A}^{2}_{\text{tree}}}{s^{\prime}_{2}}}{s_{2}} - \dfrac{1}{2}\int_{\vb{s}_{2}\vb{s}^{\prime}_{2}} \int_{\vb{s}_{1}\vb{s}_{2}} \disc{\disc{a_{1}a_{2}\mathcal{A}^{2}_{\text{tree}}}{s^{\prime}_{2}}}{s_{2}} \nonumber \\ 
    & \phantom{i\disc{i\mathcal{A}^{2}_{\text{loop}}}{\vb{s}_{1}\vb{s}_{2}}=} + \dfrac{1}{4}\int_{\vb{s}_{1}\vb{s}^{\prime}_{1}\vb{s}_{2}\vb{s}^{\prime}_{2}} \int_{\vb{s}_{1}\vb{s}_{2}} \disc{\disc{a_{1}\mathcal{A}^{1}_{\text{tree}}}{s_{2}}}{s_{1}} \disc{\disc{a_{2}\mathcal{A}^{1}_{\text{tree}}}{s^{\prime}_{2}}}{s^{\prime}_{1}} \nonumber \\
    & \phantom{i\disc{i\mathcal{A}^{2}_{\text{1-loop}}}{\vb{s}_{1}\vb{s}_{2}}=} +\dfrac{1}{4} \int_{\vb{s}_{1}\vb{s}^{\prime}_{1}\vb{s}_{2}\vb{s}^{\prime}_{2}} \int_{\vb{s}_{1}\vb{s}_{2}} i\disc{i^{2}\disc{i\mathcal{A}^{1}_{\text{tree}}}{s_{2}}}{s_{1}} i\disc{i^{2}\disc{i\mathcal{A}^{1}_{\text{tree}}}{s^{\prime}_{2}}}{s^{\prime}_{1}} \nonumber \\
    & \phantom{i\disc{i\mathcal{A}^{2}_{\text{loop}}}{\vb{s}_{1}\vb{s}_{2}}=} + \dfrac{1}{4}\int_{\vb{s}_{1}\vb{s}^{\prime}_{1}\vb{s}_{2}\vb{s}^{\prime}_{2}} \int_{\vb{s}_{1}\vb{s}_{2}} \disc{i\disc{ia_{1}\mathcal{A}^{1}_{\text{tree}}}{s_{2}}}{s_{1}} \disc{i\disc{ia_{2}\mathcal{A}^{1}_{\text{tree}}}{s^{\prime}_{2}}}{s^{\prime}_{1}} \nonumber \\
    & \phantom{i\disc{i\mathcal{A}^{2}_{\text{loop}}}{\vb{s}_{1}\vb{s}_{2}}=} + \dfrac{1}{4}\int_{\vb{s}_{1}\vb{s}^{\prime}_{1}\vb{s}_{2}\vb{s}^{\prime}_{2}} \int_{\vb{s}_{1}\vb{s}_{2}} i\disc{i\disc{a_{1}\mathcal{A}^{1}_{\text{tree}}}{s_{2}}}{s_{1}} i\disc{i\disc{a_{2}\mathcal{A}^{1}_{\text{tree}}}{s^{\prime}_{2}}}{s^{\prime}_{1}}.
\end{align}
This loop example makes explicit that the discontinuity of a one-loop exchange with $V$ vertices is determined entirely by tree correlators, displaying the same algebraic structure at tree level, once internal momenta are integrated over. Similarly, a two-loop discontinuity should be reduced to products of one-loop and tree pieces.  This recursive structure mirrors the flat-space unitarity method and provides a systematic route to develop a bootstrap program to compute higher-loop cosmological correlators from lower-order data.
 
\section{Conclusions}

In this work, we studied Schwinger-Keldysh cutting rules, motivated by unitarity, acting directly on in-in cosmological correlation functions. We first examined a single tree-level four-point exchange diagram and showed that its discontinuity factorises into products of three-point correlators. We then generalised this result to an arbitrary linear-chain graph configuration with $V$ vertices, arguing that more elaborate topologies inherit the same factorisation pattern: once a propagator is cut, the two resulting subgraphs behave analogously to the four-point case and factorise recursively. Finally, we analysed a one-loop diagram, verifying that the factorisation persists after integration over the loop momentum, thereby confirming that the cutting rules extend beyond tree level.

In all cases, the inspection of discontinuities can be carried out diagrammatically, that is, by verifying that the result can be written, according to the Feynman rules, as products of lower-point diagrams, as illustrated in Section~\ref{sec:toy-model}. However, in order to express these results explicitly as discontinuities of lower-point correlators, a new algebraic object is required, referred to in this work as \textit{barred} correlators. The need for these objects arises from the presence of vertex sign indices on the right-hand side of Eq.~(\ref{real-property-with-two-a}), which are represented diagrammatically in Step~\ref{step:4} of the cutting recipe. These barred correlators, which are simply linear combinations of diagrams that do not contribute to physical correlation functions, provide an intuitive framework for organising computations, allowing discontinuity operations to be decomposed into fundamental building blocks. In this way, our framework offers a streamlined formulation of cosmological cutting rules that can be systematically applied to higher-order and higher-loop calculations. Moreover, barred correlators and their relation to in-in correlators, particularly through the interchange of real and imaginary parts, hint at a deeper structural insight that warrants further investigation. In particular, their algebraic structure bears a resemblance to features of the $r/a$ (Keldysh) basis, in which advanced fields play an analogous role. Exploring this possible connection may provide new insights into inflationary cosmological correlators.

\section*{Acknowledgments}
The authors would like to thank Javier Huenupi, Ellie Hughes, Benjamín Navarrete, Nicolás Parra, Edgardo Rosas, Joaquín Silva, Jorge Sossa, and Spyros Sypsas for  helpful discussions. G.M. has been supported by ANID/ACT 210100 Anillo grant. F.R. is supported by FONDECYT grants 1221920 and 1230853 and by ANID Anillo grant ANID/ACT 210100. GAP was supported by a FONDECYT grant 1251511. FCM was supported by a FONDECYT grant 1210876.

\appendix

\section{Discontinuity Operation}
\label{sec:disc}

In this section we will define and provide some examples on our version of the discontinuity operation.

\textit{Definition:} For a function $f$ depending on a set of momenta $\{ k \}$ (these can be either internal or external), spatial momenta $\{ \vb{k} \}$, and sign indices $\{ a \}$. We define the algebraic operation, where $I$ is an index set:
\begin{equation}
        \disc{f(\{ k \}, \{ a \}; \{ \vb{k} \})}{k_{I}} \equiv \\
        f(\{ k \}, \{ a \}; \{ \vb{k} \}) + f^{\ast}( \{ - k_{I^{C}}, k_{I} \}, \{ a \}; \{ -\vb{k} \}).
        \label{def:disc-op}
\end{equation}
That is, adding the complex conjugate of $f$ with a minus sign to every $k$ except to those appearing in the subscript of the operator (given by the index set $I^{C}$), while applying a parity transformation to all spatial momenta $\{ \vb{k} \}$.

Note that the discontinuity operator introduced in \cref{def:disc-op} carries a plus sign rather than a minus one commonly used in flat-space amplitude works. The choice is made in such a way that its action on the interaction picture evolution operator reproduces the left-hand side of the cosmological unitarity relation \cref{cot_raw}. The sign convention is purely a matter of bookkeeping: acting with $i\mathrm{disc}(i \,\cdot\,)$ on $f(\{ k \}, \{ a \}; \{ \vb{k} \})$ provides the usual convention with a minus sign \cite{Cespedes:2020xqq,Goodhew:2020hob,Melville:2021lst,Goodhew:2021oqg,Jazayeri:2021fvk}, up to an overall sign. A similar notation was considered in \cite{Baumann:2021fxj}. It should also be noted that, when dealing with the observable, namely correlation functions, the complex conjugation operation is redundant since correlators are real quantities. Nevertheless, we will maintain this definition.

Additionally, for convenience, we will define the following notation when taking multiple discontinuity operations:
\begin{equation}
    \Disc{f}{k_{i},\dots,k_{j}} \equiv \disc{\disc{\,\cdots\,\disc{f}{k_{j}}}{k_{i+1}}}{k_{i}}.
    \label{eq:mayusc-disc-def}
\end{equation}
Note that each discontinuity operation is commutable.

As a useful example, let us compute the discontinuity operation for a one-vertex correlation function. Since we seek generality, we will consider the following interactive Lagrangian:
\begin{equation}
    \mathcal{L}_{\text{int}} = \dfrac{\lambda_{N}}{N!}\varphi^{N},
\end{equation}
with arbitrary $N$. The one-vertex correlation function for $B_{V}+1$ external particles is then written as:
\begin{equation}
    \expval{ \prod_{l=1}^{B_{V}} \varphi(\vb{k}^{(V)}_{l}) \varphi(\vb{p}) }^{\prime} = 
    \sum_{a}\int_{-\infty}^{0}d\tau(-i\lambda^{(V)})\,a \prod_{l=1}^{B_{V}}G_{a}\left(k^{(V)}_{l};\tau\right)G_{a}(p;\tau).
\end{equation}
In this notation we are taking $\lambda^{(V)} \equiv \lambda_{B_{V}+1}$. Two possible discontinuity operations for the one-vertex correlation with respect to $p$ are:
\begin{align}
    & \disc{\expval{\cdots\varphi(\vb{p}) }}{p} =\sum_{a=\pm}\int_{-\infty}^{0}d\tau(-i\lambda^{(V)})a\prod_{l=1}^{B_{V}}G_{a}\left(k^{(l)};\tau\right)\left[ G_{a}(p;\tau) + (-1)^{B_{V}+1}G^{\ast}_{a}(p;\tau) \right], \\ 
    & i\disc{i\expval{\cdots\varphi(\vb{p}) }}{p} =-\sum_{a=\pm}\int_{-\infty}^{0}d\tau(-i\lambda^{(V)})a\prod_{l=1}^{B_{V}}G_{a}\left(k^{(l)};\tau\right)\left[ G_{a}(p;\tau) - (-1)^{B_{V}+1}G^{\ast}_{a}(p;\tau) \right],
\end{align}
where the ellipsis stand for $\prod_{l=1}^{B_{V}} \varphi(\vb{k}^{(V)}_{l})$. Hence, up to some constants, the two discontinuity operations differs by a real or imaginary part of the bulk-to-boundary propagator with momenta $p$, depending on the parity of $B_{V}$, yielding:
\begin{equation}
    \disc{ \expval{\cdots\varphi(\vb{p})}^{\prime} }{p} =
    \begin{cases} 
        \displaystyle\sum_{a}\int_{-\infty}^{0}d\tau(-i\lambda^{(V)})a\prod_{l=1}^{B_{V}}G_{a}\left(k^{(V)}_{l};\tau\right)2\Re\left( G_{a}(p;\tau)\right), & \text{odd } B_{V}, \\
        \displaystyle\sum_{a}\int_{-\infty}^{0}d\tau(-i\lambda^{(V)})a\prod_{l=1}^{B_{V}}G_{a}\left(k^{(V)}_{l};\tau\right)2i\Im\left( G_{a}(p;\tau)\right), & \text{even } B_{V}.
    \end{cases}
    \label{disc-1-vertex}
\end{equation}
\begin{equation}
    i\disc{i\expval{\cdots\varphi(\vb{p})}^{\prime} }{p} =
    \begin{cases} 
        \displaystyle-\sum_{a}\int_{-\infty}^{0}d\tau(-i\lambda^{(V)})a\prod_{l=1}^{B_{V}}G_{a}\left(k^{(V)}_{l};\tau\right)2i\Im\left( G_{a}(p;\tau)\right), & \text{odd } B_{V}, \\
        \displaystyle-\sum_{a}\int_{-\infty}^{0}d\tau(-i\lambda^{(V)})a\prod_{l=1}^{B_{V}}G_{a}\left(k^{(V)}_{l};\tau\right)2\Re\left( G_{a}(p;\tau)\right), & \text{even } B_{V}.
    \end{cases}
    \label{idisci-1-vertex}
\end{equation}
Two distinct cases are obtained when computing the discontinuity operation for a \textit{barred} correlator with respect to $p$,
\begin{equation}
    \disc{ \expvalbar{\cdots\varphi(\vb{p})}^{\prime}_{a} }{p} =
    \begin{cases} 
        \displaystyle\sum_{a}\int_{-\infty}^{0}d\tau(-i\lambda^{(V)})\prod_{l=1}^{B_{V}}G_{a}\left(k^{(l)};\tau\right)2\Re\left( G_{a}(p;\tau)\right), & \text{odd } B_{V}, \\
        \displaystyle\sum_{a}\int_{-\infty}^{0}d\tau(-i\lambda^{(V)})\prod_{l=1}^{B_{V}}G_{a}\left(k^{(l)};\tau\right)2i\Im\left( G_{a}(p;\tau)\right), & \text{even } B_{V},
    \end{cases}
    \label{disc-1-vertex-a}
\end{equation}
\begin{equation}
    i\disc{ i\expvalbar{\cdots\varphi(\vb{p})}^{\prime}_{a} }{p} =
    \begin{cases} 
        \displaystyle-\sum_{a}\int_{-\infty}^{0}d\tau(-i\lambda^{(V)})\prod_{l=1}^{B_{V}}G_{a}\left(k^{(l)};\tau\right)2i\Im\left( G_{a}(p;\tau)\right), & \text{odd } B_{V}, \\
        \displaystyle-\sum_{a}\int_{-\infty}^{0}d\tau(-i\lambda^{(V)})\prod_{l=1}^{B_{V}}G_{a}\left(k^{(l)};\tau\right)2\Re\left( G_{a}(p;\tau)\right). & \text{even } B_{V},
    \end{cases}
    \label{idisci-1-vertex-a}
\end{equation}
Where we have used the fact $a^{2}=1$. We will recurrently use \cref{disc-1-vertex,idisci-1-vertex,disc-1-vertex-a,idisci-1-vertex-a} in section \ref{sec:general-tree-lvl-rule} when analysing one-vertex cuts. Now consider a $(V-1)-$vertex correlator,
\begin{multline}
    \expval{ \prod_{j=1}^{V-1}\prod_{l=1}^{B_{j}}\varphi(\vb{k}_{l}^{(j)}) \varphi(\vb{p})} = \\
    \sum_{a_{1}, \dots, a_{V-1} }\prod_{j=1}^{V-1}\left[\int_{-\infty}^{0}d\tau_{j}(-i\lambda^{(j)})a_{j}\prod_{l=1}^{B_{j}}G_{a_{j}}\left(k^{(l)}_{j};\tau_{j}\right)G_{a_{V-1}}\left(p;\tau_{V-1}\right)\right]\times \\
    \times \prod_{m=1}^{V-1}G_{a_{m}a_{m+1}}(s_{m};\tau_{m},\tau_{m+1}).
\end{multline}
The discontinuity operation with respect to one of the external momenta attached to $a_{V-1}$ yields,
\begin{multline}
    \disc{\expval{ \cdots\varphi_{\vb{p}} } }{p} = \sum_{a_{1}, \dots, a_{V-1}}\prod_{j=1}^{V-1}\left[\int_{-\infty}^{0}d\tau_{j}(-i\lambda^{(j)})a_{j}\prod_{l=1}^{B_{j}}G_{a_{j}}\left(k^{(l)}_{j};\tau_{j}\right)\right] \times \\
    \times \prod_{m=1}^{V-2}G_{a_{m}a_{m+1}}(s_{m};\tau_{m},\tau_{m+1})\left( G_{a_{V-1}}(p;\tau_{V-1}) + (-1)^{\sum_{j=1}^{V-1}B_{j}+3V-4}G^{\ast}_{a_{V-1}}(p;\tau_{V-1}) \right),
\end{multline}
where the ellipsis on the correlators stand for $\prod_{j=1}^{V-1}\prod_{l=1}^{B_{j}}\varphi(\vb{k}_{l}^{(j)})$. The $(-1)^{\sum_{j=1}^{V-1}B_{j}+2V-3}$ sign at the second line is provided by the conjugation of $V-1$ imaginary numbers, followed by $V-2$ momentum parity and conjugation to the bulk-to-bulk propagators, and $\sum_{j=1}^{V-1}B_{j}$ momentum parity and conjugation to the bulk-to-boundary propagators. The final sign can be rewritten simply as $(-1)^{B-B_{V}-1}$. Then, similar to the one-vertex case, depending on the parity of the number $\mathcal{B} \equiv B-B_{V}-1$ we will have real or imaginary operators.

Analogue to the one-vertex case, we can compute the two discontinuities:
\begin{equation}
\disc{\expval{ \cdots\varphi_{\vb{p}} }}{p} =
    \begin{cases} 
        \displaystyle\sum_{a_{1}, \dots, a_{V-1}}\prod_{j=1}^{V-1}\left[\int_{-\infty}^{0}d\tau_{j}(-i\lambda^{(j)})a_{j}\prod_{l=1}^{B_{j}}G_{a_{j}}\left(k^{(l)}_{j};\tau_{j}\right)\right] \times \\
        \phantom{\sum\prod}\displaystyle\times \prod_{m=1}^{V-2}G_{a_{m}a_{m+1}}(s_{m};\tau_{m},\tau_{m+1})2\Re\left( G_{a_{V-1}}(p;\tau_{V-1}) \right), & \text{even } \mathcal{B}, \\
        \displaystyle\sum_{a_{1}, \dots, a_{V-1}}\prod_{j=1}^{V-1}\left[\int_{-\infty}^{0}d\tau_{j}(-i\lambda^{(j)})a_{j}\prod_{l=1}^{B_{j}}G_{a_{j}}\left(k^{(l)}_{j};\tau_{j}\right)\right] \times \\
        \phantom{\sum\prod}\displaystyle\times \prod_{m=1}^{V-2}G_{a_{m}a_{m+1}}(s_{m};\tau_{m},\tau_{m+1})2i\Im\left( G_{a_{V-1}}(p;\tau_{V-1}) \right), & \text{odd } \mathcal{B}.
    \end{cases}
    \label{disc-(v-1)-vertex}
\end{equation}
\begin{equation}
i\disc{i\expval{ \cdots\varphi_{\vb{p}} }}{p} =
    \begin{cases} 
        -\displaystyle\sum_{a_{1}, \dots, a_{V-1}}\prod_{j=1}^{V-1}\left[\int_{-\infty}^{0}d\tau_{j}(-i\lambda^{(j)})a_{j}\prod_{l=1}^{B_{j}}G_{a_{j}}\left(k^{(l)}_{j};\tau_{j}\right)\right] \times \\
        \phantom{-\sum\prod}\displaystyle\times \prod_{m=1}^{V-2}G_{a_{m}a_{m+1}}(s_{m};\tau_{m},\tau_{m+1})2i\Im\left( G_{a_{V-1}}(p;\tau_{V-1}) \right), & \text{even } \mathcal{B}, \\
        -\displaystyle\sum_{a_{1}, \dots, a_{V-1}}\prod_{j=1}^{V-1}\left[\int_{-\infty}^{0}d\tau_{j}(-i\lambda^{(j)})a_{j}\prod_{l=1}^{B_{j}}G_{a_{j}}\left(k^{(l)}_{j};\tau_{j}\right)\right] \times \\
        \phantom{-\sum\prod}\displaystyle\times \prod_{m=1}^{V-2}G_{a_{m}a_{m+1}}(s_{m};\tau_{m},\tau_{m+1})2\Re\left( G_{a_{V-1}}(p;\tau_{V-1}) \right), & \text{odd } \mathcal{B}.
    \end{cases}
    \label{idisci-(v-1)-vertex}
\end{equation}
Similarly, a \textit{barred} correlator,
\begin{equation}
\disc{ \expvalbar{\cdots\varphi_{\vb{p}}}^{\prime}_{a_{I}} }{p} =
    \begin{cases} 
        \displaystyle\sum_{a_{1}, \dots, a_{V-1}}\prod_{j=1}^{V-1}\left[\int_{-\infty}^{0}d\tau_{j}(-i\lambda^{(j)})a_{j}\prod_{l=1}^{B_{j}}G_{a_{j}}\left(k^{(l)}_{j};\tau_{j}\right)\right] \times \\
        \displaystyle\times \prod_{m=1}^{V-2}G_{a_{m}a_{m+1}}(s_{m};\tau_{m},\tau_{m+1})\prod_{i\in I}a_{i}2\Re\left( G_{a_{V-1}}(p;\tau_{V-1}) \right), & \text{even } \mathcal{B}, \\
        \displaystyle\sum_{a_{1}, \dots, a_{V-1}}\prod_{j=1}^{V-1}\left[\int_{-\infty}^{0}d\tau_{j}(-i\lambda^{(j)})a_{j}\prod_{l=1}^{B_{j}}G_{a_{j}}\left(k^{(l)}_{j};\tau_{j}\right)\right] \times \\
        \displaystyle\times \prod_{m=1}^{V-2}G_{a_{m}a_{m+1}}(s_{m};\tau_{m},\tau_{m+1})\prod_{i\in I}a_{i}2i\Im\left( G_{a_{V-1}}(p;\tau_{V-1}) \right), & \text{odd } \mathcal{B}.
    \end{cases}
    \label{disc-(v-1)-vertex-a}
\end{equation}
\begin{equation}
i\disc{ i\expvalbar{\cdots\varphi_{\vb{p}}}^{\prime}_{a_{I}} }{p} =
    \begin{cases} 
        -\displaystyle\sum_{a_{1}, \dots, a_{V-1}}\prod_{j=1}^{V-1}\left[\int_{-\infty}^{0}d\tau_{j}(-i\lambda^{(j)})a_{j}\prod_{l=1}^{B_{j}}G_{a_{j}}\left(k^{(l)}_{j};\tau_{j}\right)\right] \times \\
        \displaystyle\times \prod_{m=1}^{V-2}G_{a_{m}a_{m+1}}(s_{m};\tau_{m},\tau_{m+1})\prod_{i\in I}a_{i}2i\Im\left( G_{a_{V-1}}(p;\tau_{V-1}) \right), & \text{even } \mathcal{B}, \\
        -\displaystyle\sum_{a_{1}, \dots, a_{V-1}}\prod_{j=1}^{V-1}\left[\int_{-\infty}^{0}d\tau_{j}(-i\lambda^{(j)})a_{j}\prod_{l=1}^{B_{j}}G_{a_{j}}\left(k^{(l)}_{j};\tau_{j}\right)\right] \times \\
        \displaystyle\times \prod_{m=1}^{V-2}G_{a_{m}a_{m+1}}(s_{m};\tau_{m},\tau_{m+1})\prod_{i\in I}a_{i}2\Re\left( G_{a_{V-1}}(p;\tau_{V-1}) \right), & \text{odd } \mathcal{B}.
    \end{cases}
    \label{idisci-(v-1)-vertex-a}
\end{equation}
We will use \cref{disc-(v-1)-vertex,idisci-(v-1)-vertex,disc-(v-1)-vertex-a,idisci-(v-1)-vertex-a} when discussing multi-vertex cuts.

\section{Cosmological Optical Theorem validation}\label{chap:cot} 

In this section, we verify the relationship between the results obtained from the cosmological optical theorem (COT) and the discontinuity operator defined in section \ref{sec:toy-model}. For this purpose, consider a Hamiltonian\footnote{With symmetry factors absorbed into the coupling constants.} of the following form:
\begin{equation}
     H_{\text{int}}(\tau) = \int d^3x \left(g_Aa^2(\tau)\varphi(\partial_i \varphi)^2 + g_Ba(\tau)^4\varphi^3 \right),
\end{equation}
considering interactions without time derivatives. This section partially reproduces the results obtained by \cite{Goodhew:2020hob}, now using the SK formalism. We begin by writing \( H_{\text{int}} \) in Fourier space\footnote{
The integrals conventions are taken to be,
\begin{equation}
    \int \frac{d^3k}{(2\pi)^3}=\int_{\mathbf{k}}, \ \int dx = \int_{\mathbf{x}}.
    \label{kintegral}
\end{equation}},
\begin{align}
    H_{\text{int}}(\tau) & = g_A \int_{\mathbf{p}_1,\mathbf{p}_2,\mathbf{p}_3} \mathcal{K}(\mathbf{p_{1}},\mathbf{p_{2}},\mathbf{p_{3}}) \varphi(\mathbf{p}_1,\tau)\varphi(\mathbf{p}_2,\tau)\varphi(\mathbf{p}_3,\tau) (2\pi)^3 \delta^{(3)}(\mathbf{p}_1+\mathbf{p}_2+\mathbf{p}_3) \nonumber \\
    & \phantom{=} + g_B \int_{\mathbf{q}_1,\mathbf{q}_2,\mathbf{q}_3} \varphi(\mathbf{q}_1,\tau)\varphi(\mathbf{q}_2,\tau)\varphi(\mathbf{q}_3,\tau) (2\pi)^3 \delta^{(3)}(\mathbf{q}_1+\mathbf{q}_2+\mathbf{q}_3) \nonumber \\
    & = H_A(\tau) + H_B(\tau),
\end{align}
where,
\begin{equation}
    \mathcal{K}(\mathbf{p_{1}},\mathbf{p_{2}},\mathbf{p_{3}}) = \mathbf{p}_1 \cdot \mathbf{p}_2 + \mathbf{p}_1 \cdot \mathbf{p}_3 + \mathbf{p}_3 \cdot \mathbf{p}_2.
\end{equation}
According to \cref{DysonSeries}, we write \( \delta\mathcal{U} \) perturbatively for the Hamiltonian,
\begin{subequations}
    \begin{equation}
    \begin{split}
\delta\mathcal{U}_{g}(0,-\infty) &= -i\int_{-\infty(1-i\epsilon)}^{0} d\tau H_{\text{int}}(\tau) = -i\int_{-\infty(1-i\epsilon)}^{0} d\tau \big(H_A(\tau) + H_B(\tau)\big),
    \end{split}
    \label{orden1a}
    \end{equation}
    \begin{equation}
    \begin{split}
    \delta\mathcal{U}^{\dagger}_{g}(0,-\infty) &= i\int_{-\infty(1-i\epsilon)}^{0} d\tau H_{\text{int}}^{\dagger}(\tau) = i\int_{-\infty(1-i\epsilon)}^{0} d\tau \big(H_A^{\dagger}(\tau) + H_B^{\dagger}(\tau)\big),
    \end{split}
    \label{orden1b}
    \end{equation}
    \label{orden1}
\end{subequations}
where the \( i\epsilon \) prescription is considered. At order $g^{2}$,
\begin{align}
    & \delta\mathcal{U}_{g^2}(0, -\infty) = -\int_{-\infty(1-i\epsilon)}^{0} d\tau d\tau' H_{\text{int}}(\tau)H_{\text{int}}(\tau')\Theta(\tau-\tau') \nonumber \\
        & \phantom{\delta\mathcal{U}_{g^2}(0, -\infty)} = -\int_{-\infty(1-i\epsilon)}^{0}  d\tau d\tau' \big( H_A(\tau)H_A(\tau') + H_A(\tau)H_B(\tau') \nonumber \\
        & \phantom{\delta\mathcal{U}_{g^2}(0, -\infty)}\phantom{= -\int_{-\infty(1-i\epsilon)}^{0}  d\tau d\tau'\big(} + H_B(\tau)H_A(\tau') + H_B(\tau)H_B(\tau') \big)\Theta(\tau-\tau'), \label{orden2a}
\end{align}
\begin{align}
     &\delta\mathcal{U}^{\dagger}_{g^2}(0, -\infty) = -\int_{-\infty(1-i\epsilon)}^{0} d\tau d\tau' H_{\text{int}}^{\dagger}(\tau')H_{\text{int}}^{\dagger}(\tau)\Theta(\tau-\tau') \nonumber \\
    &\phantom{\delta\mathcal{U}^{\dagger}_{g^2}(0, -\infty)} = -\int_{-\infty(1-i\epsilon)}^{0} d\tau d\tau' \big( H_A^{\dagger}(\tau')H_A^{\dagger}(\tau) + H_A^{\dagger}(\tau')H_B^{\dagger}(\tau) \nonumber \\ 
    &\phantom{\delta\mathcal{U}^{\dagger}_{g^2}(0, -\infty)}\phantom{= -\int_{-\infty(1-i\epsilon)}^{0} d\tau d\tau' \big(} + H_B^{\dagger}(\tau')H_A^{\dagger}(\tau) + H_B^{\dagger}(\tau')H_B^{\dagger}(\tau) \big)\Theta(\tau-\tau'),
        \label{orden2b}
\end{align}
\begin{align}
    & \delta\mathcal{U}_{g}\delta \mathcal{U}^{\dagger}_{g}(0,-\infty)= -i^2\int_{-\infty(1-i\epsilon)}^{0} d\tau d\tau' H_{\text{int}}(\tau)H_{\text{int}}^{\dagger}(\tau') \nonumber \\
        & \phantom{\delta\mathcal{U}_{g}\delta \mathcal{U}^{\dagger}_{g}(0,-\infty)} = -i^2\int_{-\infty(1-i\epsilon)}^{0} d\tau d\tau' \big( H_A(\tau)H_A^{\dagger}(\tau') + H_A(\tau)H_B^{\dagger}(\tau') \nonumber \\ 
        & \phantom{\delta\mathcal{U}_{g}\delta \mathcal{U}^{\dagger}_{g}(0,-\infty)}\phantom{= -i^2\int_{-\infty(1-i\epsilon)}^{0} d\tau d\tau' \big(} + H_B(\tau)H_A^{\dagger}(\tau') + H_B(\tau)H_B^{\dagger}(\tau') \big).
        \label{orden2c}
\end{align}
Moreover, it is possible to determine the matrix elements in \eqref{cot} by inserting the above in \eqref{orderbyorder}. For this purpose, we express the fields in terms of ladder operators, obtained by quantising the free theory. This process can be carried out perturbatively in the coupling.

\subsection{Order \texorpdfstring{$g$}{g}}

From \eqref{orderbyorder}, by inserting \( |\Omega(\tau = -\infty (1-i\varepsilon))\rangle \) and \( |\{ \mathbf{k}_1, \varphi; \mathbf{k}_2, \varphi; \mathbf{k}_3, \varphi; \mathbf{k}_4, \varphi \} (\tau = 0)\rangle = |\prod \mathbf{k}_i\rangle \) on the right and left, respectively, we obtain:
\begin{align}
    \left\langle \prod \mathbf{k}_i|\delta\mathcal{U}_{g}(0, -\infty)|\Omega\right\rangle &= -i\langle \Omega|\a_{\mathbf{k}_1}\a_{\mathbf{k}_2}\a_{\mathbf{k}_3}\int d\tau \big(H_A(\tau) + H_B(\tau)\big)|\Omega\rangle \nonumber \\
    &=-i g_A \int d\tau \langle \Omega|\a_{\mathbf{k}_1}\a_{\mathbf{k}_2}\a_{\mathbf{k}_3}\int_{\mathbf{p}_i}(2\pi)^3\delta^{(3)}\left(\sum_{i}\mathbf{p}_i\right)\mathcal{K}(\mathbf{p_{1}},\mathbf{p_{2}},\mathbf{p_{3}}) \times \nonumber \\ 
    &\phantom{=-}\times \ophi(\mathbf{p}_1,\tau)\ophi(\mathbf{p}_2,\tau)\ophi(\mathbf{p}_3,\tau)|\Omega\rangle + (H_A \leftrightarrow H_B).
    \label{orden1bcot}
\end{align}
Non-trivial contributions are given by products with an equal number of raising and lowering operators, restricting the number of operators to compute after expanding the products in \eqref{ladderOg2}, corresponding to,
\begin{equation}
\a_{\mk_1}\a_{\mk_2}\a_{\mk_3}\ad_{-\mmp_1}\ad_{-\mmp_2}\ad_{-\mmp_3}.
\end{equation}
Additionally, it will be used that the state \( |\Omega\rangle \), at \( \tau = 0 \), is equivalent to $|\Omega\rangle$ defined at \( \tau = -\infty (1 \pm i\varepsilon) \). Furthermore, the normalisation condition is:
\begin{equation}
    \langle \Omega|\Omega\rangle=1.
    \label{eq:norm-cond}
\end{equation}
Thus, equation \eqref{orden1bcot} yields:
\begin{multline}
    \left\langle \prod \mathbf{k}_i |\delta\mathcal{U}_{g}|\Omega \right\rangle =-i g_A \int d\tau \int_{\mathbf{p}_i}(2\pi)^3\delta^{(3)}\left(\sum_{i}\mathbf{p}_i\right)\mathcal{K}(\mathbf{p_{1}},\mathbf{p_{2}},\mathbf{p_{3}}) \times \\ 
    \times u^*_{\mathbf{p}_1}(\tau)u^*_{\mathbf{p}_2}(\tau)u^*_{\mathbf{p}_2}(\tau)\langle \Omega|\a_{\mathbf{k}_1}\a_{\mathbf{k}_2}\a_{\mathbf{k}_3}\ad_{-\p_1}\ad_{-\p_2}\ad_{-\p_3}|\Omega\rangle + (H_A\leftrightarrow H_B). \label{order1cc}
\end{multline}
The above can be reduced by commuting ladder operators,
\begin{align}
    & \a_{\mk_1}\a_{\mk_2}\a_{\mk_3}\ad_{-\mmp_1}\ad_{-\mmp_2}\ad_{-\mmp_3} = \a_{\mk_1}\a_{\mk_2}[\a_{\mk_3},\ad_{-\mmp_1}]\ad_{-\mmp_2}\ad_{-\mmp_3} - \a_{\mk_1}\a_{\mk_2}\ad_{-\mmp_1}\a_{\mk_3}\ad_{-\mmp_2}\ad_{-\mmp_3} \nonumber \\
    & \phantom{\a_{\mk_1}\a_{\mk_2}\a_{\mk_3}\ad_{-\mmp_1}\ad_{-\mmp_2}\ad_{-\mmp_3}} = (2\pi)^3\delta^{(3)}(\mk_3-\mmp_1)\a_{\mk_1}\a_{\mk_2}\ad_{-\mmp_2}\ad_{-\mmp_3} - \cdots
\end{align}
Repeating this procedure for all products of \( a_{\mathbf{k}} \) and $a_{-\mathbf{k}}^{\dagger}$, such that the normal ordering $:a_{\mathbf{k}}^{\dagger} \cdots a_{\mathbf{k}}:$ generates the Dirac delta combination:
\begin{equation}
    \a_{\mk_1}\a_{\mk_2}\a_{\mk_3}\ad_{-\mmp_1}\ad_{-\mmp_2}\ad_{-\mmp_3} = (2\pi)^9\delta^{(3)}(\mk_3-\mmp_1)\delta^{(3)}(\mk_2-\mmp_2)\delta^{(3)}(\mk_1-\mmp_3) + \text{perms}.
    \label{schannelg1}
\end{equation}
Replacing into \eqref{order1cc},  
\begin{multline}
    \left\langle \prod \mathbf{k}_i|\delta\mathcal{U}_{g}|\Omega\right\rangle =i g_A \int d\tau \int_{\mathbf{p}_i}(2\pi)^3\delta^{(3)}\left(\sum_{i}\mathbf{p}_i\right)\mathcal{K}(\mathbf{p_{1}},\mathbf{p_{2}},\mathbf{p_{3}}) \\
    \phantom{=} \times u^*_{\mathbf{p}_1}(\tau)u^*_{\mathbf{p}_2}(\tau)u^*_{\mathbf{p}_3}(\tau)\big[(2\pi)^9\delta^{(3)}(\mk_3-\mmp_1)\delta^{(3)}(\mk_2-\mmp_2)\delta^{(3)}(\mk_1-\mmp_3) + \text{perms} \big] + (H_A\leftrightarrow H_B). \label{order1ccc}
\end{multline}
Hence, by solving the momentum integrals:
\begin{multline}
    \left\langle \prod \mathbf{k}_i|\delta\mathcal{U}_{g}|\Omega\right\rangle =i g_A (2\pi)^3\delta^{(3)}\left(\sum_{i}\mathbf{k}_i\right)\mathcal{K}(\mathbf{k_{1}},\mathbf{k_{2}},\mathbf{k_{3}}) \times \nonumber \\
    \times \int d\tau \left[u^*_{\mathbf{k}_1}(\tau) u^*_{\mathbf{k}_2}(\tau) u^*_{\mathbf{k}_3}(\tau) + \text{perms} \right] + (H_A\leftrightarrow H_B).
\end{multline}
Note that each permutation will lead to the same mode functions integral, absorbed by the symmetry factor.

\subsection{Order \texorpdfstring{$g^{2}$}{g squared}}
\subsubsection{Correlator Discontinuity}\label{cot_lhs}

We then proceed to compute $g^{2}$ contributions given by \cref{orden2a,orden2b,orden2c}. The associated matrix element is
\begin{align}
    & \left\langle \prod \mathbf{k}_i|\delta\mathcal{U}_{g^2}|\Omega\right\rangle = -\langle \Omega | \prod_{n=1}^{4} \a_{\mathbf{k}_{n}}\int d\tau d\tau' \big( H_A(\tau)H_A(\tau') + H_A(\tau)H_B(\tau') \nonumber \\
    &\phantom{\left\langle \prod \mathbf{k}_i(0)|\delta\mathcal{U}_{g^2}(0,-\infty)|\Omega\right\rangle =} + H_B(\tau)H_A(\tau') + H_B(\tau)H_B(\tau') \big) \Theta(\tau-\tau') | \Omega \rangle \nonumber \\
    & \phantom{=} = -\int d\tau d\tau'\langle \Omega|\prod_{n=1}^{4} \a_{\mathbf{k}_{n}} \big(H_A(\tau)H_B(\tau')+H_B(\tau)H_A(\tau')\big)\Theta(\tau-\tau')+\cdots|\Omega\rangle.
\label{cruzado1}
\end{align}
The products \(H_A(\tau)H_A(\tau')\) and \(H_B(\tau)H_B(\tau')\) are expressed using temporal ordering and the symmetry of the temporal integration variables. The $g_Ag_B$ contribution is given by:
\begin{align}
    & \left\langle \prod \mathbf{k}_i | \delta\mathcal{U}_{g_A g_B} | \Omega \right \rangle 
    = -g_A g_B \int d\tau d\tau' \int_{\mathbf{p}_i} \int_{\mathbf{q}_i} \mathcal{K}(\mathbf{p_{1}},\mathbf{p_{2}},\mathbf{p_{3}}) (2\pi)^6 \delta^{(3)}\bigg(\sum_{i=1}^{3} \mathbf{p}_i\bigg) \delta^{(3)}\bigg(\sum_{i=1}^{3} \mathbf{q}_i\bigg) \times \nonumber \\
    & \phantom{ \left\langle \prod \mathbf{k}_i | \delta\mathcal{U}_{g_A g_B} | \Omega \right \rangle 
    = -} \times \bigg( \langle \Omega | \prod_{n=1}^{4} \a_{\mathbf{k}_{n}} \prod_{i=1}^{3} \ophi(\mathbf{p}_{i},\tau) \prod_{j=1}^{3} \ophi(\mathbf{q}_{j},\tau^{\prime}) | \Omega \rangle \Theta(\tau - \tau') \nonumber \\
    & \phantom{ \left\langle \prod \mathbf{k}_i | \delta\mathcal{U}_{g_A g_B} | \Omega \right \rangle 
    = - \times \Big( } + \langle \Omega | \prod_{n=1}^{4} \a_{\mathbf{k}_{n}} \prod_{j=1}^{3} \ophi(\mathbf{q}_{j},\tau^{\prime}) \prod_{i=1}^{3} \ophi(\mathbf{p}_{i},\tau) | \Omega \rangle \Theta(\tau' - \tau) \bigg).
\end{align}
Replacing field operators, we obtain,
\begin{align}
    & \left\langle \prod \mathbf{k}_i | \delta\mathcal{U}_{g_A g_B} | \Omega \right \rangle 
    = - g_A g_B \int d\tau d\tau' \int_{\mathbf{p}_i} \int_{\mathbf{q}_i} 
    \mathcal{K}(\mathbf{p_{1}},\mathbf{p_{2}},\mathbf{p_{3}}) (2\pi)^6 \delta^{(3)}\bigg(\sum_{i=1}^{3} \mathbf{p}_i\bigg) \delta^{(3)}\bigg(\sum_{i=1}^{3} \mathbf{q}_i\bigg) \times \nonumber \\    
    & \times \Bigg( \langle \Omega | \prod_{n=1}^{4} \a_{\mathbf{k}_{n}} \prod_{i=1}^{3} (\a_{\mathbf{p}_{i}} u_{\mathbf{p}_{i}}(\tau) + \a_{-\mathbf{p}_{i}}^{\dagger} u_{\mathbf{p}_{i}}^*(\tau)) \prod_{j=1}^{3} (\a_{\mathbf{q}_{j}} u_{\mathbf{q}_{j}}(\tau') + \a_{-\mathbf{q}_{j}}^{\dagger} u_{\mathbf{q}_{j}}^*(\tau'))
    | \Omega \rangle \Theta(\tau - \tau') \nonumber \\
    &+ \langle \Omega | \prod_{n=1}^{4} \a_{\mathbf{k}_{n}} \prod_{j=1}^{3} (\a_{\mathbf{q}_{j}} u_{\mathbf{q}_{j}}(\tau') + \a_{-\mathbf{q}_{j}}^{\dagger} u_{\mathbf{q}_{j}}^*(\tau')) \prod_{i=1}^{3} (\a_{\mathbf{p}_{i}} u_{\mathbf{p}_{i}}(\tau) + \a_{-\mathbf{p}_{i}}^{\dagger} u_{\mathbf{p}_{i}}^*(\tau)) 
    | \Omega \rangle \Theta(\tau' - \tau) \Bigg).
    \label{ladderOg2}
\end{align}
Some examples of the non-zero contributions are given by:
\begin{subequations}
    \begin{equation}
        \hat{A}_{\mathbf{kpq}} \equiv \a_{\mathbf{k}_1} \a_{\mathbf{k}_2} \a_{\mathbf{k}_3} \a_{\mathbf{k}_4} 
        \a_{\mathbf{p}_1} \a_{-\mathbf{p}_2}^{\dagger} \a_{-\mathbf{p}_3}^{\dagger} 
        \a_{-\mathbf{q}_1}^{\dagger} \a_{-\mathbf{q}_2}^{\dagger} \a_{-\mathbf{q}_3}^{\dagger},
        \label{exa1}
    \end{equation}
    \begin{equation}
        \hat{A}_{\mathbf{kqp}} \equiv \a_{\mathbf{k}_1} \a_{\mathbf{k}_2} \a_{\mathbf{k}_3} \a_{\mathbf{k}_4} 
        \a_{\mathbf{q}_1} \a_{-\mathbf{q}_2}^{\dagger} \a_{-\mathbf{q}_3}^{\dagger} 
        \a_{-\mathbf{p}_1}^{\dagger} \a_{-\mathbf{p}_2}^{\dagger} \a_{-\mathbf{p}_3}^{\dagger}.
        \label{exa2}
    \end{equation}
\end{subequations}
When \eqref{exa1} and \eqref{exa2} are present, which we will refer to as \( \langle \mathbf{k}_i(0)|\delta\mathcal{U}_{g_Ag_B}|\Omega\rangle_{\mathbf{s}} \),
\begin{align}
    &\left\langle \prod \mathbf{k}_i | \delta\mathcal{U}_{g_A g_B} | \Omega \right \rangle 
    = - g_A g_B \int d\tau d\tau' \int_{\mathbf{p}_i} \int_{\mathbf{q}_i} 
    \mathcal{K}(\mathbf{p_{1}},\mathbf{p_{2}},\mathbf{p_{3}}) (2\pi)^6 \delta^{(3)}\bigg(\sum_{i=1}^{3} \mathbf{p}_i\bigg) \delta^{(3)}\bigg(\sum_{i=1}^{3} \mathbf{q}_i\bigg) \times \nonumber \\
    & \phantom{ \langle \mathbf{k}_i | \delta\mathcal{U}_{g_A g_B} | } \times \big[u_{\mathbf{p}_1}(\tau) u^*_{\mathbf{p}_2}(\tau) u^*_{\mathbf{p}_3}(\tau) u^*_{\mathbf{q}_1}(\tau') u^*_{\mathbf{q}_2}(\tau') u^*_{\mathbf{q}_3} (\tau') \Theta(\tau - \tau') \langle \Omega | \hat{A}_{\mathbf{kpq}} | \Omega \rangle \nonumber \\
    & \phantom{ \langle \mathbf{k}_i | \delta\mathcal{U}_{g_A g_B} | \times\big[} + u_{\mathbf{q}_1}(\tau') u^*_{\mathbf{q}_2}(\tau') u^*_{\mathbf{q}_3} (\tau') u^*_{\mathbf{p}_1}(\tau) u^*_{\mathbf{p}_2}(\tau) u^*_{\mathbf{p}_3}(\tau) \Theta(\tau' - \tau) \langle \Omega | \hat{A}_{\mathbf{kqp}} | \Omega \rangle \big].
\end{align}
By applying commutation relations and ordering of ladder operators, we obtain,
\begin{subequations}
    \begin{equation}
        \langle \Omega | \hat{A}_{\mathbf{kpq}} | \Omega \rangle 
        = (2\pi)^{15} \delta^{(3)}(\mathbf{p}_1 - \mathbf{q}_1) \delta^{(3)}(\mathbf{k}_4 - \mathbf{p}_3) \delta^{(3)}(\mathbf{k}_3 - \mathbf{p}_2) \delta^{(3)}(\mathbf{k}_2 - \mathbf{q}_2) \delta^{(3)}(\mathbf{k}_1 - \mathbf{q}_3) + \cdots,
    \end{equation}
    \begin{equation}
        \langle \Omega | \hat{A}_{\mathbf{kqp}} | \Omega \rangle = (2\pi)^{15} \delta^{(3)}(\mathbf{p}_1 - \mathbf{q}_1) \delta^{(3)}(\mathbf{k}_4 - \mathbf{p}_3) \delta^{(3)}(\mathbf{k}_3 - \mathbf{p}_2) \delta^{(3)}(\mathbf{k}_2 - \mathbf{q}_2) \delta^{(3)}(\mathbf{k}_1 - \mathbf{q}_3) + \cdots,
    \end{equation}
\end{subequations}
where the ellipsis refers to non-zero permutations from the ladder operators appearing in \eqref{ladderOg2}. Thus, using the normalisation condition \eqref{eq:norm-cond}, the matrix element reads,
\begin{align}
    & \left\langle \prod \mathbf{k}_i | \delta\mathcal{U}_{g_A g_B} | \Omega \right \rangle 
    = - g_A g_B \int d\tau d\tau' \int_{\mathbf{p}_i} \int_{\mathbf{q}_i} 
    \mathcal{K}(\mathbf{p_{1}},\mathbf{p_{2}},\mathbf{p_{3}}) (2\pi)^6 \delta^{(3)}\bigg(\sum_{i=1}^{3} \mathbf{p}_i\bigg) \delta^{(3)}\bigg(\sum_{i=1}^{3} \mathbf{q}_i\bigg) \times \nonumber \\
    & \phantom{\langle \mathbf{k}_i | \delta\mathcal{U}_{g_A g_B} | \Omega \rangle 
    = - g_A g_B} \times \big[ u_{\mathbf{p}_1}(\tau) u^*_{\mathbf{p}_2}(\tau) u^*_{\mathbf{p}_3}(\tau) u^*_{\mathbf{q}_1}(\tau') u^*_{\mathbf{q}_2}(\tau') u^*_{\mathbf{q}_3} (\tau') \Theta(\tau - \tau') \times \nonumber \\
    & \times \left\{ (2\pi)^{15} \delta^{(3)}(\mathbf{p}_1 - \mathbf{q}_1) \delta^{(3)}(\mathbf{k}_4 - \mathbf{p}_3) \delta^{(3)}(\mathbf{k}_3 - \mathbf{p}_2) \delta^{(3)}(\mathbf{k}_2 - \mathbf{q}_2) \delta^{(3)}(\mathbf{k}_1 - \mathbf{q}_3)  + \langle \Omega | \cdots | \Omega \rangle \right\} \nonumber \\
    & \phantom{\langle \mathbf{k}_i | \delta\mathcal{U}_{g_A g_B} | \Omega \rangle 
    = - g_A g_B \times \big[} + u_{\mathbf{q}_1}(\tau') u^*_{\mathbf{q}_2}(\tau') u^*_{\mathbf{q}_3} (\tau') u^*_{\mathbf{p}_1}(\tau) u^*_{\mathbf{p}_2}(\tau) u^*_{\mathbf{p}_3}(\tau) \Theta(\tau' - \tau) \times \nonumber \\
    & \times \left\{(2\pi)^{15} \delta^{(3)}(\mathbf{p}_1 - \mathbf{q}_1) \delta^{(3)}(\mathbf{k}_4 - \mathbf{p}_3) \delta^{(3)}(\mathbf{k}_3 - \mathbf{p}_2) \delta^{(3)}(\mathbf{k}_2 - \mathbf{q}_2) \delta^{(3)}(\mathbf{k}_1 - \mathbf{q}_3)  + \langle \Omega | \cdots | \Omega \rangle \right\}\big].
    \label{schannelg3}
\end{align}
Solving the integrals and using \eqref{bulk-to-bulk} whilst dropping the ellipsis we obtain:
\begin{multline}
    \left\langle \prod \mathbf{k}_i | \delta\mathcal{U}_{g_A g_B} | \Omega \right \rangle 
    = (2\pi)^3 \delta^{(3)}\big(\mathbf{k}_1 + \mathbf{k}_2 + \mathbf{k}_3 + \mathbf{k}_4 \big) g_A g_B (\mathbf{s} \cdot \mathbf{k}_3 + \mathbf{s} \cdot \mathbf{k}_4 + \mathbf{k}_3 \cdot \mathbf{k}_2) \times \\
    \times \int d\tau d\tau' u^*_{\mathbf{k}_1}(\tau) u^*_{\mathbf{k}_2}(\tau) G_{++}(s, \tau, \tau') u^*_{\mathbf{k}_3}(\tau') u^*_{\mathbf{k}_4}(\tau').
    \label{eq:cot-lhs}
\end{multline}
Note that the previously omitted contributions will yield external momentum permutations, in other words the $t$ and $u$ channels.

Furthermore, the analysis for $\delta \mathcal{U}^{\dagger}$ is similar. Similarly, some examples of the non-zero contributions for the ladder operators are:
\begin{subequations}
    \begin{equation} 
        \hat{A}_{\mathbf{pqk}} \equiv \a_{\mathbf{p}_1} \a_{\mathbf{p}_2} \a_{\mathbf{p}_3}
        \a_{\mathbf{q}_1}\a_{\mathbf{q}_2} \ad_{-\mathbf{q}_3}\ad_{\mathbf{k}_1} \ad_{\mathbf{k}_2} \ad_{\mathbf{k}_3} \ad_{\mathbf{k}_4},
        \label{exa1_1}
    \end{equation}
    \begin{equation}
        \hat{A}_{\mathbf{qpk}} \equiv \a_{\mathbf{q}_1} \a_{\mathbf{q}_2} \a_{\mathbf{q}_3}
        \a_{\mathbf{p}_1} \ad_{-\mathbf{p}_2} \a_{\mathbf{p}_3} \ad_{\mathbf{k}_1} \ad_{\mathbf{k}_2} \ad_{\mathbf{k}_3} \ad_{\mathbf{k}_4}.
        \label{exa2_1}
    \end{equation}
\end{subequations}
Then, the matrix element is:
\begin{multline}
    \left\langle \Omega |\delta\mathcal{U}_{g_A g_B}|\prod\mathbf{k}_i\right\rangle = (2\pi)^3 \delta^{(3)}\big(\mathbf{k}_1 + \mathbf{k}_2 + \mathbf{k}_3 + \mathbf{k}_4 \big) g_A g_B (\mathbf{s} \cdot \mathbf{k}_3 + \mathbf{s} \cdot \mathbf{k}_4 + \mathbf{k}_3 \cdot \mathbf{k}_2) \\
    \times \int d\tau d\tau' u^*_{\mathbf{k}_1}(\tau) u^*_{\mathbf{k}_2}(\tau) G^{*}_{++}(s, \tau, \tau') u^*_{\mathbf{k}_3}(\tau') u^*_{\mathbf{k}_4}(\tau') \label{deltaU2ss}
\end{multline}
Moreover, when analysing the right-hand side of \cref{cot} we will be able to recover the necessary factors in order to rewrite the mode functions in \cref{deltaU2ss} as bulk-to-boundary propagators, in accordance to \cref{bulk-to-bound-plus,bulk-to-bound-minus}.

\subsubsection{Diagrammatic Cuts} \label{cot_rhs}

The matrix element on the right of \cref{cot} can be written as:
\begin{multline}
    \left\langle \prod \mathbf{k}_i| \delta\mathcal{U}_{g}\delta\mathcal{U}^{\dagger}_{g}|\Omega \right\rangle = \Big\langle \prod\mathbf{k}_{i} | \int d\tau d\tau' \bigg( H_A(\tau)H_A^{\dagger}(\tau')  + H_B(\tau)H_B^{\dagger}(\tau') \\
    + H_A(\tau)H_B^{\dagger}(\tau') + (A \leftrightarrow B) \bigg) |\Omega \Big\rangle.
    \label{orden2rhs} 
\end{multline}
The $g_{A}g_{B}$ order contributions are given by:
\begin{align}
    \left\langle \prod \mathbf{k}_i | \delta\mathcal{U}_{g_A}\delta\mathcal{U}^{\dagger}_{g_B} + (A \leftrightarrow B)|\Omega \right\rangle = & \int d\tau d\tau' \left\langle \prod \mathbf{k}_i |\left( H_A(\tau)H_B^{\dagger}(\tau') + (A \leftrightarrow B) \right) |\Omega \right\rangle \nonumber \\
    =&  \int d\tau d\tau' \bigg[\sum_{m=0}^{\infty}\ \int \frac{d^3\mathbf{l}_1}{(2\pi)^3} \cdots \frac{d^3\mathbf{l}_m}{(2\pi)^3} \left\langle \prod \mathbf{k}_i | H_A(\tau)| \mathbf{l}_{m} \right\rangle \times \nonumber \\ 
    & \phantom{\int d\tau}\times \langle \mathbf{l}_{m}|H_B^{\dagger}(\tau')|\Omega \rangle +  (A \leftrightarrow B) \bigg],
    \label{orden2rhs2} 
\end{align}
where a complete basis was introduced. Since \( m \) spans all possible multi-particle states, it is necessary to determine which ones will contribute non-trivially. Furthermore, performing a similar analysis as in \ref{cot_lhs} we arrive at:
\begin{multline}
    \Big\langle \prod \mathbf{k}_i(0) | \delta \mathcal{U}_{g_{A}} \delta \mathcal{U}^{\dagger}_{g_B} | \Omega \Big\rangle = (2\pi)^3 \delta^{(3)}(\mathbf{k}_1 + \mathbf{k}_2 + \mathbf{k}_3 + \mathbf{k}_4) g_A g_B (\mathbf{k}_1 \cdot \mathbf{k}_2 + \mathbf{k}_1 \cdot \mathbf{s} + \mathbf{s} \cdot \mathbf{k}_2) \\
    \times \int d\tau d\tau' \left(u^*_{\mathbf{k}_1}(\tau) u^*_{\mathbf{k}_2}(\tau) u_{\mathbf{s}}(\tau) u^*_{\mathbf{s}}(\tau') u^*_{\mathbf{k}_3}(\tau') u^*_{\mathbf{k}_4}(\tau') \right) + (A \leftrightarrow B) + \text{perms}.
    \label{eq:cot-rhs}
\end{multline}
Furthermore, bringing together \cref{eq:cot-lhs} and \cref{eq:cot-rhs} whilst multiplying both sides by the overall late-time factor,
\begin{equation}
    u_{\mathbf{k}_{1}}(\tau_{f})u_{\mathbf{k}_{2}}(\tau_{f})u_{\mathbf{k}_{3}}(\tau_{f})u_{\mathbf{k}_{4}}(\tau_{f}),
\end{equation}
will make each mode function appears in the combination $u^{\ast}_{\mathbf{k}_{i}}(\tau)u_{\mathbf{k}_{i}}(\tau_{f})$. By \cref{bulk-to-bound-plus} and \cref{bulk-to-bound-minus}, they correspond to bulk-to-boundary propagators. Moreover, to \cref{eq:cot-rhs} we insert the unity in the form $u_{\mathbf{s}}(\tau_{f})u^{\ast}_{\mathbf{s}}(\tau_{f})/u_{\mathbf{s}}(\tau_{f})u^{\ast}_{\mathbf{s}}(\tau_{f})$. Such term will provide the correct factors in order to write it in terms of bulk-to-boundary propagators, the compensating factor $|u_{\mathbf{s}}(\tau_{f})|$ is used in addition to a delta function in order to write $f(s,s')$ defined in \cref{def-f-factor}. The final expression is equivalent to the cutting rule shown in section \ref{sec:toy-model}.

\section{Time Derivatives on Internal Propagators}

In this section, we show how time derivatives act on bulk-to-bulk propagators under temporal integrals. For this, we only consider the $G_{++}$ case, as $G_{--}$ will be the complex conjugate of the result. Moreover, the propagators $G_{+-}$ and $G_{-+}$ do not require this analysis, since there are no Heaviside step functions present.

\subsection{Single Derivative}\label{sec:single-deriv}

Consider the following integral:
\begin{align}
    & \int_{\tau_0}^{\tau_f} d\tau f(\tau) \int_{\tau_0}^{\tau_f} d\tau' g(\tau') \partial_{\tau_1} G_{++}(k, \tau_1, \tau_2) \nonumber \\
    & \phantom{ \int_{\tau_0}^{\tau_f} d\tau f(\tau) } = \int_{\tau_0}^{\tau_f} d\tau f(\tau) \int_{\tau_0}^{\tau_f} d\tau' g(\tau') \partial_{\tau} \left[ u_\mathbf{k}(\tau) u^{*}_\mathbf{k}(\tau') \theta(\tau-\tau') + u^{*}_\mathbf{k}(\tau) u_\mathbf{k}(\tau') \theta(\tau' - \tau) \right] \nonumber \\
    & \phantom{ \int_{\tau_0}^{\tau_f} d\tau f(\tau) }= \int_{\tau_0}^{\tau_f} d\tau f(\tau) \int_{\tau_0}^{\tau_f} d\tau' g(\tau') \Big[ \partial_{\tau} u_\mathbf{k}(\tau) u^{*}_\mathbf{k}(\tau') \theta(\tau-\tau') + u_\mathbf{k}(\tau) u^{*}_\mathbf{k}(\tau') \partial_{\tau} \theta(\tau-\tau') \nonumber \\
    & \phantom{ \int_{\tau_0}^{\tau_f} d\tau f(\tau) } \phantom{ \int_{\tau_0}^{\tau_f} d\tau f(\tau)\int_{\tau_0}^{\tau_f}d } \quad + \partial_{\tau} u^{*}_\mathbf{k}(\tau) u_\mathbf{k}(\tau') \theta(\tau' - \tau) + u^{*}_\mathbf{k}(\tau) u_\mathbf{k}(\tau') \partial_{\tau} \theta(\tau' - \tau) \Big],
\end{align}
where \( f(\tau) \) and \( g(\tau') \) are arbitrary functions. Additionally, note that \( \partial_{\tau} \theta(\tau - \tau') = \delta(\tau - \tau') \) and \( \partial_{\tau} \theta(\tau' - \tau) = -\delta(\tau - \tau') \). Replacing it in the above, we obtain:
\begin{align}
    & \int_{\tau_0}^{\tau_f} d\tau f(\tau) \int_{\tau_0}^{\tau_f} d\tau' g(\tau') \partial_{\tau_1} G_{++}(k, \tau_1, \tau_2) \nonumber \\
    & \phantom{ \int_{\tau_0}^{\tau_f} d\tau f(\tau) }= \int_{\tau_0}^{\tau_f} d\tau f(\tau) \int_{\tau_0}^{\tau_f} d\tau' g(\tau') \Big[ \partial_{\tau} u_\mathbf{k}(\tau) u^{*}_\mathbf{k}(\tau') \theta(\tau - \tau') + u_\mathbf{k}(\tau) u^{*}_\mathbf{k}(\tau') \delta(\tau - \tau') \nonumber \\
    & \phantom{ \int_{\tau_0}^{\tau_f} d\tau f(\tau) } \phantom{ \int_{\tau_0}^{\tau_f} d\tau f(\tau)\int_{\tau_0}^{\tau_f}d\tau' }\quad + \partial_{\tau} u^{*}_\mathbf{k}(\tau) u_\mathbf{k}(\tau') \theta(\tau' - \tau) - u^{*}_\mathbf{k}(\tau) u_\mathbf{k}(\tau') \delta(\tau - \tau') \Big].
    \label{eq:single-deriv-pre-integration}
\end{align}
The mode functions accompanied by the Dirac delta functions are going to cancel after performing the integral, leaving only the Heaviside function dependent terms,
\begin{multline}
    \int_{\tau_0}^{\tau_f} d\tau f(\tau) \int_{\tau_0}^{\tau_f} d\tau' g(\tau') \partial_{\tau_1} G_{++}(k, \tau_1, \tau_2) \\
    = \int_{\tau_0}^{\tau_f} d\tau f(\tau) \int_{\tau_0}^{\tau_f} d\tau' g(\tau') \Big[ \partial_{\tau} u_\mathbf{k}(\tau) u^{*}_\mathbf{k}(\tau') \theta(\tau - \tau') + \partial_{\tau} u^{*}_\mathbf{k}(\tau) u_\mathbf{k}(\tau') \theta(\tau' - \tau) \Big].
\end{multline}
Thus, for a single temporal derivative, $G_{++}$ reads
\begin{equation}
    \partial_{\tau} G_{++}(k, \tau_1, \tau_2) = \partial_{\tau} u_\mathbf{k}(\tau) u^{*}_\mathbf{k}(\tau') \theta(\tau - \tau') + \partial_{\tau} u^{*}_\mathbf{k}(\tau) u_\mathbf{k}(\tau') \theta(\tau' - \tau).
\end{equation}

\subsection{Two Derivatives}\label{sec:two_dt}

Let us analyse the case when two temporal derivatives are present,
\begin{multline}
    \int_{\tau_0}^{\tau_f} d\tau_1 f(\tau_1) \int_{\tau_0}^{\tau_f} d\tau_2 g(\tau_2) \partial_{\tau_1} \partial_{\tau_2} G_{++}(k; \tau_1, \tau_2) \\
    = \int_{\tau_0}^{\tau_f} d\tau f(\tau) \int_{\tau_0}^{\tau_f} d\tau' g(\tau') \partial_{\tau} \partial_{\tau'} \left[ u_\mathbf{k}(\tau) u^{*}_\mathbf{k}(\tau') \theta(\tau - \tau') + u^{*}_\mathbf{k}(\tau) u_\mathbf{k}(\tau') \theta(\tau' - \tau) \right].
\end{multline}
It will be notationally convenient to use \( G_>(k, \tau, \tau') =u_\mathbf{k}(\tau) u^{*}_\mathbf{k}(\tau') \) and \( G_<(k, \tau, \tau') = u^{*}_\mathbf{k}(\tau) u_\mathbf{k}(\tau') \). Thus, a direct computation will yield,
\begin{align}
    & \int_{\tau_0}^{\tau_f} d\tau f(\tau) \int_{\tau_0}^{\tau_f} d\tau' g(\tau') \partial_{\tau} \partial_{\tau'} G_{++}(k, \tau, \tau') \nonumber \\
    & \phantom{\int_{\tau_0}^{\tau_f} d\tau f(\tau)} = \int_{\tau_0}^{\tau_f} d\tau f(\tau) \int_{\tau_0}^{\tau_f} d\tau' g(\tau') \left[ \theta(\tau - \tau') \partial_{\tau} \partial_{\tau'} G_{>} + \theta(\tau' - \tau) \partial_{\tau} \partial_{\tau'} G_{<} \right. \nonumber \\
    & \phantom{\int_{\tau_0}^{\tau_f} d\tau f(\tau)} \phantom{ \int_{\tau_0}^{\tau_f} d\tau f(\tau) \int_{\tau_0}^{\tau_f} } \left. - \delta(\tau - \tau') (\partial_{\tau} - \partial_{\tau'}) (G_{>} - G_{<}) - (G_{>} - G_{<}) \partial_{\tau} \delta(\tau - \tau') \right] \nonumber \\
    & \phantom{\int_{\tau_0}^{\tau_f} d\tau f(\tau)} = \int_{\tau_0}^{\tau_f} d\tau f(\tau) \int_{\tau_0}^{\tau_f} d\tau' g(\tau') \Bigg[ \theta(\tau - \tau') \partial_{\tau} \partial_{\tau'} G_{>} + \theta(\tau' - \tau) \partial_{\tau} \partial_{\tau'} G_{<} \nonumber \\
    & \phantom{\int_{\tau_0}^{\tau_f} d\tau f(\tau)} \phantom{ = \Bigg[ \int_{\tau_0}^{\tau_f} d\tau f(\tau) \int_{\tau_0}^{\tau_f}d\tau' g(\tau') } + \left( u^{*}_{\mathbf{k}}(\tau) u_{\mathbf{k}}^{ \prime}(\tau') - u_{\mathbf{k}}(\tau) u_{\mathbf{k}}^{* \prime}(\tau') \right) \delta(\tau - \tau') \nonumber \\
    & \phantom{\int_{\tau_0}^{\tau_f} d\tau f(\tau)} \phantom{ = \int_{\tau_0}^{\tau_f} d\tau f(\tau) \int_{\tau_0}^{\tau_f}d } + \frac{f'(\tau)}{f(\tau)} g(\tau') \left( u^{*}_{\mathbf{k}}(\tau) u_{\mathbf{k}}(\tau') - u_{\mathbf{k}}(\tau) u^{*}_{\mathbf{k}}(\tau') \right) \delta(\tau - \tau') \Bigg],
\end{align}
where an integration by parts was done to obtain the last expression. Similarly to \cref{eq:single-deriv-pre-integration}, we can simplify the above by performing one integral 
\begin{align}
    & \int_{\tau_0}^{\tau_f} d\tau f(\tau) \int_{\tau_0}^{\tau_f} d\tau' g(\tau') \left[ \theta(\tau - \tau') \partial_{\tau} \partial_{\tau'} G_{>} + \theta(\tau' - \tau) \partial_{\tau} \partial_{\tau'} G_{<} \right] \nonumber \\
    & \phantom{ \int_{\tau_0}^{\tau_f} d\tau f(\tau) \int_{\tau_0}^{\tau_f} d\tau' g(\tau')[ } + \frac{i}{c} \int_{\tau_0}^{\tau_f} d\tau f(\tau) g(\tau) \nonumber \\
    = &  \int_{\tau_0}^{\tau_f} d\tau f(\tau) \int_{\tau_0}^{\tau_f} d\tau' g(\tau') \left[ \theta(\tau - \tau') \partial_{\tau} \partial_{\tau'} G_{>} + \theta(\tau' - \tau) \partial_{\tau} \partial_{\tau'} G_{<} \right] \nonumber \\
    & \phantom{ \int_{\tau_0}^{\tau_f} d\tau f(\tau) \int_{\tau_0}^{\tau_f} d\tau' g(\tau')[ } + \frac{i}{c} \int_{\tau_0}^{\tau_f} d\tau \int_{\tau_0}^{\tau_f} d\tau' f(\tau) g(\tau') \delta(\tau - \tau').
\end{align}
Then, ignoring normalisation constants, double temporal derivatives on bulk-to-bulk propagators reads
\begin{equation}
    \partial_{\tau} \partial_{\tau'} G_{++}(k, \tau, \tau') = \partial_{\tau} u_\mathbf{k}(\tau) \partial_{\tau'} u^{*}_\mathbf{k}(\tau') \theta(\tau - \tau') + \partial_{\tau} u^{*}_\mathbf{k}(\tau) \partial_{\tau'} u_\mathbf{k}(\tau') \theta(\tau' - \tau) + i \delta(\tau - \tau').
\end{equation}

\section{Integrated \texorpdfstring{$V=2$}{V=2} example}
In this section, we provide an example for the $V=2$ cutting rule studied in \cref{sec:toy-model} for a $\phi^{3}$ theory for a massless minimally coupled scalar in de Sitter. The mode function is
\begin{equation}
    G_{>}(k;\tau_{1},\tau_{2}) = \dfrac{H^{2}}{2k^{3}}\left( 1+ik\tau_{1} \right)\left( 1-ik\tau_{2} \right)e^{-ik\left( \tau_{1}-\tau_{2} \right)}.
\end{equation}
Furthermore, we write the one-vertex temporal integral following the convention of \cite{Palma:2025oux} for an $n-$point function,
\begin{equation}
    I(\tau;k_{1},\dots,k_{n}) \equiv \int_{-\infty}^{\tau}\dfrac{d\tau^{\prime}}{{\tau^{\prime}}^{4}}\prod_{j=1}^{n}(1-ik\tau^{\prime})e^{i\tau^{\prime}K},
    \label{eq:I-integral}
\end{equation}
where $K=k_{1}+\dots+k_{n}$. Then, $\mathcal{A}(k_{1},\dotsc,k_{n},a=\pm) \equiv \mathcal{A}_{\pm}$ may be recast as
\begin{equation}
    \mathcal{A}_{+} = -i\dfrac{\lambda_{n}}{H^{4}}\dfrac{H^{2n}}{2^{n}k_{1}^{3}\cdots k_{n}^{3}}I, \qquad \mathcal{A}_{-} = i\dfrac{\lambda_{n}}{H^{4}}\dfrac{H^{2n}}{2^{n}k_{1}^{3}\cdots k_{n}^{3}}I^{\ast},
\end{equation}
where $I^{\ast}$ is the complex conjugate of $I(\tau;k_{1},\dots,k_{n})$. Hence, correlators and barred correlators can be written as
\begin{align}
    \expval{\phi_{k_{1}}\cdots \phi_{k_{n}}}^{\prime} & = \mathcal{A}_{+} + \mathcal{A}_{-} = -\dfrac{\lambda_{n}}{H^{4}}\dfrac{H^{2n}}{2^{n-1}k_{1}^{3}\cdots k_{n}^{3}}\Im(I), \label{eq:integrated-corr-ex} \\
    \expvalbar{\phi_{k_{1}}\cdots \phi_{k_{n}}}^{\prime}_{a} & = \mathcal{A}_{+} - \mathcal{A}_{-} = -\dfrac{i\lambda_{n}}{H^{4}}\dfrac{H^{2n}}{2^{n-1}k_{1}^{3}\cdots k_{n}^{3}}\Re(I),\label{eq:integrated-barred-corr-ex}
\end{align}
where we have stripped the Dirac delta function and $2\pi$ factors. Furthermore, the relevant discontinuity operations of \cref{eq:integrated-corr-ex,eq:integrated-barred-corr-ex} are, for $n=3$ external lines,
\begin{align}
    i\disc{i\expval{\phi_{k_{1}}\phi_{k_{2}}\phi_{s}}^{\prime}}{s} & = -\dfrac{\lambda_{3}H^{2}}{4k_{1}^{3}k_{2}^{3}s^{3}}\left[\Im(I(\tau,k_{1},k_{2},s)) + \Im(I(\tau,k_{1},k_{2},-s)) \right], \label{eq:disc-integrated-corr-ex} \\
    \disc{\expvalbar{\phi_{k_{1}}\phi_{k_{2}}\phi_{s}}^{\prime}_{a}}{s} & = -\dfrac{i\lambda_{3}H^{2}}{4k_{1}^{3}k_{2}^{3}s^{3}}\left[\Re(I(\tau,k_{1},k_{2},s)) - \Re(I(\tau,k_{1},k_{2},-s)) \right].\label{eq:disc-integrated-barred-corr-ex}
\end{align}
The integral of \cref{eq:I-integral} is then,
\begin{equation}
    I(\tau;k_{1},k_{2},k_{3}) = \dfrac{e^{iK\tau}}{3}\left( -\dfrac{1}{\tau^{3}} + \dfrac{iK}{\tau^{2}} + \dfrac{K^{2}-3K_{2}}{2\tau} \right) - \dfrac{\pi}{3}K_{3} + \dfrac{i}{3}K_{3}\text{Ei}\left( -iK\abs{\tau} \right),
\end{equation}
where $K_{N} \equiv \sum_{j=1}^{3}k_{j}^{N}$. The above can be put in the approximate form
\begin{multline}
    I(\tau;k_{1},k_{2},k_{3}) = \dfrac{1}{3}\left( -\dfrac{\cos(K\tau)}{\tau^{3}} - \dfrac{K\sin(K\tau)}{\tau^{2}} + \dfrac{\left(K^{2}-K_{2}\right)\cos(K\tau)}{\tau} \right) - \dfrac{\pi}{6}K_{3} + \Re\left(\mathcal{O}\left(\tau\right)\right) \\
    + \dfrac{i}{3}\left( -\dfrac{\sin(K\tau)}{\tau^{3}} + \dfrac{K\sin(K\tau)}{\tau^{2}} + \dfrac{\left(K^{2}-K_{2}\right)\sin(K\tau)}{\tau} \right) + \dfrac{i}{3}K_{3}\left(\ln\left(-K\tau\right)+\gamma\right) + \Im\left(\mathcal{O}\left(\tau\right)\right),
\end{multline}
where $\gamma$ is the Euler–Mascheroni constant. In the limit $\abs{k_{i}\tau}\ll 1$, the above reads
\begin{equation}
    I(\tau;k_{1},k_{2},k_{3}) =\dfrac{i}{3}K_{3}\ln\left( -\tau K \right) - \dfrac{1}{3\tau^{3}} - \dfrac{K_{2}}{2\tau} + \dfrac{iK\left(K^{2}-9K_{2}\right)}{18} + \dfrac{K_{3}}{3}\left( i\gamma - \dfrac{\pi}{2} \right) +  \mathcal{O}(\tau).
\end{equation}
Thus, the discontinuity of the correlator of \cref{eq:disc-integrated-corr-ex} in the limit $\abs{k_{i}\tau}\ll 1$ will read,
\begin{multline}
    i\disc{i\expval{\phi_{k_{1}}\phi_{k_{2}}\phi_{s}}^{\prime}}{s} = -\dfrac{\lambda_{3}H^{2}}{4k_{1}^{3}k_{2}^{3}s^{3}}\Bigg[ \dfrac{K_{3}}{3}\ln\left( -\tau K\right) + \dfrac{K_{3}-2s^{3}}{3}\ln\left( -\tau \left( K - 2s \right) \right) \\ 
    + \dfrac{K\left(K^{2}-9K_{2}\right)}{18} + \dfrac{\left( K - 2s \right)\left[ \left( K - 2s \right)^{2} - 9K_{2} \right]}{18} + \dfrac{\left(2K_{3}-2s^{3}\right)\gamma}{3} + \mathcal{O}(\tau)\Bigg].
    \label{eq:disc-corr-v-2-example}
\end{multline}
Similarly, for the barred correlator of \cref{eq:disc-integrated-barred-corr-ex} in the same limit is,
\begin{equation}
    \disc{\expvalbar{\phi_{k_{1}}\phi_{k_{2}}\phi_{s}}^{\prime}_{a}}{s} = -\dfrac{i\lambda_{3}H^{2}}{4k_{1}^{3}k_{2}^{3}s^{3}}\left[-\dfrac{\pi}{3}s^{3}+ \mathcal{O}(\tau)\right] =  \dfrac{i\pi\lambda_{3}H^{2}}{12k_{1}^{3}k_{2}^{3}}+\mathcal{O}(\tau).
    \label{eq:disc-corr-v-2-example-barred}
\end{equation}
Therefore, the $V=2$ tree-level $s-$channel discontinuity is provided by the lower-point discontinuity products of \cref{eq:disc-corr-v-2-example,eq:disc-corr-v-2-example-barred} according to \cref{eq:v-2-tree-level-cutting-rule}. Furthermore, the $\vb{ss}^{\prime}$ integral may be recast as,
\begin{equation}
    \int_{\vb{s}\vb{s'}}f(s,s') = \int d^{3}s\, d^{3}s'\, \dfrac{\delta^{(3)}(\vb{s}-\vb{s'})}{u(\tau_{f};s)u^{\ast}(\tau_{f};s')} = \int d^{3}s\left( \dfrac{H^{2}}{2s^{3}}\left( 1 + s^{2}\tau_{f}^{2} \right) \right)^{-1}.
\end{equation}
Such integral can collapse when removing the prime from the correlators, or equivalently, restoring the three-momentum Dirac delta.

\printbibliography

\end{document}